\documentclass[apj]{emulateapj}

\newcommand{\be}{\begin{equation}}
\usepackage{threeparttable}
\usepackage{booktabs}
\newcommand{\ee}{\end{equation}}
\newcommand{\bea}{\begin{eqnarray}}
\newcommand{\eea}{\end{eqnarray}}

\usepackage{graphicx,times}
\usepackage{subfigure}
\usepackage{amsmath}
\usepackage{cases}
\usepackage{bm}
\usepackage{longtable}
\usepackage{hyperref}
\usepackage{epstopdf}
\usepackage{amsmath,bm}
\usepackage{amssymb}
\usepackage{natbib}
\usepackage{morefloats}
\usepackage{multirow}
\usepackage{array}
\usepackage{verbatim}
\usepackage[stable]{footmisc}

\begin{document} 

\title{Linewidth differences of neutrals and ions induced by MHD turbulence}
\author{Siyao Xu\altaffilmark{1}, A. Lazarian\altaffilmark{2} and Huirong Yan\altaffilmark{3}}

\altaffiltext{1}{Department of Astronomy, School of Physics, Peking University, Beijing 100871, China}
\altaffiltext{2}{Department of Astronomy, University of Wisconsin, 475 North Charter Street, Madison, WI 53706, USA}
\altaffiltext{3}{Kavli Institute for Astronomy and Astrophysics, Peking University, Beijing 100871, China; hryan@pku.edu.cn}

\begin{abstract}
We address the problem of the difference of line widths of neutrals and ions observed from molecular clouds and explore whether this difference can arise from the effects of magnetohydrodynamic (MHD) turbulence acting on partially ionized gas. 
Among the three fundamental modes of MHD turbulence, we find fast modes do not contribute to linewidth differences, 
whereas slow modes can have an effect on different line widths for certain parameters. 
We focus on Alfv\'{e}nic component because they contain most of the turbulent energy, 
and consider the damping of this component taking into account both neutral-ion collisions and neutral viscosity. 
We consider different regimes of turbulence corresponding to different media magnetizations and turbulent drivings. 
In the case of super-Alfv\'{e}nic turbulence, when the damping scale of Alfv\'{e}nic turbulence is below $l_A$, 
where $l_A$ is the injection scale of anisotropic GS95-type turbulence, 
the linewidth difference does not depend on the magnetic field strength. 
While for other turbulent regimes, the dependence is present.  
For instance, the difference between the squares of the neutral and ion velocity dispersions 
in strong sub-Alfv\'{e}nic turbulence allows evaluation of magnetic field. 
We discuss earlier findings on the neutral-ion linewidth differences in the literature and compare the expressions for magnetic field we obtain with those published earlier. 
\end{abstract}


\section{Introduction}

The interstellar medium (ISM) is turbulent and magnetized
(see \citealt{Armstrong95, CheL10}). 
As the densest part of ISM, 
molecular clouds are typical environments where many observational phenomena can only be correctly understood 
in the framework of MHD turbulence
(see \citealt{Mckee_Ostriker2007} and references therein).

The turbulence in molecular clouds takes place in partially ionized gas. This makes turbulence more
complicated and induces new effects related to the relative motion of neutrals and ions. 
The difference of neutral and ion line widths has been detected by a number of observations of turbulent molecular clouds
\citep{Houde00a, Houde00b, Lai03}.
The narrower line profiles of ions have been explained by modeling bulk motions of neutral flows 
and their frictions with ions, which are trapped along magnetic field lines
\citep{Houde04}.
This process was interpreted as ambipolar diffusion (see \citealt{Shu92}). 
Moreover, \citet{Houde02} argued the ion-to-neutral line width ratio is related to the orientation of magnetic field, which would open
a new way to study magnetic fields.

An important study that explored the line width differences was by 
\citet{LH08} (henceforth LH08)
who for the first time related these differences to the different turbulence truncation for neutrals and ions. 
In fact, they attributed the difference between the turbulent velocity dispersion spectra of coexistent neutrals and ions
to their different turbulent energy dissipation scales, relating
it to the ambipolar diffusion concept.
Their approach provided for the first time a plausible explanation of the linewidth differences based on the concept of ubiquitous interstellar turbulence. 
Referring to some of the available studies of turbulence in partially ionized gas, LH08 proposed a technique to determine the neutral-ion decoupling 
scale and also the strength of the plane of-the-sky component of magnetic field.
Their work served as a cookbook for the follow-up studies attempting to measure from observations magnetic field strength in molecular clouds 
(see \citealt{Hez10, Hez14}).

While we agree with the interpretation of the neutral-ion linewidth differences as arising from the differential damping of the turbulence cascade, 
however, in the absence of a theoretically justified and numerically tested treatment of MHD turbulence, 
the approach employed in LH08 can be problematic as far as the quantitative conclusions and derived analytical expressions are concerned.
MHD turbulence has been a focus of intensive investigations in the last decade, which change the subject considerably 
(see reviews by \citealt{Lazssrv12, BraL14} and references therein). 
For instance, the anisotropy of Alfv\'{e}nic turbulence is an essential part of MHD turbulent cascade 
(\citealt{GS95}, hereafter GS95) 
and corroborated by the numerical studies by mode decomposition 
\citep{CL02_PRL, CL03, KowL09}.
Ignoring this is known to be erroneous for many applications of turbulence, e.g.
the acceleration and propagation of cosmic rays 
(see \citealt{YL04, Yan15}),
propagation of heat
(see \citealt{Lazarian06}), 
turbulent magnetic reconnection 
\citep{LV99, Eyin13}, 
and other astrophysical problems. 
As the essencial theoretical ingredient, 
we believe that the proper treatment of MHD turbulence is required to address the problem of neutral-ion difference in linewidths.

Studies that treat MHD turbulence in the partially ionized gas include 
\citet{LG01}
and 
\citet{LVC04}.
The former studies focused on only one damping mechanism, e.g. neutral-ion collisions in 
\citet{LG01}
and viscosity in neutrals in 
\citet{LVC04}.
\citet{LG01} dealt with a high-$\beta$ and ion dominated medium, and derived an isotropic damping rate. 
Since the neutral fraction under their consideration is sufficiently small, they argued the cascade of Alfv\'{e}n modes can survive the neutral-ion collisional damping 
and is truncated at the transverse length scale of the proton gyroradius. 
And slow modes are damped at the proton diffusion scale, at which protons can diffuse across an eddy during its turnover time.
\citet{LVC04} 
analyzed turbulence damping in view of the magnetic reconnection, and discussed the effects of neutral viscosity in high Prandtl number turbulence,
i.e. the turbulence in the media with viscosity much larger than resistivity 
(see numerical simulations in \citealt{CLV_newregime}).
These approaches can only be applied in particular media. 
To achieve a comprehensive picture of the damping process, both damping effects should be taken into account, without any restrictions imposed on environment parameters.

Another worry on LH08 is that they adopted the lower envelope of the velocity dispersion spectra to represent the actual three-dimensional (3-D) one. The correspondence between the 
3-D velocity dispersions and minima of the two-dimensional (2-D) ones has later been studied in
\citet{FalLaz10}.
Using a number of MHD simulations with different sonic and Alfv\'{e}nic Mach numbers, 
they found the limitations associated with the procedure of obtaining the 3-D velocity dispersion from observations.
They showed that the discrepancy can be significant in some particular cases (see figure 2 in their work) and therefore
the accuracy of the technique that makes use of these dispersions is also limited.

Our main goal is to explain neutral-ion linewidth differences based on the physically motivated and numerically tested picture of MHD turbulence as a 
composition of the cascades of Alfv\'{e}n, slow and fast modes.
We consider mostly Alfv\'{e}nic modes in this paper
and provide the detailed treatment of their damping as well as decoupling of neutral fluid from the Alfv\'{e}nic motions.
We consider the effects of scale-dependent anisotropy associated with the cascade and find that it is very important for understanding the
physics of neutral-ion interactions at sufficiently small scales. 
To gain a general solution, we will study super- and sub-Alfv\'{e}nic turbulent plasmas separately. To study the damping process in partially 
ionized plasma, we treat ion-electron and neutral fluids separately.
This two-fluid approach is fully
described in 
\citet{Zaqa11} 
and studied numerically by 
\citet{TilBal10}.
We will also compare our results with those in earlier studies, e.g. in 
\citet{LG01}.
This understanding of turbulence we will use to obtain expressions for the difference in neutral and ion squared velocity dispersions in various turbulent regimes. 

We organize this paper as follows. 
We briefly present the scaling laws of MHD turbulence cascade in Section 2. Section 3 contains our investigation on damping process in partially ionized plasma
and explicit damping scales in different turbulence regimes of Alfv\'{e}nic turbulence. 
Section 4 briefly discusses the damping of fast and slow modes.
Following that, in Section 5 we illustrate the effect of distinctive damping scales of ions and neutrals on their different spectral linewidths in various situations. 
Some important results are extracted and summarized in Section 6. 
Section 7 introduces methods to determine magnetic field which can be applicable to both super- and sub-Alfv\'{e}nic turbulent molecular clouds.  
Discussions are given in Section 8. Finally, Section 9 summarizes our results. 

\section{Properties of MHD cascade}
\label{sec: proscal}
It is known that small scale MHD perturbations can be decomposed into Alfv\'{e}n, slow and fast modes.
(see \citealt{Dob80}).
However, there exists an opinion that such a decomposition is not meaningful within the strong compressible MHD turbulence 
due to the high coupling of the modes. 
(see \citealt{Stone98}).
Numerical simulations show that the cascade of Alfv\'{e}n modes can be treated independently due to the weak back-reaction from slow and 
fast modes
\citep{CL03}. 
This also agrees with the theoretical arguments in the pioneering 
\citet{GS95}
study 
(see also \citealt{LG01}). 
The decomposition was usually discussed in literature for the case of a strong background magnetic field with infinitesimal fluctuations.
\citet{CL02_PRL, CL03} dealt with perturbations of substantial amplitude and clearly showed the statistical nature of the procedure. 
Potentially a more accurate decomposition was suggested by 
\citet{KowL10}. 
In addition to Fourier transformations, they introduced wavelet transformations which follow the local magnetic field direction. Their study 
confirmed the results in 
\citet{CL03}. 

Next we first discuss the Alfv\'{e}nic cascade, which is expected to carry most of the MHD turbulence energy  
(see \citealt{CL05}). 

MHD turbulence can be subdivided into super- and sub-Alfv\'{e}nic regimes, determined by the initial turbulent energy relative to the magnetic energy
(see \citealt{Brad13} for more details). 
When we are dealing with super-Alfv\'{e}nic turbulence, i.e. Alfv\'{e}nic Mach number $M_A=V_L/V_A>1$, where $V_L$ is turbulent velocity at the injection scale of turbulence $L$, and $V_A=\frac{B}{\sqrt{4\pi\rho}}$ is 
Alfv\'{e}n velocity\footnote{The notations used in this paper are summarized in Appendix \ref{app:c}. }, 
we have 
(see \citealt{Lazarian06})
\begin{equation}\label{eq: scarsuphy}
  k_\|  \sim k_\perp,
\end{equation}
\begin{equation}
  v_l \sim V_L(\frac{l}{L})^{1/3}, 
\end{equation}
for $l_A<1/k<L$, and 
\begin{equation} \label{eq: supscal}
  k_\|\sim l_A^{-1}(k_\perp l_A)^{2/3}, 
\end{equation}
\begin{equation}
  v_l \sim v_A(\frac{l_\perp}{l_A})^{1/3}=V_L(\frac{l_\perp}{L})^{1/3}, 
\end{equation}
for $1/k<l_A$. Here $l_A=LM_A^{-3}$ is the injection scale of GS95 turbulence, 
where magnetic field becomes dynamically important and turbulence anisotropy develops
\citep{Lazarian06}.

The cascading rate is given by 
\begin{subnumcases}
 {\tau_{cas}^{-1}=\label{eq: supcara}}
k^{2/3}L^{-1/3}V_L,~~~~~~~l_A<1/k<L, \label{eq: supcaraa}\\
k_{\perp}^{2/3}L^{-1/3}V_L,~~~~~~~1/k<l_A. \label{eq: supcarab}
\end{subnumcases}
where the rate given by Eq. \eqref{eq: supcaraa} is a usual Kolmogorov cascading rate for hydrodynamic turbulence, while Eq. \eqref{eq: supcarab} corresponds 
to GS95 cascading of a strong balanced cascade of Alfv\'{e}nic turbulence\footnote{The cascade is balanced when the flux of energy in one direction is equal to the flux in the opposite direction. The theories of imbalanced turbulence are more complicated than the GS95 theory 
(\citealt{LG03, LGS07, ChaB08, PB08}, etc. ).
The model by 
\citet{BL08}
was tested numerically in 
\citet{BL09}.}.

We then turn to sub-Alfv\'{e}nic case ($M_A<1$).
Weak turbulence exists in a range $l_{tr}<1/k<L$, where $l_{tr}=LM_A^2$ is defined as the transition scale from weak to strong turbulence
\citep{LV99}. 
In this paper, we spare the discussion on weak turbulence because of its very limited spatial range.

When we arrive at strong turbulence region with scales smaller than $l_{tr}$, scalings become
\begin{equation}
 k_\|\sim L^{-1}(k_\perp L)^{2/3}M_A^{4/3},
  \label{eq: subscal}
 \end{equation}
 and 
 \begin{equation}
 v_l\sim v_{tr}(\frac{l_\perp}{l_{tr}})^{1/3}=
 V_L (\frac{l_\perp}{L})^{1/3}M_A^{1/3}.
 \end{equation}

The corresponding cascading rate is 
\begin{equation}
\tau_{cas}^{-1}=v_l/l_\perp=k_\perp^{\frac{2}{3}}L^{-\frac{1}{3}}V_LM_A^{\frac{1}{3}}. \label{eq: subcarab}
\end{equation}
We see the cascade proceeds faster at smaller scales.

The other two basic modes in MHD turbulence are fast and slow modes, which are compressible. 
The cascade of slow modes evolves passively and follows the same GS95 scaling as described above
(GS95, \citealt{LG03, CL03}). 
Fast modes have weak coupling with Alfv\'{e}n modes and show isotropic distribution. 
The cascade of fast modes is radial in Fourier space and have scaling relations compatible with acoustic turbulence
\citep{CL02_PRL}.
The cascading rate of fast modes is 
\citep{YL04}, 
\begin{equation}\label{eq: carfm}
  \tau_{cas}^{-1}=(\frac{k}{L})^{\frac{1}{2}}\frac{V_L^2}{V_f}, 
\end{equation}
where $V_f$ is the phase speed of fast modes. 

The damping analysis in the following of the paper will be put on the basis of the properties of MHD turbulence described above.

\section{Damping of Alfv\'{e}nic cascade in partially ionized plasma}
To study turbulence damping we will compare the rate of turbulence cascading with the rate of wave damping.
Our study of the damping process of turbulence is based on the linear analysis of MHD perturbations. 
In this section we first discuss the decoupling scales, then we present the damping scales in different turbulent regimes. 

\subsection{Decoupling scale}
\label{ssec: decscale}
Decoupling can happen when neutrals decouple from ions or contrariwise. 
In mostly neutral medium, neutrals decouple at a larger scale compared to ions. 
In what follows, the decoupling scale we consider 
is the scale where neutrals decouple from ions.
It is determined by the condition that the frequency of Alfv\'{e}n waves is equal to neutral-ion collision frequency $\nu_{ni}$, namely
\begin{equation}
 k_\|V_A=\nu_{ni}.
 \label{eq: dam}
\end{equation}
Here $\nu_{ni}=\gamma_d\rho_{i}$. $\gamma_d$ is the drag coefficient defined in 
\citet{Shu92}.
It is related to the ion-neutral collision frequency $\nu_{in}$ by $\nu_{ni}\rho_{n}=\nu_{in}\rho_{i}$.

{\it Super-Alfv\'{e}nic turbulence}. In the case of super-Alfv\'{e}nic turbulence, turbulence performs isotropic Kolmogorov cascade until reaching $l_A$. Then turbulent eddies get 
more and more elongated along magnetic field lines. 
For the decoupling scale in MHD turbulence regime, 
in combination with Eq. \eqref{eq: supscal}, Eq. \eqref{eq: dam} yields 
\begin{equation}
k_{\text{dec}}= \nu_{ni}^{3/2}L^{1/2}V_L^{-3/2} \sqrt{1+\frac{V_A}{\nu_{ni}l_A}},~~~~~1/k_\text{dec}<l_A.  \label{eq: supsdecom}
\end{equation}
If $k_\text{dec}^{-1}\ll l_A$, $\frac{V_A}{\nu_{ni}l_A}$ becomes much smaller than $1$. Then one can approximately get, 
\begin{equation}\label{eq: supsdecs}
k_{\text{dec}}\sim \nu_{ni}^{3/2}L^{1/2}V_L^{-3/2}.
\end{equation}
Notice that, the component of $k_\text{dec}$ perpendicular to magnetic field is 
\begin{equation}
\label{eq: sudecper}
k_{\text{dec},\perp}=\nu_{ni}^{3/2}L^{1/2}V_L^{-3/2}.
\end{equation}
It shows $k_{\text{dec}}\sim k_{\text{dec},\perp}$ when $k_\text{dec}^{-1}\ll l_A$ due to increasing anisotropy with decreasing scales. 

{\it Sub-Alfv\'{e}nic turbulence}. For the sub-Alfv\'{e}nic case, anisotropy applies to all scales below the injection scale and is prominent in strong turbulence regime. 
In strong turbulence regime, by inserting Eq. \eqref{eq: subscal} in Eq. \eqref{eq: dam}, we can obtain 
\begin{equation}
 k_{\text{dec}}=\nu_{ni}^{3/2}L^{1/2}V_L^{-2}V_A^{1/2} \sqrt{1+\frac{V_A M_A^4}{\nu_{ni}L}}, ~~~~1/k_\text{dec} < l_{tr}, \label{eq: subsdecom}
\end{equation}
Similarly, for a small decoupling scale, the second term in the square root can be neglected. 
Then $k_\text{dec}$ can be approximated by its perpendicular component, that is, 
\begin{equation}
\label{eq: subdec}
\begin{aligned}
k_{\text{dec}}\sim k_{\text{dec},\perp}&=\nu_{ni}^{3/2}L^{1/2}V_L^{-2}V_A^{1/2} \\
                                                           &=\nu_{ni}^{3/2}L^{1/2}V_L^{-3/2}M_A^{-1/2}.
\end{aligned}
\end{equation}
It is different from $k_{\text{dec},\perp}$ in super-Alfv\'{e}nic case (Eq. \eqref{eq: sudecper}) by $M_A^{-1/2}$.

Below the decoupling scale, the interactions between neutrals begin to overtake those with ions. Therefore, neutral fluid starts to evolve along the hydrodynamic 
cascade, while ions still experience the collisional friction with neutrals and MHD cascade proceeds in ions. 
The Alfv\'{e}n waves in ions below the decoupling scale still propagate with the speed $V_{A}=\frac{B}{\sqrt{4\pi\rho}}$, where $\rho$ is total density. Until reaching the 
scale where ions decouple from neutrals and the two species are completely decoupled, then the Alfv\'{e}n speed in ions becomes $V_{Ai}=\frac{B}{\sqrt{4\pi\rho_i}}$, where
$\rho_i$ is ion density. 

\subsection{Damping scale of ions}
\label{subsec: damsi}
In molecular cloud context, the general dispersion relation of Alfv\'{e}n waves incorporating damping through both neutral-ion collisions and neutral viscosity takes the form 
\begin{equation}
\begin{aligned}
\label{eq: gends}
& \omega^3+i(\tau_\upsilon^{-1}+(1+\chi)\nu_{ni})\omega^2-(k^2\cos^2\theta V_{Ai}^2+\chi\tau_\upsilon^{-1}\nu_{ni})\omega \\
& -i(\tau_\upsilon^{-1}+\nu_{ni})k^2\cos^2\theta V_{Ai}^2=0,
\end{aligned}
 \end{equation} 
where $\tau_\upsilon^{-1}\equiv k^2 \nu_{n}$, representing collision frequency of neutrals
\citep{LVC04}.
Here $\nu_{n}$ is the kinematic viscosity in neutrals, 
 \begin{equation}
 \nu_{n}=l_n c_{sn}=\frac{1}{n_n\sigma_{nn}} c_{sn}, ~~~c_{sn}=\sqrt{\gamma\frac{P_n}{\rho_n}}.
 \end{equation}
$\chi$ is defined as $\rho_n/\rho_i$. 
If we set $\tau_\upsilon^{-1}=0$, the above equation recovers the classic dispersion relation of Alfv\'{e}n waves found by 
e.g., \citet{Pidd56, Sol13}.

The complex wave frequency is expressed as $\omega=\omega_R+i\omega_I$. By assuming weak damping, i.e. $|\omega_I|\ll|\omega_R|$, we obtain 
the approximate analytic solution
 \begin{subequations}
 \begin{align}
  &\omega_R^2=\frac{F_1(\tau_\upsilon^{-1}, \nu_{ni})}{F_2(\tau_\upsilon^{-1}, \nu_{ni})}, \\
  &\omega_I=-\frac{\left[\tau_\upsilon^{-1}(\tau_\upsilon^{-1}+(1+\chi)\nu_{ni})+k^2\cos^2\theta V_{Ai}^2\right]  \chi\nu_{ni} }{2[k^2\cos^2\theta V_{Ai}^2+\chi \tau_\upsilon^{-1}\nu_{ni}+(\tau_\upsilon^{-1}+(1+\chi)\nu_{ni})^2]}, \label{eq:2dp}
  \end{align}
 \end{subequations}
 where 
 \begin{equation}
 \begin{aligned}
    F_1(\tau_\upsilon^{-1}, \nu_{ni})=&(k^2\cos^2\theta V_{Ai}^2+\chi \tau_\upsilon^{-1} \nu_{ni} )^2+ \\
    &(\tau_\upsilon^{-1}+(1+\chi)\nu_{ni})(\tau_\upsilon^{-1}+\nu_{ni})k^2\cos^2\theta V_{Ai}^2,  \\
    F_2(\tau_\upsilon^{-1}, \nu_{ni})=&\chi \tau_\upsilon^{-1}\nu_{ni}+k^2\cos^2\theta V_{Ai}^2+ \\
    &(\tau_\upsilon^{-1}+(1+\chi)\nu_{ni})^2.
 \end{aligned}
 \end{equation}

 The absolute value of the imaginary component $|\omega_I|$ is the rate of damping. By equaling the damping rate and the cascading rate of turbulence, we can 
 determine the scale where the cascade of Alfv\'{e}nic turbulence in ions is truncated, i.e., the damping scale of ions $k_\text{dam}^{-1}$.
 And in what follows, we consider "damping scales" to be the damping scales in ions, unless turbulence is damped in strong coupling regime, i.e., motions in both fluids are damped 
simultaneously. 
 
 Derivation of $k_\text{dam}$ directly from Eq. \eqref{eq:2dp} is not easy and the resulting expression may be too complicated to illuminate the physical meaning. 
 We found at weakly coupled limit, $\nu_{ni}, \nu_{in}\ll \omega$, the wave frequency can be reduced to 
\begin{subequations} \label{eq: genwcsol}
 \begin{align}
 & \omega_R^2=k^2 \cos^2\theta V_{Ai}^2 \label{eq: genwcsolr} \\
 & \omega_I=-\frac{\nu_{in}}{2}. \label{eq: damrwd}
  \end{align}
 \end{subequations}
 It signifies neutral viscosity can only influence the behavior of Alfv\'{e}n waves over the scales when neutrals and ions are 
 strongly coupled. 
Hence the damping scales incorporating two damping effects can be calculated in the strong coupling regime. 
Although we adopt the assumption of strong coupling, it turns out the analytical wave frequencies derived at this limit align consistently with 
the real values even at scales smaller than $1/k_\text{dec}$ (see Section \ref{ssec: appli}). 
Therefore, we are able to employ $\omega_I$ for coupled two fluids to derive $k_\text{dam}$ analytically, even for a situation where $k_\text{dam}>k_\text{dec}$.

We first rewrite Eq. \eqref{eq: gends} as 
\begin{equation}
\begin{aligned}
& \omega^3-\omega_k^2\omega+\nu_{ni}\left[ i(1+\chi)\omega^2-\chi \tau_\upsilon^{-1} \omega- i \omega_k^2 \right] \\
&+\tau_\upsilon^{-1} (i\omega^2-i\omega_k^2)=0,
\end{aligned}
\end{equation}
where $\omega_k=k\cos\theta V_{Ai}$.
We consider strongly coupled regime, namely, $\nu_{ni} \gg \omega$. After some simplifications, the above equation becomes 
\begin{equation}
\label{eq: simgends}
(1+\chi)\nu_{ni}\omega^2+i(\chi \tau_\upsilon^{-1} \nu_{ni}+\omega_k^2) \omega -\nu_{ni} \omega_k^2=0.
\end{equation}
The approximate solutions are then 
\begin{subequations} \label{eq: genstratt}
 \begin{align}
& \omega_R^2=k^2 \cos^2\theta V_{A}^2, \label{eq: genstratr} \\
& \omega_I=-\frac{\xi_n}{2}\left(\tau_\upsilon^{-1}+\frac{\omega_k^2}{\nu_{in}}\right). \label{eq: genstrat}
\end{align}
\end{subequations}
We see the real part of the wave frequency (Eq. \eqref{eq: genstratr} and \eqref{eq: genwcsolr}) corresponds to the classic Alfv\'{e}n wave under the 
 weak damping assumption. 
 Given this simplified expression of $\omega_I$, we are able to obtain $k_\text{dam}$ analytically. 
 
 (1) \emph{Super-Alfv\'{e}nic turbulence}~~~~
 In Kolmogorov turbulence regime, the equation between $|\omega_I|$ (Eq. \eqref{eq: genstrat}) and $\tau_{cas}^{-1}$ (Eq. \eqref{eq: supcaraa}) yields
 \begin{equation}
 \label{eq: suphyds2d}
   k_\text{dam}=2^\frac{3}{2}\xi_n^{-\frac{3}{4}}L^{-\frac{1}{4}}V_L^\frac{3}{4}(2\nu_n+\frac{V_{Ai}^2}{\nu_{in}})^{-\frac{3}{4}}, ~~~~~l_A<1/k_\text{dam}<L.
 \end{equation}
 Here $\cos\theta$ in $\omega_k$ is just $k_\|/k$. We adopt the scaling relation given by Eq. \eqref{eq: scarsuphy}.
 
 In MHD turbulence regime, according to the critical balance condition given by GS95, $k_\| V_A \sim k_\perp v_k$, 
i.e. $k_\| V_A \sim  \tau_{cas}^{-1}$.  
Due to $|\omega_R|=k_\| V_A$ (Eq. \eqref{eq: genstratr}), 
the damping condition $\tau_{cas}^{-1}=|\omega_I|$ is equivalent to 
\begin{equation}
\label{eq: gsbc}
    |\omega_R|=|\omega_I|.
\end{equation}
By taking advantage of Eq. \eqref{eq: supscal}, the above equation gives the corresponding damping scale
\begin{subequations}
\label{eq: dssup2dam}
\begin{align}
 & k_{\text{dam},\|}=\frac{-(\nu_n+\frac{V_{Ai}^2}{\nu_{in}})+\sqrt{(\nu_n+\frac{V_{Ai}^2}{\nu_{in}})^2+\frac{8V_A\nu_n l_A}{\xi_n}}}{2\nu_n l_A},  \label{eq: supkpar}\\
 & k_\text{dam}=k_{\text{dam},\|} \sqrt{1+l_A k_{\text{dam},\|}}, ~~~~~~~~~1/k_\text{dam}<l_A.
\end{align}
\end{subequations}
The damping of Alfv\'{e}nic turbulence depends on the angle between $k$ and $B$. Here we assume the parallel and perpendicular components of $k_\text{dam}$ with 
respect to the local magnetic field are related by GS95 scaling relation (Eq. \eqref{eq: supscal} for super-Alfv\'{e}nic turbulence), which describes the scale-dependent 
anisotropy of turbulent eddies, and has been proved by numerical simulations 
\citep{CL02_PRL, CL03}.
The accuracy of this approximation is discussed in Appendix \ref{app:a} by providing more detailed calculations. We find the approximation used here is sufficiently accurate and therefore 
we use it for the rest of the paper.  

(2) \emph{Sub-Alfv\'{e}nic turbulence}~~~~
In strong MHD turbulence regime, by using Eq. \eqref{eq: genstrat} and \eqref{eq: subscal}, 
the condition $|\omega_R|=|\omega_I|$ yields 
\begin{equation}
\begin{split}
\label{eq: subgsbc}
& k_{\text{dam},\|}=\frac{-(\nu_n+\frac{V_{Ai}^2}{\nu_{in}})+\sqrt{(\nu_n+\frac{V_{Ai}^2}{\nu_{in}})^2+\frac{8V_A\nu_n LM_A^{-4}}{\xi_n}}}{2\nu_n L M_A^{-4}}, \\
& k_\text{dam}=k_{\text{dam},\|} \sqrt{1+LM_A^{-4} k_{\text{dam},\|}}, ~~~~~~~~~1/k_\text{dam}<l_{tr}.
\end{split}
\end{equation}
It's very similar to Eq. \eqref{eq: dssup2dam}. In fact, we find Eq. \eqref{eq: subgsbc} can be directly obtained from Eq. \eqref{eq: dssup2dam} by replacing $l_A$ with $L M_A^{-4}$. 
This difference comes from the different cascading rates of super- and sub-Alfv\'{e}nic turbulence (see
Eq. \eqref{eq: supcarab} and \eqref{eq: subcarab}).

The damping scales presented here can be used in a general situation when both neutral-ion collisions and neutral viscosity act on turbulence damping. 
Next, we will first discuss the relative importance of the two damping effects in strongly coupled regime, where neutral viscosity can play a crucial role (Section \ref{sec: rat}), 
Then we study the simplified dispersion relations at limit cases with only one dominant damping effect (Section \ref{sssec: ni} and \ref{sssec: nv}). 
The damping scale introduced in each case has a simpler form and applies to different situations.

\subsection{Relative importance of two damping effects}
\label{sec: rat}
When it comes to astrophysical applications, it's important to evaluate the relative importance of the two damping effects.
The starting point is the damping rate derived from the simplified general dispersion relation in strongly coupled regime (Eq. \eqref{eq: simgends}).
The two terms in Eq. \eqref{eq: genstrat} represent the contributions from the two damping effects. 
 Their ratio, 
 \begin{equation}
 \label{eq: rati}
 r=\frac{\tau_\upsilon^{-1} \nu_{in}}{\omega_k^2}, 
 \end{equation}
 reflects the relative role of neutral viscosity, as compared to neutral-ion collisional damping.
This expression can be further evaluated by taking into account different turbulent regimes. For super-Alfv\'{e}nic turbulence, it becomes 
\begin{subnumcases} 
 {r\approx}
   0.8 (\frac{T}{10\rm K})^{\frac{1}{2}}(\frac{B}{10\rm \mu G})^{-2}\frac{\rho}{10^{-20} \rm g cm^{-3}} \xi_i, \nonumber \\
   ~~~~~~~~~~~~~~~~~~~~~~~~~~~~~~~~~~~~~~~~~~~~~~l_A<1/k<L, \\
   0.4(\frac{T}{10\rm K})^{\frac{1}{2}}(\frac{B}{10\rm \mu G})^{-2}\frac{\rho}{10^{-20} \rm g cm^{-3}} \xi_i (\frac{l}{l_A})^{-\frac{2}{3}}, \nonumber \\
   ~~~~~~~~~~~~~~~~~~~~~~~~~~~~~~~~~~~~~~~~~~~~~~~~~~~~~1/k<l_A. \label{eq: supgam} 
\end{subnumcases}
And for sub-Alfv\'{e}nic turbulence, when $1/k<l_{tr}$, 
we have 
\begin{equation}
 r\approx
   0.4(\frac{T}{10\rm K})^{\frac{1}{2}}(\frac{B}{10\rm \mu G})^{-2}\frac{\rho}{10^{-20} \rm g cm^{-3}} \xi_i (\frac{l}{L})^{-\frac{2}{3}}M_A^{-\frac{8}{3}}. \label{eq: subgam} 
\end{equation}
Here $l=k^{-1}$, and $\cos\theta$ in $\omega_k$ is derived from the scalings presented in Section \ref{sec: proscal}. 
Specifically, $r>1$ indicates neutral 
viscosity is the dominant damping effect. Conversely, it can be safely neglected.
Notice that $r$ increases with decreasing length scales. It can always exceed one at a sufficiently small scale. 
But in fact, due to the assumption of strong coupling, 
the validity of this criteria is restricted to large scales. As is discussed earlier, neutral viscosity has no effect on Alfv\'{e}n waves at small 
scales when neutrals are decoupled. Therefore, the criteria is applicable to determine the relative importance of the two 
damping mechanisms at a certain scale in strong coupling regime, where neutral viscosity can be important. 
By comparing Eq. \eqref{eq: supgam} and \eqref{eq: subgam}, we find with the same $T, B, \rho, \xi_i$ and $l$, sub-Alfv\'{e}nic turbulence is more likely to have $r>1$.
We will explore the applicability of the criteria for selected models of molecular clouds in Section \ref{ssec: appli}. 

The damping scales for a joint damping can be further simplified when only one damping effect is taken into account. 
We again perform the analysis in super- and sub-Alfv\'{e}nic turbulence separately. 

(1) \emph{Super-Alfv\'{e}nic turbulence}~~~~
$k_\text{dam}$ in Kolmogorov turbulence (Eq. \eqref{eq: suphyds2d}) can be reduced to 
\begin{subnumcases}  
 {k_\text{dam}=}
    2^\frac{3}{2}\nu_{ni}^\frac{3}{4}L^{-\frac{1}{4}}V_L^\frac{3}{4}V_A^{-\frac{3}{2}}, ~~~~~~~r<1, \\
    2^\frac{3}{4}\xi_n^{-\frac{3}{4}}\nu_n^{-\frac{3}{4}}L^{-\frac{1}{4}}V_L^\frac{3}{4},~~~~~~r>1. \label{eq: suphyddamnvt}
\end{subnumcases}

In MHD turbulence regime, when $r<1$, $k_{\text{dam},\|}$ (Eq. \eqref{eq: supkpar}) becomes 
\begin{equation}
\label{eq: kparsuprl}
  k_{\text{dam},\|}=\frac{2\nu_{ni}}{V_A}. 
\end{equation}
The resulting $k_\text{dam}$ is
\begin{equation}
 k_\text{dam}=(2\nu_{ni})^\frac{3}{2} L^\frac{1}{2}V_L^{-\frac{3}{2}} \sqrt{1+\frac{V_A}{2\nu_{ni}l_A}}.
\end{equation}
If we take into account $k_\| \ll k_\perp$ and $k\sim k_\perp$ at scales much smaller than $l_A$, 
$k_\text{dam}$ can be approximated by its perpendicular component 
\begin{equation}
\label{eq: supkdamperpgen}
  k_\text{dam}\sim k_{\text{dam},\perp}=(2\nu_{ni})^{\frac{3}{2}}L^{\frac{1}{2}}V_L^{-\frac{3}{2}}.
\end{equation}
Notice that different from total $k_\text{dam}$, its perpendicular component $k_{\text{dam},\perp}$ doesn't have a dependence on $V_A$, or $B$. 
In addition, going back to the approximate expression of $k_\text{dec}$ at $k_\text{dec}^{-1}\ll l_A$ (Eq. \eqref{eq: supsdecs}), we find 
\begin{equation}
\label{eq: damdecrel}
  k_\text{dam} \approx 2^{\frac{3}{2}}k_\text{dec}.
\end{equation}

In the opposite situation when $r>1$, damping scale in Eq. \eqref{eq: dssup2dam} has the form
\begin{subequations}
\begin{align}
   &k_{\text{dam},\|}=\sqrt{\frac{2V_A}{\xi_n \nu_n l_A}}, \\
  & k_\text{dam}=(\frac{2}{\xi_n \nu_n})^\frac{3}{4}L^{-\frac{1}{4}}V_L^\frac{3}{4}\sqrt{1+\left(\frac{\xi_n \nu_n}{2 l_A V_A}\right)^\frac{1}{2}}.
\end{align}
\end{subequations}
With $l_n\ll l_A$ taken into account, $ k_\text{dam}$ is approximated by 
\begin{equation}
\label{eq: supdamnvt}
   k_\text{dam}\approx2^\frac{3}{4}\xi_n^{-\frac{3}{4}} \nu_n^{-\frac{3}{4}}L^{-\frac{1}{4}}V_L^\frac{3}{4}, 
\end{equation}
which is the same as that given by Eq. \eqref{eq: suphyddamnvt}. It shows $k_\text{dam}$ due to neutrals' viscous damping 
is uniform over all scales in super-Alfv\'{e}nic turbulence.

(2) \emph{Sub-Alfv\'{e}nic turbulence}~~~~
In strong turbulence regime, at $r<1$, Eq. \eqref{eq: subgsbc} takes the form 
\begin{equation}
\begin{split}
& k_{\text{dam},\|}=\frac{2\nu_{ni}}{V_A}, \\
& k_\text{dam}=(2\nu_{ni})^\frac{3}{2} L^\frac{1}{2}V_L^{-2}V_A^\frac{1}{2}\sqrt{1+\frac{V_AM_A^4}{2\nu_{ni}L}}.
\end{split}
\end{equation}
$k_\text{dam}$ can be simplified to 
\begin{equation}
\begin{aligned}
\label{eq: subnidamappgen}
k_\text{dam}\sim k_{\text{dam}, \perp}&= (2\nu_{ni})^{\frac{3}{2}}L^{\frac{1}{2}}V_L^{-2}V_A^{\frac{1}{2}} \\
                                                            &=  (2\nu_{ni})^{\frac{3}{2}}L^{\frac{1}{2}}V_L^{-\frac{3}{2}} M_A^{-\frac{1}{2}},
\end{aligned}
\end{equation} 
due to strong anisotropy at small scales. 
Compared with $k_{\text{dam}, \perp}$ in super-Alfv\'{e}nic case (Eq. \eqref{eq: supkdamperpgen}), the only difference is $M_A^{-1/2}$ in above expression. 
It can also be related to $k_\text{dec}$ in Eq. \eqref{eq: subdec} by 
\begin{equation}
k_\text{dam} \approx 2^{\frac{3}{2}}k_\text{dec}.
\end{equation}

Also, we find $k_{\text{dam},\|}$ remains the same as that in super-Alfv\'{e}nic case (Eq. \eqref{eq: kparsuprl}). 
The different turbulence properties between super- and sub-Alfv\'{e}nic turbulence can only affect $k_{\text{dam},\perp}$ in this case.

At $r>1$, Eq. \eqref{eq: subgsbc} is simplified to, 
\begin{equation}
\begin{split}
 &  k_{\text{dam},\|}=\sqrt{\frac{2V_A}{\xi_n \nu_n LM_A^{-4}}}, \\
 & k_\text{dam}=(\frac{2}{\xi_n \nu_n})^\frac{3}{4}L^{-\frac{1}{4}}V_LV_A^{-\frac{1}{4}} \sqrt{1+\left(\frac{\xi_n \nu_n M_A^5}{2LV_L}\right)^\frac{1}{2}}.
\end{split}
\end{equation}
Under the consideration of $l_n \ll L$,  $k_\text{dam}$ can be approximated by 
\begin{equation}
 k_\text{dam}\approx2^\frac{3}{4}\xi_n^{-\frac{3}{4}} \nu_n^{-\frac{3}{4}}L^{-\frac{1}{4}}V_L^\frac{3}{4}M_A^\frac{1}{4}. \\
\end{equation}
 
The damping scales and their approximate expressions presented above are derived by applying the assumption of strong coupling to 
the general dispersion relation (Eq. \eqref{eq: simgends}). 
In most cases, only one damping mechanism plays the dominant role. 
Knowing the relative importance of the frictional and viscous damping, we only need to deal with a simplified dispersion relation considering one damping effect. 
In this spirit, we will perform the analysis with focus on only the dominant damping process in the following subsections. 

\subsection{Damping of Alfv\'{e}nic cascade due to neutral-ion collisions }
\label{sssec: ni}
In the case that damping due to neutral viscosity is negligible, i.e. $r<1$, by setting $\tau_\upsilon^{-1}=0$ in Eq. \eqref{eq: gends},
we can obtain the well-known dispersion relation in the presence of 
only neutral-ion collisions
(see e.g., \citealt{Sol13}),
\begin{equation}\label{eq:dp}
    \omega^3+i(1+\chi)\nu_{ni}\omega^2-k^2\cos^2\theta V_{Ai}^2\omega-i\nu_{ni}k^2\cos^2\theta V_{Ai}^2=0.
\end{equation}    
We obtain the damping rate $|\omega_I|$ by approximately solving the above equation under the weak-damping assumption. The approximate solutions are 
 \begin{subequations} \label{eq: nicsol}
 \begin{align}
 &\omega_R^2=\frac{k^2\cos^2\theta V_{Ai}^2((1+\chi)\nu_{ni}^2+k^2\cos^2\theta V_{Ai}^2)}{(1+\chi)^2\nu_{ni}^2+k^2\cos^2\theta V_{Ai}^2},  \\
 &\omega_I=-\frac{\nu_{ni}\chi k^2\cos^2\theta V_{Ai}^2}{2((1+\chi)^2\nu_{ni}^2+k^2\cos^2\theta V_{Ai}^2)}. \label{eq: anasol} 
 \end{align}
 \end{subequations}
The solutions can be further simplified when neutrals and ions are strongly coupled, i.e., $k<k_\text{dec}$ where $k_\text{dec}$ is described in Section \ref{ssec: decscale},
\begin{subequations} \label{eq: nisc}
 \begin{align}
 &\omega_R^2=k^2\cos^2\theta V_{A}^2, \\
 &\omega_I=-\frac{\xi_n \omega_R^2}{2\nu_{ni}}  \label{eq: anasolsc} .
  \end{align}
 \end{subequations}

By comparing with $\tau_{cas}^{-1}$, we find the ratio $|\omega_I|/\tau_{cas}^{-1}$ depends on both the coupling of two fluids 
and turbulence properties,

(1) \emph{Super-Alfv\'{e}nic turbulence}~~~~
We first consider super-Alfv\'{e}nic turbulence. 
We find (Eq. \eqref{eq: anasolsc}, \eqref{eq: scarsuphy}, \eqref{eq: supscal} and \eqref{eq: supcara})
\begin{subnumcases} 
{\frac{|\omega_I|}{\tau_{cas}^{-1}}\sim}
k^{4/3},~~~~~~~~~~~~~~l_A<k_\text{dec}^{-1}<L, \\
k^{2/3}, ~~~~~~~~~~~~~~k_\text{dec}^{-1}<l_A,
\end{subnumcases} 
in strongly coupled regime, and (Eq. \eqref{eq: damrwd}, \eqref{eq: supcara})
\begin{equation}
 \frac{|\omega_I|}{\tau_{cas}^{-1}}\sim k^{-2/3}
\end{equation}
in weakly coupled regime.
It means the damping may take place either in strongly coupled regime or in the vicinity of $k_\text{dec}$. 
Actually we will show in Section \ref{ssec: appli} that Eq. \eqref{eq: anasolsc} can still serve as a good approximation of the actual value below the decoupling scale.
Therefore we can use Eq. \eqref{eq: anasolsc} to calculate the damping scale. 

By equalizing $|\omega_I|$ (Eq. \eqref{eq: anasolsc}) and $\tau_{cas}^{-1}$ (Eq. \eqref{eq: supcara}), the damping scale is given by 
\begin{subnumcases} 
{k_\text{dam}=\label{eq: supkdam}} 
2^{\frac{3}{2}}\nu_{ni}^{\frac{3}{4}}\xi_n^{-\frac{3}{4}}L^{-\frac{1}{4}}V_L^{\frac{3}{4}}V_A^{-\frac{3}{2}}, \nonumber \\
~~~~~~~~~~~~~~~~~~~~~~~~~~~~~~~~l_A<1/k_\text{dam}<L, \label{eq: supkdama} \\
(\frac{2\nu_{ni}}{\xi_n})^{\frac{3}{2}}L^{\frac{1}{2}}V_L^{-\frac{3}{2}}\sqrt{1+\frac{\xi_nV_A}{2\nu_{ni}l_A}}, \nonumber \\
~~~~~~~~~~~~~~~~~~~~~~~~~~~~~~~~~~~~~~~~1/k_\text{dam}< l_A, \label{eq: supkdamb} 
\end{subnumcases}
where $\xi_n=\rho_n/\rho$.
The scaling relations between $k_\perp$ and $k_\|$ we use are taken from Eq. \eqref{eq: scarsuphy} for $[L, l_A]$ and Eq. \eqref{eq: supscal} for $k^{-1}< l_A$.
The perpendicular component of $k_\text{dam}$ in Eq. \eqref{eq: supkdamb} is
\begin{equation}
\label{eq: supkdamperp}
  k_{\text{dam}, \perp}=(\frac{2\nu_{ni}}{\xi_n})^{\frac{3}{2}}L^{\frac{1}{2}}V_L^{-\frac{3}{2}}, 
\end{equation}
which is independent of magnetic field strength. 

(2) \emph{Sub-Alfv\'{e}nic turbulence}~~~~
We then move to sub-Alfv\'{e}nic turbulence. 
When $k_\text{dec}$ is in strong turbulence regime, $|\omega_I|/\tau_{cas}^{-1}$ in strong and weak coupling regimes are 
(Eq. \eqref{eq: anasolsc}, \eqref{eq: damrwd}, \eqref{eq: subcarab}, \eqref{eq: subscal})
\begin{subnumcases} 
{\frac{|\omega_I|}{\tau_{cas}^{-1}}\sim}
k^{2/3},~~~~~~~~~~~~~~k_\text{dec}^{-1}\ll k^{-1}<l_{tr}, \\
k^{-2/3}, ~~~~~~~~~~~k^{-1}\ll k_\text{dec}^{-1}.
\end{subnumcases} 
Analogously, we equate $|\omega_I|$ (Eq. \eqref{eq: anasolsc}) and $\tau_{cas}^{-1}$ (Eq. \eqref{eq: subcarab}), in combination with Eq. \eqref{eq: subscal}, 
and derive
\begin{equation} 
k_\text{dam}=
(\frac{2\nu_{ni}}{\xi_n})^{\frac{3}{2}}L^{\frac{1}{2}}V_L^{-2}V_A^{\frac{1}{2}}\sqrt{1+\frac{\xi_nV_AM_A^4}{2\nu_{ni}L}}.\label{eq: subkdamb} 
\end{equation}
At small scales, it can be approximated by $k_{\text{dam},\perp}$,
\begin{equation}
\label{eq: subnidamapp}
   k_{\text{dam},\perp}=(\frac{2\nu_{ni}}{\xi_n})^{\frac{3}{2}}L^{\frac{1}{2}}V_L^{-2}V_A^{\frac{1}{2}}
                                   =(\frac{2\nu_{ni}}{\xi_n})^{\frac{3}{2}}L^{\frac{1}{2}}V_L^{-\frac{3}{2}}M_A^{-\frac{1}{2}}.
\end{equation}

Another issue needs to be stressed is when solving the dispersion relation (Eq. \eqref{eq:dp}), we find the properties of the solutions depend on the value of $\chi$. 
\citet{Sol13} 
pointed out, 
when $\chi<8$, we can always get a complex wave frequency, while when $\chi>8$, 
there is a interval of parallel wavenumbers $(k_\|^+, k_\|^-)$ where only purely imaginary solutions exist. 
This "cutoff" region has been identified earlier by
\citet{Kulsrud_Pearce}, corresponding to the range of no propagation of Alfv\'{e}n waves.
The physical meaning has been discussed phenomenally, without taking anisotropy of turbulence into account
(see \citealt{KalN98, Mous87}).
It is necessary to reexamine the cutoff region from the point of view of scale-dependent anisotropy. 
By using the scaling relations given in Section \ref{sec: proscal}, we derive the 
full expressions of $k^+$ and $k^-$ for both super- and sub-Alfv\'{e}nic turbulence (see Appendix \ref{app:b}). 

In fact, the "cutoff" region can also be approximately located by setting 
\begin{equation} \label{eq: rieq}
|\omega_R|=|\omega_I|. 
\end{equation}
In strongly coupled regime, from the expressions in Eq. \eqref{eq: nisc}, Eq. \eqref{eq: rieq} gives 
\begin{equation}\label{eq: cbalt}
k_{c,\|}^+=\frac{2\nu_{ni}}{ V_A \xi_n}.
\end{equation}
In weakly coupled regime, the solutions in Eq. \eqref{eq: nicsol} are reduced to Eq. \eqref{eq: genwcsol}. 
And Eq. \eqref{eq: rieq} yields 
\begin{equation}\label{eq: cbaltw}
k_{c,\|}^-=\frac{\nu_{in}}{ 2 V_{Ai} }.
\end{equation}
The maximum ionization fraction required for the existence of "cutoff" wavelengths can also be roughly estimated by setting $k_{c,\|}^+=k_{c,\|}^-$, 
yielding, $\chi_c \sim 16$.

Then we turn to the expressions of $k^\pm_\|$ 
\citep{Sol13}, 
\begin{equation}
\label{eq: sol13}
k^\pm_\|=\frac{\nu_{ni}}{V_{Ai}}\left[\frac{\chi^2+20\chi-8}{8(1+\chi)^3} \pm \frac{\chi^{1/2}(\chi-8)^{3/2}}{8(1+\chi)^3}\right]^{-1/2}.
\end{equation}
They stand when $\chi>8$. At the limit of large $\chi$, the above expressions approximate to 
\begin{subequations} 
 \begin{align}
 & k^+_\| \approx \frac{2\nu_{ni}}{V_{A}}\xi_n^{1/2}, \\
 & k^-_\| \approx 0.6 \frac{\nu_{in}}{V_{Ai}},
  \end{align}
 \end{subequations}
which coincide with Eq. \eqref{eq: cbalt} and \eqref{eq: cbaltw}. It confirms that the "cutoff" set by $|\omega_R|=|\omega_I|$ provides 
a good approximation of nonpropagating region with $\omega_R=0$. 

Meanwhile, as described in Section \ref{subsec: damsi}, $|\omega_R|=|\omega_I|$ in strong coupling and strong MHD turbulence 
regime is equivalent to the damping condition and provides the damping scale. 
Notice unlike $\omega_R$, which has a change of phase speed at $k_\text{dec}$ from $V_A$ to $V_{Ai}$, 
$\tau_{cas}^{-1}$ is only determined by the conditions at $L$ (Section \ref{sec: proscal}). 
Therefore, $k_\text{dam}$ corresponds to the lower limit wavenumber $k_{c}^+$ of the cutoff. 
That explains $k_{c,\|}^+$ in Eq. \eqref{eq: cbalt} is the same as $k_{\text{dam}, \|}$ in Eq. \eqref{eq: kparsuprl} at $\xi_n \sim 1$.

Moreover, the consistency between the expressions of $k^+$ (Appendix \ref{app:b}) and $k_\text{dam}$ indicates the calculation of the cutoff region provides an alternative approach of determining damping scales in the case of neutral-ion collisional damping. We will numerically compare $k^+$ and $k_\text{dam}$ in Section \ref{ssec: appli}.
However, it's worthwhile to mention that the same as the critical balance, this approach only applies when $k_\text{dam}$ is in strong MHD turbulence. 
Another limitation is that it requires a low ionization degree, i.e. $\chi>8$, which is usually true in molecular clouds. 
Although our approach doesn't impose any restriction on $\chi$, we recall that $r$ increases with $\xi_i$ (see Section \ref{sec: rat}). 
It means viscosity of neutrals tends to dominate damping in a highly ionized medium.

The cutoff arises due to the linear interaction between two fluids. The main difference of the damping process from cutoff is the 
involvement of nonlinear turbulence cascade. 
But indeed, we see the correlation between the boundary of the cutoff $k_c^+$ and $k_\text{dam}$. 
The physical reason is that Alfv\'{e}nic turbulence has its Alfv\'{e}n rate ($k_\| V_A$) equal to the eddy turnover rate ($k_\perp v_l$).
The critical balance between the wave-like motions parallel to magnetic field and mixing motions of magnetic field lines in the perpendicular direction 
bridges the linear waves and nonlinear turbulence cascade 
(GS95, \citealt{CLV_incomp, CL03}).
To better seek the physical connections between $k_{c}^+$ and $k_\text{dam}$, in the following of this paper, 
we use the condition $|\omega_R|=|\omega_I|$ to confine the cutoff region $(k_c^+, k_c^-)$, which is also shown by numerical results to have 
marginal difference from the non-propagation region $(k^+, k^-)$ with $\omega_R=0$. 
We list the expressions of $k_c^\pm$ for different turbulence regimes in Appendix \ref{app:b}.

\subsection{Damping of Alfv\'{e}nic cascade due to viscosity of neutrals }
\label{sssec: nv}
When neutral viscosity is the dominant damping effect, i.e. $r>1$, the general dispersion relation (Eq. \eqref{eq: gends}) can be simplified to 
\begin{equation}
 \omega^3+i\tau_\upsilon^{-1}\omega^2-k^2\cos^2\theta V_{Ai}^2\omega-i\tau_\upsilon^{-1}k^2\cos^2\theta V_{Ai}^2=0. \label{eq: nvlu}
 \end{equation} 
The approximation $\nu_{ni}=0$ used here artificially removes the effect of neutral-ion collisions, as well as their coupling. 
Recall that neutral viscosity can only affect Alfv\'{e}n waves when neutrals and ions are coupled (Section \ref{subsec: damsi}). 
Thus Eq. \eqref{eq: nvlu} has a limited applicability and is only used here for obtaining a simplified damping rate, 
\begin{equation}
 |\omega_I|=\tau_\upsilon^{-1}, \label{eq: nvsappb}
\end{equation}
which enables us to achieve a concise expression of $k_\text{dam}$ in this limit case. 

(1) \emph{Super-Alfv\'{e}nic turbulence}~~~~
Following the same method described above, we equalize $|\omega_I|$ (Eq. \eqref{eq: nvsappb}) and $\tau_{cas}^{-1}$ (Eq. \eqref{eq: supcara}). The damping scale of super-Alfv\'{e}nic turbulence 
becomes 
\begin{subnumcases} 
{k_\text{dam}=\label{eq: supnvkd}}
\nu_{n}^{-3/4}L^{-1/4}V_L^{3/4},~~~~~~~~l_A<1/k_\text{dam}<L, \label{eq: supnvkda}  \\
\sqrt{\frac{\sqrt{l_A^{-\frac{4}{3}}+4\nu_{n}^{-1}L^{-\frac{1}{3}}V_L}-l_A^{-\frac{2}{3}}}{2}}\nu_{n}^{-\frac{1}{2}}L^{-\frac{1}{6}}V_L^{\frac{1}{2}},\nonumber \\
~~~~~~~~~~~~~~~~~~~~~~~~~~~~~~~~~~~~~~~~~~~~~~~1/k_\text{dam}< l_A. \label{eq: supnvkdb} 
\end{subnumcases}
If we consider $l_n$ is much smaller than $l_A$, Eq. \eqref{eq: supnvkdb} becomes 
\begin{equation}
\label{eq: supnvapp}
 k_\text{dam}\approx \nu_n^{-\frac{3}{4}}L^{-\frac{1}{4}}V_L^\frac{3}{4},
\end{equation}
the same as Eq. \eqref{eq: supnvkda}.  

(2) \emph{Sub-Alfv\'{e}nic turbulence}~~~~
The damping scale of sub-Alfv\'{e}nic turbulence can be given by 
\begin{equation}
\begin{aligned}
k_\text{dam}=&
\sqrt{\frac{\sqrt{L^{-\frac{4}{3}}M_A^{\frac{16}{3}}+4\nu_{n}^{-1}L^{-\frac{1}{3}}V_LM_A^{\frac{1}{3}}}-L^{-\frac{2}{3}}M_A^{\frac{8}{3}}}{2}}  \\
&\nu_{n}^{-\frac{1}{2}}L^{-\frac{1}{6}}V_L^{\frac{1}{2}}M_A^{\frac{1}{6}}, ~~~~~~~~~~~~~~~~~1/k_\text{dam}< l_{tr}. \label{eq: subnvkdb}
\end{aligned}
\end{equation}
If we take $l_n \ll L$ into account, Eq. \eqref{eq: subnvkdb} can be simplified to 
\begin{equation}
\label{eq: subnvapp}
k_\text{dam} \approx \nu_{n}^{-3/4}L^{-1/4}V_L^{3/4}M_A^{1/4}.
\end{equation}

Based on Eq. \eqref{eq: genwcsol} and \eqref{eq: genstratt}, 
for MHD turbulence regime ($1/k<l_A$) in super-Alfv\'{e}nic turbulence and strong turbulence regime ($1/k<l_{tr}$) in sub-Alfv\'{e}nic turbulence, 
the cutoff region in this case has boundaries as 
$k_{c}^+=k_\text{dam}$, 
and the same $k_{c,\|}^-$ as in the case of neutral-ion collisional damping (Eq. \eqref{eq: cbaltw}).
The expressions of total $k_{c}^-$ are given by Eq. \eqref{eq: alfkcmsup} and \eqref{eq: alfkcm}.

Table \ref{Tab: Dams} summarizes the damping scales derived in Section \ref{sssec: ni} and \ref{sssec: nv}. 
In comparison with the results in Section \ref{sec: rat} for the corresponding limit situations, we find that in an almost neutral plasma, i.e. $\xi_n \sim 1$, the two approaches come 
to the same results. The expressions in Table \ref{Tab: Dams} provide us a convenient way to evaluate $k_\text{dam}$ in different turbulent regimes. 
We will numerically test the accuracy of these analytical  
$k_\text{dam}$ from Table \ref{Tab: Dams}
in Section \ref{ssec: appli}.
\begin{table*}
\centering
\begin{threeparttable}
\caption[]{Damping scales in different turbulent regimes
}\label{Tab: Dams} \setlength\tabcolsep{2.3pt}
\begin{tabular}{|c|c|c|c|c|c|c|}
 \toprule
    Processes & \multicolumn{3}{c}{Neutral-ion collisions}                      &    \multicolumn{3}{|c|}{Neutral viscosity} \\
      \hline
  Turbulence & \multicolumn{2}{c}{Super} & Sub      &    \multicolumn{2}{|c|}{Super} & Sub     \\
     \hline
  Scales & $[L, l_A]$ & $< l_A$                 & $< l_{tr}$     &     $[L, l_A]$ & $< l_A$ &  $< l_{tr}$  \\
    \hline
  $k_\text{dam}$ & \eqref{eq: supkdama} & \eqref{eq: supkdamb} or \eqref{eq: supcfkp} &  \eqref{eq: subkdamb} or \eqref{eq: subcfkp} & \eqref{eq: supnvkda} & \eqref{eq: supnvkdb} &  \eqref{eq: subnvkdb} \\
\bottomrule
    \end{tabular}
 \end{threeparttable}
\end{table*}

\subsection{Damping scale of neutrals}
\label{subsec: dasneu}
After neutrals decouple from ions, i.e. $k>k_\text{dec}$, hydrodynamic turbulence starts to evolve in neutral fluid, with a cascading rate 
\begin{equation}
 \tau_{cas}^{-1}=k^{2/3}L^{-1/3}V_L.
\end{equation} 
The dissipation mechanisms of turbulence in neutrals also include both neutral-ion collisions and viscosity in neutrals. 
Their relative importance can be determined by the ratio of their 
damping rates, 
\begin{equation}
    r_n=\frac{\tau_\upsilon^{-1} }{\nu_{ni}}=\frac{\nu_{n}}{\nu_{ni}}k^2.
\end{equation}
At the decoupling scale, assuming we are in MHD turbulence regime, $\nu_{ni}$ is equal to $\tau_{cas}^{-1}$ owing to the critical balance.
And $r_n$ can be smaller than $1$ as a result of $l_n \ll k_\text{dec}^{-1}$. 
In this case neutral-ion collisions dominates damping. But since $\tau_{cas}^{-1}/\nu_{ni} \propto k^{2/3}$, hydrodynamic cascade in neutrals at scales $k>k_\text{dec}$ remains unaffected 
by their collisions with ions. 
On the other hand, $r_n$ increases with $k$, so viscosity will become the dominant damping effect at small scales when $r_n>1$. 
By equaling the viscous damping rate and turbulence cascading rate, i.e. $\tau_\upsilon^{-1} =\tau_{cas}^{-1}$, we can get the corresponding viscous scale, 
\begin{equation}
\label{eq: viscl}
 k_\nu=\nu_n^{-\frac{3}{4}}L^{-\frac{1}{4}}V_L^{\frac{3}{4}}.
 \end{equation}
$k_\nu^{-1}$ is the damping scale where the hydrodynamic cascade terminates. In a typical molecular cloud, the damping scale of 
neutrals is usually much smaller than that of ions. 

It is worthwhile to clarify that in a particular situation, damping of Alfv\'{e}nic turbulence can happen before neutrals decouple from ions. In this case the above analysis cannot apply 
since no turbulence exists in neutral fluid at $k>k_\text{dam}$. 
But the turbulence cascade in ions may 
reemerge below the damping scale, since magnetic field perturbations are not suppressed and drive velocity fluctuations in the damped regime
\citep{LVC04}. 
We will not perform a detailed discussion about this situation in this work. 
We refer the reader to  
\citet{LVC04}
for extensive information.

\subsection{Application to models of typical molecular clouds} \label{ssec: appli}
We list the parameters used for typical  super- and sub-Alfv\'{e}nic molecular clouds in Table \ref{Tab: MC}.
$L$ and $V_L$ are chosen to have typical values for ISM. We set different $V_L$ values for the two models of sub-Alfv\'{e}nic molecular clouds. 
$\gamma_d$ value is taken from the calculations in 
\citet{Drai83}. 
Other parameters are taken from 
\citet{LVC04}.
We adopt the mean molecular mass of ions and neutrals as $m_i = 29m_H$ and $m_n = 2.3m_H$ 
(see \citealt{Shu92}). 
We define $\beta=\frac{2c_s^2}{V_A^2}$ for the whole plasma, and $\beta_i=\frac{2c_{si}^2}{V_{Ai}^2}$ for the ion-electron fluid. 
We will term them as \emph{Model 1}, \emph{2} and \emph{3} in the following discussions. 

\begin{table*}
\centering
\begin{threeparttable}
\caption[]{Parameters used in typical molecular clouds.
}\label{Tab: MC} \setlength\tabcolsep{2.3pt}
\begin{tabular}{ccccccccccc}
      \toprule
  &   $L$& $V_L$          & $n$                 & $\xi_i$ & $T$ & $B$          & $\gamma_d$                        & $\beta$ & $\beta_{i}$ & $M_A$ \\
  &   [pc] & [km s$^{-1}$] & [cm$^{-3}$]     &            &  [K]  & [$\mu$ G] & [cm$^3$g$^{-1}$s$^{-1}$]    &              &                   &             \\
      \midrule
 Model 1 &    $30$ & $10$ & $300$ & $1.3\times10^{-3}$ & $20$ & $8.66$ & $3.5\times10^{13}$ & $0.46$ & $9.2\times10^{-5}$ & $13.9$ \\
    \hline
Model 2  & $30$ & $5$   & $300$ & $1.3\times10^{-3}$ & $20$ & $86.6$ & $3.5\times10^{13}$ & $0.0046$ & $9.2\times10^{-7}$ & $0.695$ \\
Model 3 & $30$ & $2$ & $300$ & $1.3\times10^{-3}$ & $20$ & $173.2$ &  $3.5\times10^{13}$ & $0.0012$ & $2.3\times10^{-7}$ & $0.139$ \\
 \bottomrule
    \end{tabular}
 \end{threeparttable}
\end{table*}

We then apply the analytical expressions of damping scales to molecular clouds. 
With the parameters used, we numerically solved the general dispersion relation (Eq. \eqref{eq: gends}), and compare the numerically derived damping scales with the analytical ones 
listed in Table \ref{Tab: Dams}. 

\begin{figure}[htbp]
\centering
 \includegraphics[width=6.8cm]{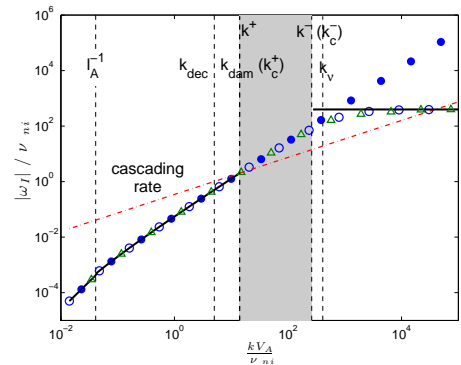}
\caption{Damping rate of Alfv\'{e}n mode in a typical super-Alfv\'{e}nic molecular cloud. Solid line shows the result by numerically 
solving the dispersion relation (Eq. \eqref{eq: gends}). Open circles are analytical solution using Eq. \eqref{eq: anasol}. 
Filled circles represent the simplified solution under strong coupling assumption (Eq. \eqref{eq: anasolsc}). Triangles represent the damping rate with
both neutral viscosity and neutral-ion collisional damping (Eq. \eqref{eq:2dp}). Dash-dotted line is the cascading rate of 
Alfv\'{e}n mode. We also indicate the scales, $1/l_A$, $k_\text{dec}$, $k_\text{dam}$ and $k_\nu$ (vertical dashed lines) using their analytical 
expressions derived in this work. The shaded area corresponds to the cutoff region, defined by $k^{+}$ and $k^{-}$.}
\label{fig:supdam}
\end{figure}

Fig. \ref{fig:supdam} illustrates the normalized damping rate as a function of normalized wave number for \emph{Model 1}.
Open circles are analytical result for neutral-ion collisional damping (Eq. \eqref{eq: anasol}). Its simplification in strong coupling regime is shown 
by filled circles (Eq. \eqref{eq: anasolsc}), which provides a good approximation over a wide range of wave numbers. 
Triangles represent the general analytical solution including two
damping processes (Eq. \eqref{eq:2dp}). 
It is consistent with the numerical damping rate (solid line).
No effect of neutral viscosity can be seen in this case. 
Purely imaginary solutions are omitted in the numerical result, corresponding to the discontinuous interval of the solid line.
The cutoff (shade region) is confined by $k^{+}$ and $k^{-} $ (Eq. \eqref{eq: supcfkptot}), 
and $k^{-}$ overlaps with $k_c^-$ (Eq. \eqref{eq: alfkcmsup}). 
Clearly, $k^{+}$ coincides with the $k_\text{dam}$ calculated using our analytical expression (Eq. \eqref{eq: supkdamb}). 
They both also coincide with the wave number where the turbulence cascade (dash-dotted line) is truncated by damping.
Other scales, $l_A^{-1}$, $k_\text{dec}$ (Eq. \eqref{eq: supsdecom}), $k_\nu$ (Eq. \eqref{eq: viscl}) are also denoted by vertical dashed lines. 

\begin{figure*}[htbp]
\centering
\subfigure[]{
   \includegraphics[width=6.5cm]{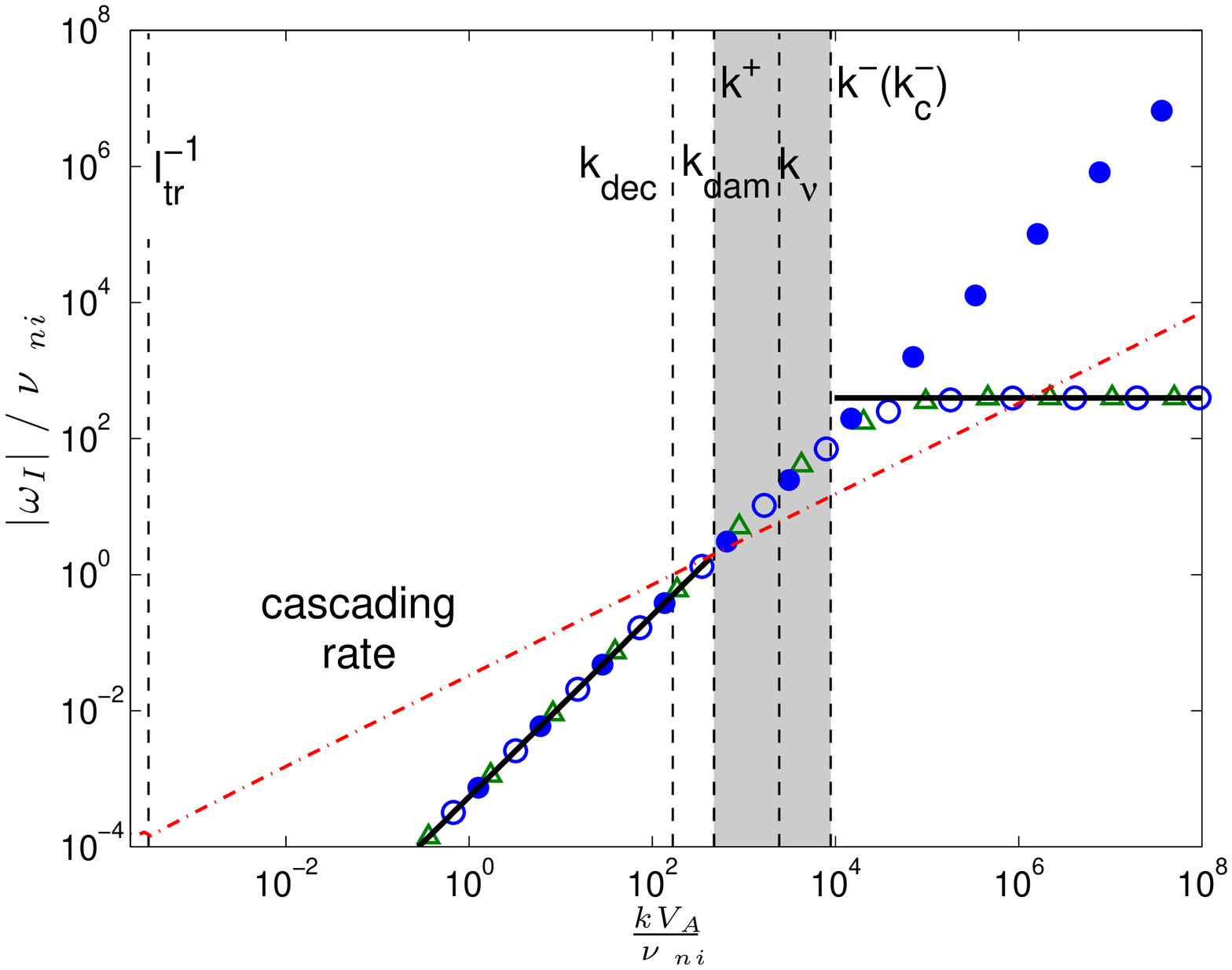}\label{fig:subdam1}}
\subfigure[]{
   \includegraphics[width=6.5cm]{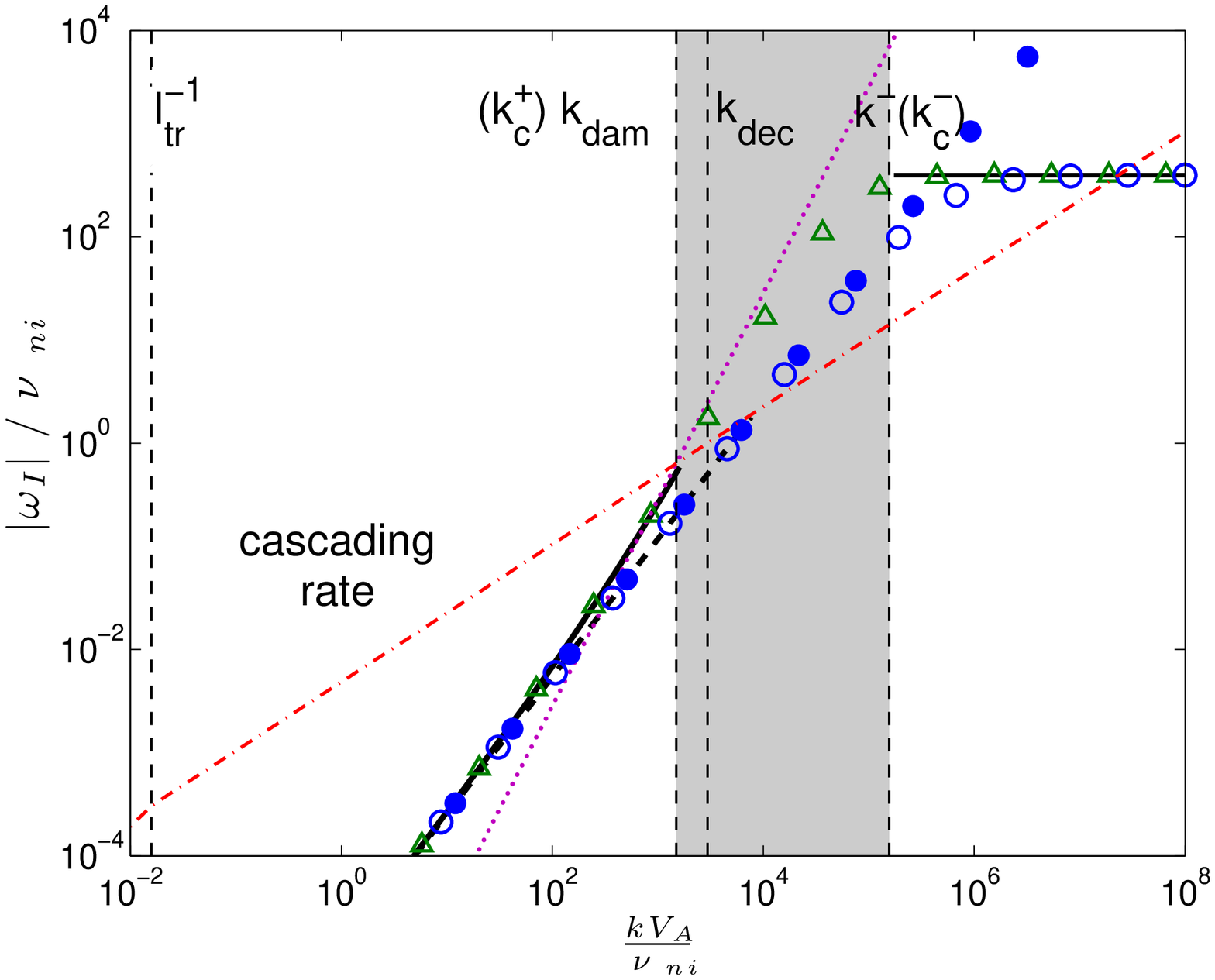}\label{fig:subdam2}}
\subfigure[]{   
   \includegraphics[width=6.5cm]{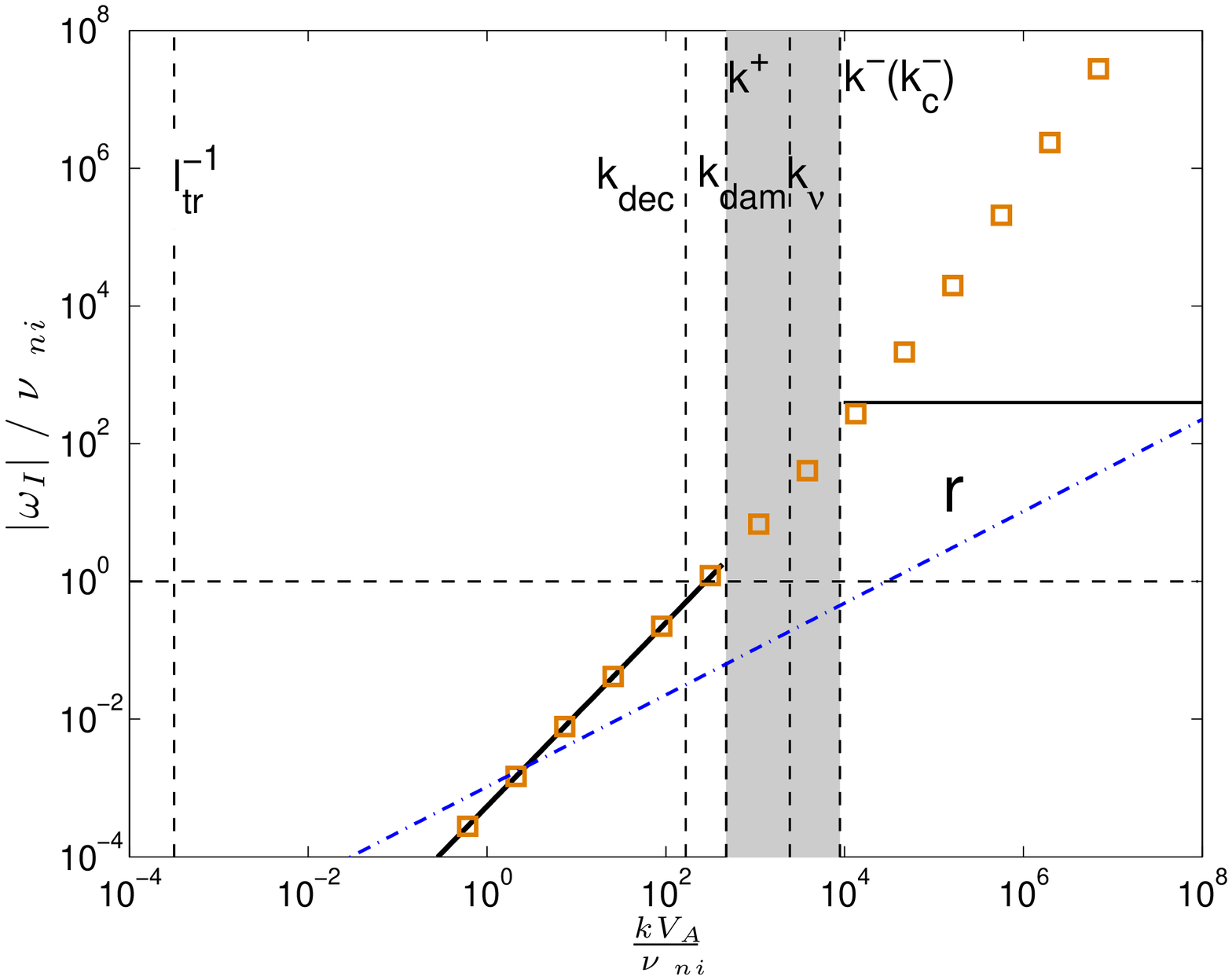}\label{fig:subdam1rat}}
\subfigure[]{
   \includegraphics[width=6.5cm]{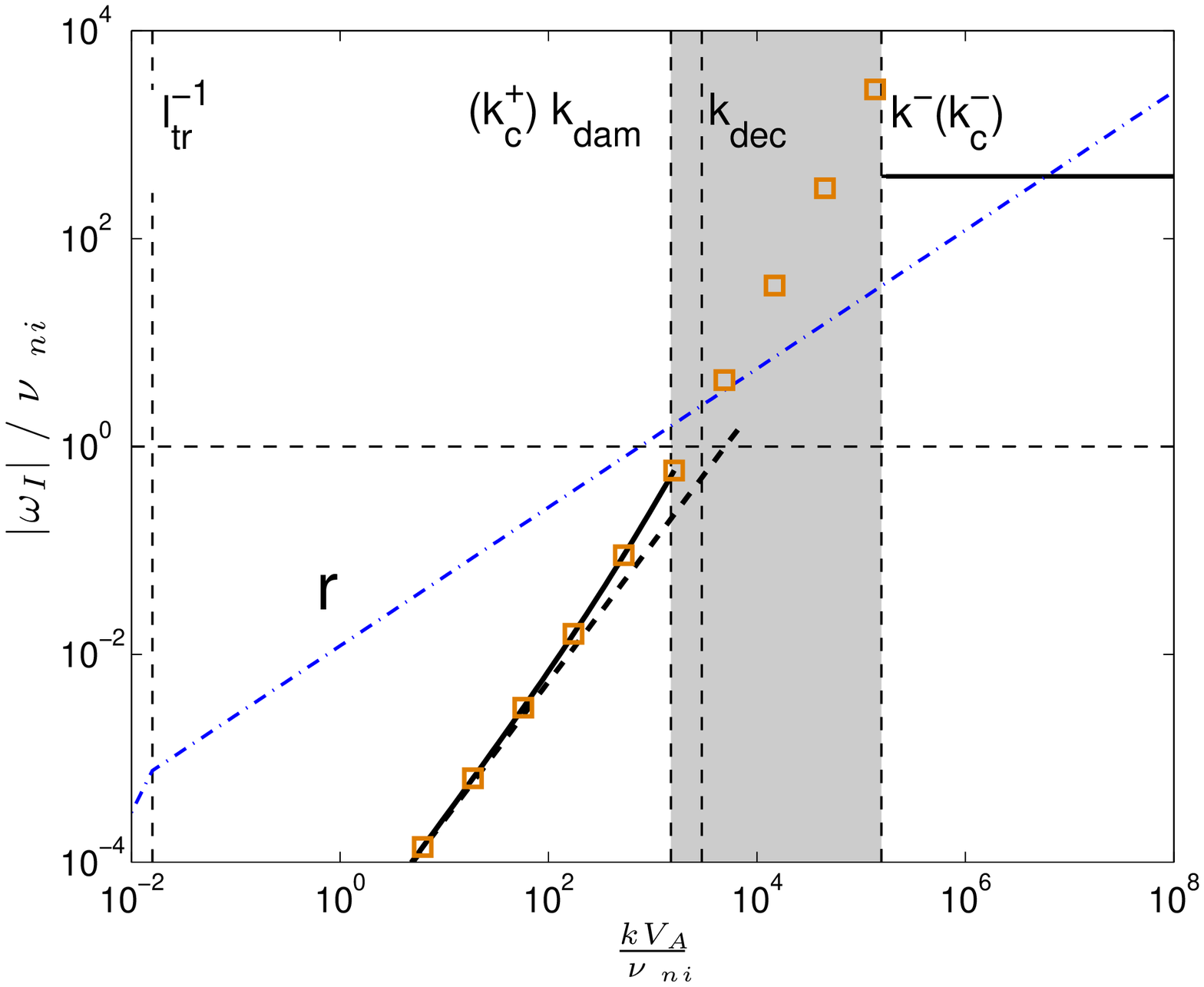}\label{fig:subdam2rat}}
\caption{Same as Fig. \ref{fig:supdam} but for sub-Alfv\'{e}nic molecular clouds, (a) \emph{Model 2}. (b) \emph{Model 3}. $l_{tr}$ is the transition scale 
from weak to strong MHD turbulence regimes. 
The critical scales indicated by vertical dashed lines are from our analytical results. 
In (b), the dashed line represents the numerical damping rate from Eq. \eqref{eq:dp} considering only frictional damping. 
The dotted line corresponds to only neutral viscous damping (Eq. \eqref{eq: nvsappb}). 
(c) and (d) A test of $r$ (Eq. \eqref{eq: rati}). 
The damping rate given Eq. \eqref{eq: genstrat} is represented by squares. The dash-dotted line is $r$. The horizontal dashed line shows the position of one.}
\label{fig:subdam}
\end{figure*}

Fig. \ref{fig:subdam1} and \ref{fig:subdam2} present the results for \emph{Model 2} and \emph{3}. 
Notice the damping both happens in strong MHD turbulence regime. 
Damping in \emph{Model 2} exhibits a similar behavior to \emph{Model 1}. Our analytical $k_\text{dam}$ (Eq. \eqref{eq: subkdamb}) agrees 
well with $k^{+}$ (Eq. \eqref{eq: subcfkp}) and numerical result. 

However, \emph{Model 3} shows distinctive features. First, neutral viscosity dominates damping. 
The analytical damping rate for both damping effects (Eq. \eqref{eq:2dp}, triangles) agrees well with the numerical result (solid line, Eq. \eqref{eq: gends}), 
which is larger than that of neutral-ion collisional damping obtained by numerically solving Eq. \eqref{eq:dp} (dashed line).
The inclusion of neutral viscosity leads to a larger damping scale and a wider cutoff region than the predictions given by neutral-ion collisional damping alone.
The actual boundaries of the cutoff region are $k_\text{dam}$ (i.e. $k_c^+$, Eq. \eqref{eq: subnvapp}) and $k_c^-$ (Eq. \eqref{eq: alfkcm}). $k_c^-$ and $k^{-}$ (Eq. \eqref{eq: subcfkpb}) overlap. 

Furthermore, \emph{Model 3} falls into the situation we discussed at the end of Section \ref{subsec: dasneu}, where
damping happens before neutrals decouple from ions, i.e. $k_\text{dam}< k_\text{dec}$. 
The dotted line shows the damping rate given by Eq. \eqref{eq: nvsappb}. It doesn't exactly align with the actual solution, but can still provide a good approximation (Eq. \eqref{eq: subnvapp}) of $k_\text{dam}$, where the damping rate intersects the cascading rate. 

On the other hand, the difference between \emph{Model 2} and \emph{3} offers us a good opportunity to examine the criteria $r$ we developed in 
Section \ref{sec: rat} (Eq. \eqref{eq: rati}). We replot Fig. \ref{fig:subdam1} and \ref{fig:subdam2} in Fig. \ref{fig:subdam1rat} and \ref{fig:subdam2rat}. Solid and dashed lines still represent the numerical damping rates 
with and without neutral viscosity. 
The squares show the joint contribution from both frictional and viscous damping, given by Eq. \eqref{eq: genstrat}.
It is a good approximation of the total damping rate at large scales. 
The dash-dotted line exhibits $r$ as a function of scales. Neutral viscosity overwhelms neutral-ion collisional effect at the scales where $r$ exceeds one. 
In strong coupling regime, $r$ is smaller than one over all scales in \emph{Model 2}. But $r$ passes one, 
together with the arising of neutral viscosity in \emph{Model 3}.
Therefore, we are convinced that $r$ is capable to benchmark the relative importance of neutral viscosity in turbulence damping.

\section{Damping of compressible modes in partially ionized plasma}
The present paper is mostly devoted to the damping of Alfv\'{e}nic turbulence. 
A comprehensive study on damping of compressible modes in a wide range of ISM conditions will be present in a later paper. 
For completeness, here we include a brief discussion on damping of fast and slow modes.

\subsection{Decoupling scale}
Fast modes are isotropic and have the decoupling scale as 
\begin{equation}\label{eq: fadecs}
   k_\text{dec}=\nu_{ni}/V_A.
\end{equation}
The decoupling scale of slow modes are the same as that of Alfv\'{e}n modes, with varying expressions in different turbulent regimes 
(see Section \ref{ssec: decscale}).
 
\subsection{Damping scale of ions}
We proceed as before for Alfv\'{e}n modes. We first focus on the derivation of 
the dispersion relation for magnetoacoustic modes in a partially ionized two-fluid plasma given by 
\citet{Zaqa11}
(also see \citealt{Soler13}).
We again assume weak damping $|\omega_I|\ll|\omega_R|$ and approximately attain the wave frequencies at limit cases of 
strongly and weakly coupled fluids. 
At the limit of low wave frequency, $\omega \ll \nu_{ni}$, we find 
\begin{subequations} \label{eq: mascgen}
\begin{align}
& \omega_R^2=\frac{1}{2}\left[(c_{s}^2+V_{A}^2)\pm\sqrt{(c_{s}^2+V_{A}^2)^2-4c_{s}^2V_{A}^2\cos^2\theta}\right]k^2, \\
& \omega_I=-\frac{k^2[\xi_nV_A^2(c_s^2k^2-\omega_R^2)+\xi_i c_s^2\omega_R^2]}{2\nu_{ni}[k^2(c_s^2+V_A^2)-2\omega_R^2]}. \label{eq: acdrgen}
\end{align}
 \end{subequations}
The sound speed $c_s$ used here is defined as 
\begin{equation}
c_s=\sqrt{c_{si}^2\xi_i+c_{sn}^2\xi_n}=\sqrt{\frac{\gamma kT (2n_i+n_n)}{\rho}}.
\end{equation}
The classic magnetosonic waves are regained in the real part, with the sign $\pm$ corresponding to fast and slow waves respectively.
In a low-$\beta$ ($\beta \lesssim 1$) environment, as commonly seen in molecular clouds, Eq. \eqref{eq: mascgen} reduces to
\begin{subequations} \label{eq: scfadrt}
\begin{align}
& \omega_R^2=V_A^2 k^2, \\
& \omega_I=-\frac{\xi_nV_A^2k^2}{2\nu_{ni}}, \label{eq: scfadr}
\end{align}
\end{subequations}
for fast modes, and 
\begin{subequations}  \label{eq: scsldrt}
\begin{align}
& \omega_R^2=c_s^2 k^2 \cos^2\theta, \\
& \omega_I=-\frac{\xi_nc_s^2k_\perp^2 }{2\nu_{ni}}, \label{eq: scsldr}
\end{align}
\end{subequations}
for slow modes.

At the converse limit $\omega \gg \nu_{ni}$, we get the approximate analytical solutions
\begin{subequations} \label{eq: mawcgen}
\begin{align}
& \omega_R^2=\frac{1}{2}\left[(c_{si}^2+V_{Ai}^2)\pm\sqrt{(c_{si}^2+V_{Ai}^2)^2-4c_{si}^2V_{Ai}^2\cos^2\theta}\right]k^2, \\
& ~~~~~~~~~~~~~~~~~~~~~~~~~~~~~~~~~~~~\omega_I=-\frac{\nu_{in}}{2}. \label{eq: wcfasldr}
\end{align}
\end{subequations}
Eq. \eqref{eq: mascgen} and \eqref{eq: mawcgen} are consistent with the earlier results in 
\citet{Ferr88}. 
At low-$\beta$ condition, $|\omega_R|$ in above solution can be simplified to $V_{Ai}k$ for fast modes and $c_{si}k\cos\theta$ for slow modes.  
Given the damping rate, we next explore the damping scales of fast and slow modes. 

(1) \emph{Damping of fast modes}~~~~
The comparison between $|\omega_I|$ (Eq. \eqref{eq: scfadr} and \eqref{eq: wcfasldr}) and $ \tau_{cas}^{-1}$ (Eq. \eqref{eq: carfm}) 
shows 
\begin{subnumcases} 
{\frac{|\omega_I|}{\tau_{cas}^{-1}}\sim}
k^{3/2},~~~~~~~~~~~~~k \ll k_\text{dec} , \\
k^{-1/2},~~~~~~~~~~~ k \gg k_\text{dec}.
\end{subnumcases} 
If the damping condition $|\omega_I|/\tau_{cas}^{-1}=1$ is satisfied in the strongly coupled regime, the corresponding damping scale is given by (Eq. \eqref{eq: acdrgen} and \eqref{eq: carfm})
\begin{equation} \label{eq: fakdsc}
 k_\text{dam}=L^{-\frac{1}{3}}\left(\frac{2\nu_{ni}V_L^2(c_s^2+V_A^2-2V_f^2)}{V_f\left[\xi_n V_A^2(c_s^2-V_f^2)+\xi_i c_s^2V_f^2\right]}\right)^{\frac{2}{3}}, 
\end{equation}
which takes form 
\begin{equation}
  k_\text{dam} = (\frac{2\nu_{ni}}{\xi_n})^{2/3} V_L^{4/3} L^{-1/3} V_A^{-2}, 
\end{equation}
when $\beta$ is small.

\begin{figure*}[htbp]
\centering
\subfigure[]{
   \includegraphics[width=6.5cm]{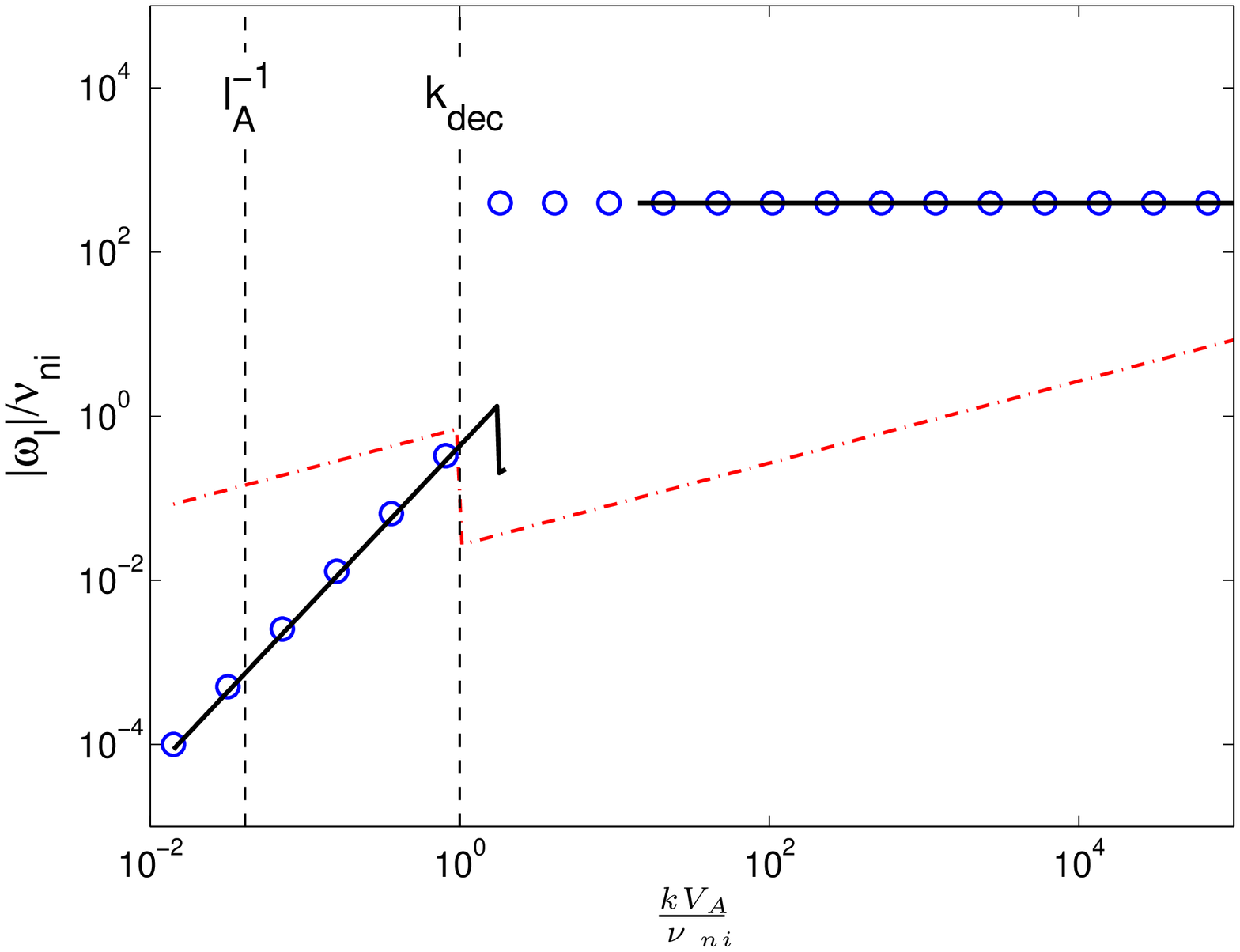}\label{fig:comexa1}}
\subfigure[]{
   \includegraphics[width=6.5cm]{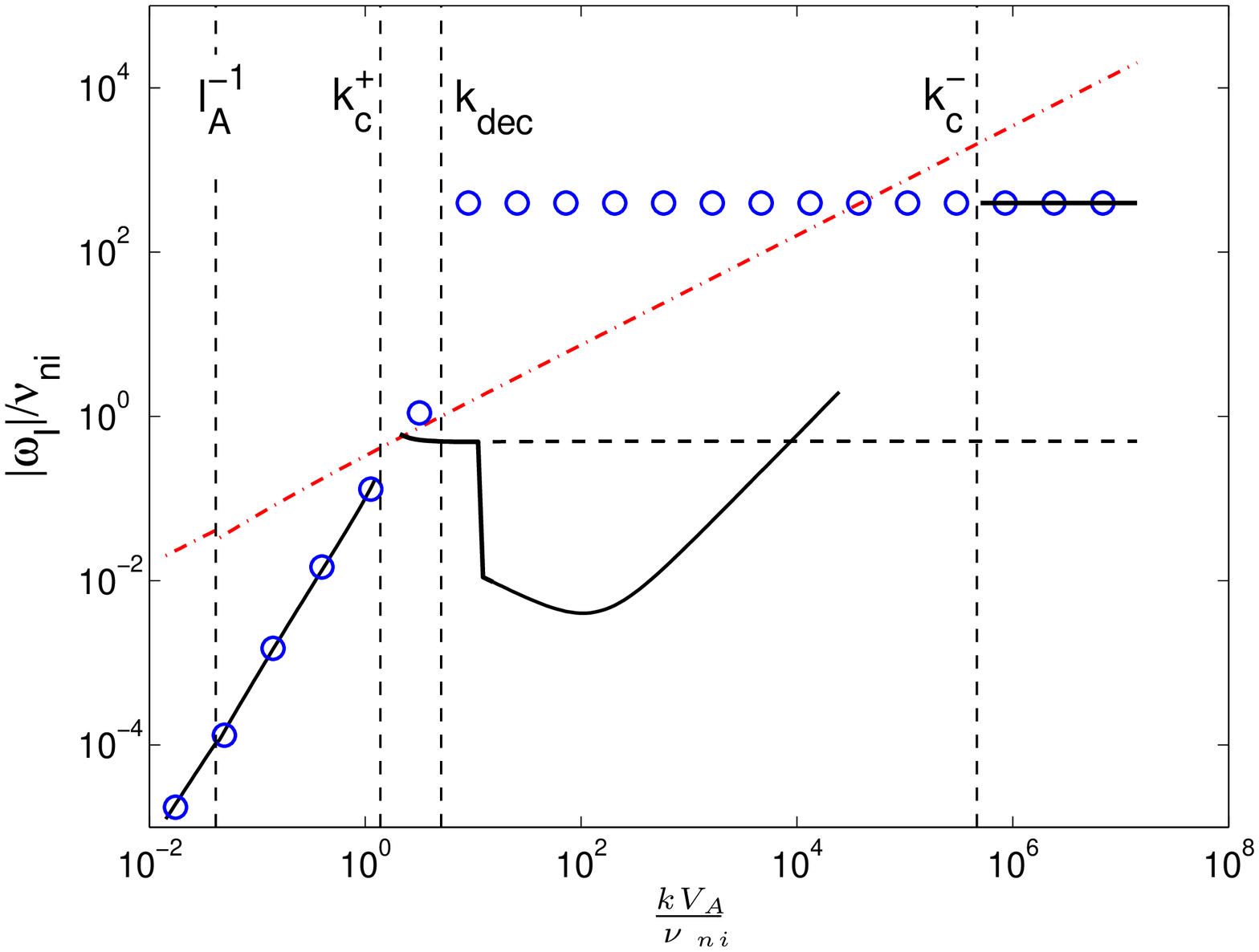}\label{fig:comexa2}}
\caption{Damping rates of (a) fast and (b) slow modes using parameters from \emph{Model 1}. 
Same symbols are used as in Fig. \ref{fig:supdam}. Solid lines are numerical solutions to the dispersion relation 
(equation (57) in \citealt{Zaqa11}).
In Fig. (a), analytical damping rate (open circles) is from Eq. \eqref{eq: scfadr} and \eqref{eq: wcfasldr}.
In Fig. (b), purely imaginary solutions are omitted in the numerical result. 
Solid and dashed lines are "ion" and "neutral" slow modes repectively. Open circles correspond to Eq. \eqref{eq: scsldr} and \eqref{eq: wcfasldr}. The outer boundaries of the two cutoff regions $k_c^\pm$ are given by Eq. \eqref{eq: slcfbdsup}. }
\label{fig:comexa}
\end{figure*}

Using the set of parameters of \emph{Model 1}, Fig. \ref{fig:comexa1} illustrates the damping of the cascade of fast modes. 
The same symbols are used as in Fig. \ref{fig:supdam}. 
The solid line is the numerical damping rate by solving the dispersion relation for compressible modes, equation (57) in
\citet{Zaqa11}. 
Open circles are the damping rate from Eq. \eqref{eq: scfadr} and \eqref{eq: wcfasldr}.
Since $\tau_{cas}^{-1}$ is above $|\omega_I|$ over the entire coupling regime, $k_\text{dam}$ in Eq. \eqref{eq: fakdsc} does not apply in this paradigmatic case. 
But there is a distinctive feature of the cascade of fast modes, that the phase speed $V_f$ is involved in $\tau_{cas}^{-1}$ (Eq. \eqref{eq: carfm}). 
In a low-$\beta$ plasma, $V_f$ experiences a change from $V_A$ in two fluids to $V_{Ai}$ in ions, resulting in a remarkable drop 
of $\tau_{cas}^{-1}$ at $k_\text{dec}$. 
Thus $|\omega_I|$ can take the advantage and get ahead of $\tau_{cas}^{-1}$ at $k_\text{dec}$. 
Therefore, we have $k_\text{dam}=k_\text{dec}$ in this case.

In fact, $k_\text{dam}$ given by Eq. \eqref{eq: fakdsc} can be larger or smaller than $k_\text{dec}$, depending on a certain combination of parameters used.
Therefore, based on both the analytical and numerical results, we come to the conclusion that the damping scale of fast modes is 
\begin{equation}\label{eq: ctnfdf}
  k_\text{dam}=\text{min}(k_\text{dam} (\text{Eq. \eqref{eq: fakdsc}}), k_\text{dec}).
\end{equation}

(2) \emph{Damping of slow modes}~~~~
Slow modes have the same $\tau_{cas}^{-1}$ as Alfv\'{e}n modes (see Section \ref{sec: proscal}). 
The ratio $|\omega_I|/\tau_{cas}^{-1}$ is (Eq. \eqref{eq: scsldr}, \eqref{eq: wcfasldr}, \eqref{eq: supcara}, \eqref{eq: subcarab}), 
\begin{subnumcases} 
{\frac{|\omega_I|}{\tau_{cas}^{-1}}\sim}
k^{4/3},~~~~~~~~~~~~~k \ll k_\text{dec} , \\
k^{-2/3},~~~~~~~~~~~ k \gg k_\text{dec}.
\end{subnumcases} 
for both super- and sub-Alfv\'{e}nic turbulence. 

Before tackling the damping of turbulence cascade, we first deal with the cutoff region of slow modes. 
We only look into regions at $1/k<l_A$ in super-Alfv\'{e}nic turbulence, and $1/k<l_{tr}$ in sub-Alfv\'{e}nic turbulence. 
From Eq. \eqref{eq: scsldrt}, Eq. \eqref{eq: supscal} and \eqref{eq: subscal}, we deduce 
\begin{equation} \label{eq: slcflbsup}
  k_{c,\|}^+=\sqrt{\frac{2\nu_{ni}}{c_s \xi_n l_A}}, ~~~~~~~~1/k_c^+<l_A, 
\end{equation}
for super-Alfv\'{e}nic turbulence, and 
\begin{equation} \label{eq: slcflbsub}
  k_{c,\|}^+=\sqrt{\frac{2\nu_{ni}}{c_s \xi_n L M_A^{-4}}}, ~~~~~~~~1/k_c^+<l_{tr}, 
\end{equation}
for sub-Alfv\'{e}nic turbulence. 
Using Eq. \eqref{eq: mawcgen} at low-$\beta$ limit, $|\omega_R|=|\omega_I|$ gives 
\begin{equation}
    k_{c,\|}^-=\frac{\nu_{in}}{2c_{si}}.
\end{equation} 
The expressions of $k_c^\pm$ are given in Appendix \ref{app:b}.

The inspection of the damping rate in Fig. \ref{fig:comexa2} shows the cutoff starts earlier before $|\omega_I|$ (solid line)
intersects with $\tau_{cas}^{-1}$ (dash-dotted line).
The dramatic boost of damping at $k_c^+$ results in $k_\text{dam}=k_c^+$. 
The analytical damping rate (open circles) are from Eq. \eqref{eq: scsldr} and \eqref{eq: wcfasldr}.
The solid and dashed lines represent two types of slow modes in ions and neutrals respectively. 
This new sort of slow modes sustained by neutrals appears at high wave frequency end, 
which was earlier revealed in 
\citet{Zaqa11}.
It has a damping rate as 
\begin{equation}
  |\omega_I|=\frac{\nu_{ni}}{2}.
\end{equation}
We see there are two nonpropagating intervals. Their outer boundaries correspond to $k_c^\pm$ (Eq. \eqref{eq: slcfbdsup}). 

We further examine $|\omega_I|$ (Eq. \eqref{eq: scsldr}) and $\tau_{cas}^{-1}$ (Eq. \eqref{eq: supcarab}, \eqref{eq: subcarab}) 
at $k_c^+$, using Eq. \eqref{eq: supscal} and \eqref{eq: subscal}, and find 
\begin{subequations} 
\begin{align}
& |\omega_I (k_c^+)|=|\omega_R (k_c^+)|=c_s k_{c,\|}^+,  \\
& \tau_{cas}^{-1}(k_c^+)=V_A k_{c,\|}^+.  \label{eq: carcol}
\end{align}
\end{subequations}
Eq. \eqref{eq: carcol} stands for both super- and sub-Alfv\'{e}nic turbulence. It shows $|\omega_I|/\tau_{cas}^{-1}=c_s/V_A<1$ at 
the scale $k_c^+$ in a low-$\beta$ plasma. 
It indicates the equality $|\omega_I|=\tau_{cas}^{-1}$ would happen at a smaller scale than $1/k_c^+$ if the cutoff were not present. 
Consequently, instead of equalizing $|\omega_I|$ from Eq. \eqref{eq: scsldr} with $\tau_{cas}^{-1}$ to get $k_\text{dam}$, 
$k_\text{dam}$ of slow modes is represented by $k_c^+$.

Another constraint on $k_\text{dam}$ of slow modes comes from Alfv\'{e}nic cascade. 
Since slow modes are slaved to Alfv\'{e}n modes and do not cascade themselves 
(GS95, \citealt{CL02_PRL}), if Alfv\'{e}n modes are damped first, the cascade of slow modes terminates subsequently. 
Hence more exactly, we have 
\begin{equation} \label{eq: slofinds}
   k_\text{dam}=\text{min}(k_c^+, k_\text{dam, Alfv\'{e}n}).
\end{equation}
as the damping scale of slow modes.

\section{Difference between the velocity dispersion spectra of neutrals and ions}
\label{sec: vdin}
We start with the Alfv\'{e}n modes. Because the energy spectra of turbulence vary in different $M_A$ domains, 
it is necessary to perform the analysis in super- and sub-Alfv\'{e}nic turbulence respectively.

\subsection{ Super-Alfv\'{e}nic turbulence}
\label{subsec: supvd}
The 3-D energy spectrum of strong MHD turbulence is given by 
\citet{CLV_incomp}, 
\begin{equation} 
\label{eq: 3denessup}
E(k_\perp, k_\|)= \frac{1}{3\pi} V_A^2 l_A^{-1/3}k_\perp^{-10/3}\exp{(-l_A^{1/3}\frac{k_\|}{k_\perp^{2/3}})}. 
\end{equation}
Here we replace the injection scale of strong MHD turbulence in the original equation with $l_A$. By integrating the above expression over $k_\|$, 
the turbulent energy spectrum density for super-Alfv\'{e}nic turbulence follows 
\begin{equation}
\label{eq: supmhdspeca}
E(k_\perp)=\frac{2}{3}L^{-2/3}V_L^2k_\perp^{-5/3}
\end{equation}
at $k^{-1}<l_A$. The Kolmogorov turbulence at $[L, l_A]$ has 
\begin{equation}
\label{eq: supmhdspecb}
E(k)=\frac{2}{3}L^{-2/3}V_L^2k^{-5/3}.
\end{equation}

Eq.\eqref{eq: supmhdspeca} and \eqref{eq: supmhdspecb} apply to both neutrals and ions when they are coupled. 
But in MHD turbulence regime ($k^{-1}<l_A$), for scales smaller than $k_\text{dec}^{-1}$, Eq. \eqref{eq: supmhdspeca} only applies to ions. 
Neutrals begin to carry out a hydrodynamic cascade independently, with $k$ being isotropic and $k=k_\perp$ at $k_\text{dec}$, 
having a energy spectrum
\begin{equation}
\label{eq: supneuesp}
E_n(k)=\frac{2}{3}L^{-2/3}V_L^2k^{-5/3}.
\end{equation}
At the scale $k_\text{dam}$,  Alfv\'{e}nic turbulence cascade terminates in ion-electron fluid, but hydrodynamic cascade proceeds in neutrals until reaching $k_\nu$. 

Since the energy spectra differ in different regimes, we discuss the following cases.

(1) \emph{$1/k_\text{dec}>1/k_\text{dam}>l_A$}~~~~
In this case, ions and neutrals have the same turbulent energy spectra, i.e. $E(k)=E_n(k)$. Since the square of velocity dispersion is proportional to the 
integration of energy spectrum in $k$ space
(LH08),
the squared velocity dispersion at $k$ is 
\begin{equation}
\sigma_i^2(k) \sim \int_{k}^{k_\text{dam}} E(k)dk=L^{-2/3}V_L^2k^{-2/3}-L^{-2/3}V_L^2k_\text{dam}^{-2/3}
\end{equation}
for ions and 
\begin{equation}
\sigma_n^2(k) \sim \int_{k}^{k_\nu} E_n(k)dk=L^{-2/3}V_L^2k^{-2/3}-L^{-2/3}V_L^2k_\nu^{-2/3}
\end{equation}
for neutrals. 
It is worth noting that in observations, $k^{-1}$ is the subcloud scale at which the corresponding 
velocity dispersion is measured. 
The difference of the squared velocity dispersions of neutrals and ions is 
\begin{equation}
\Delta \sigma^2=\sigma_n^2(k)-\sigma_i^2(k) = L^{-2/3}V_L^2(k_\text{dam}^{-2/3}-k_{\nu}^{-2/3}),
\end{equation}
It results from the different integration domains, i.e. different turbulence damping scales of neutrals and ions. 
With a small viscous scale of neutrals, the above equation becomes
\begin{equation}
\Delta \sigma^2\sim L^{-2/3}V_L^2k_\text{dam}^{-2/3}. 
\end{equation}
Thus the damping scale of ions can be determined from the measurement of $\Delta \sigma^2$.

(2) \emph{$l_A>1/k_\text{dec}>1/k_\text{dam}$}~~~~
The Alfv\'{e}nic turbulence is anisotropic in this regime. Since neutrals and ions carry the same Alfv\'{e}nic turbulence before they decouple, we only need to focus on 
the velocity dispersion at scales smaller than $1/k_\text{dec}$. From $k_\text{dec}$, isotropic turbulence arises in neutrals, thus the turbulence in the two fluids 
follow different cascades.
For ions, the squared velocity dispersion at $k(k>k_\text{dec})$ is given by 
\begin{equation}
\label{eq: supc2geni}
\begin{aligned}
\sigma_i^2(k) &\sim \int_{k_\perp}^{k_{\text{dam}, \perp}} E(k_\perp)dk_\perp \\
&=L^{-2/3}V_L^2k_\perp^{-2/3}-L^{-2/3}V_L^2k_{\text{dam}, \perp}^{-2/3}, 
\end{aligned}
\end{equation}
while for neutrals, it is 
\begin{equation}
\begin{aligned}
\label{eq: supc2genn}
\sigma_n^2(k) &\sim \int_{k}^{k_{\nu}} E_n(k)dk \\
&=L^{-2/3}V_L^2k^{-2/3}-L^{-2/3}V_L^2k_{\nu}^{-2/3}.
\end{aligned}
\end{equation}

To obtain the exact expression of $\Delta \sigma^2$, we start the integration from $k_\text{dec}$. Considering at $k_\text{dec}$, neutrals and ions still share the same 
energy spectrum, we get 
\begin{equation}
\label{eq: supsvdi}
\begin{aligned}
\sigma_i^2(k_\text{dec}) & \sim \int_{k_\text{dec}}^{k_{\text{dam}, \perp}} E(k_\perp)dk_\perp  \\
&=L^{-2/3}V_L^2k_\text{dec}^{-2/3}-L^{-2/3}V_L^2k_{\text{dam}, \perp}^{-2/3}.
\end{aligned}
\end{equation}
When the integration applies to neutrals, we have
\begin{equation}
\label{eq: supsvdn}
\begin{aligned}
\sigma_n^2(k_\text{dec}) & \sim \int_{k_\text{dec}}^{k_{\nu}} E_n(k)dk \\
&=L^{-2/3}V_L^2k_\text{dec}^{-2/3}-L^{-2/3}V_L^2k_{\nu}^{-2/3}.
\end{aligned}
\end{equation}
Thus the difference of the squared velocity dispersions of neutrals and ions can be obtained, 
\begin{equation}
\label{eq: supsvddf}
\begin{aligned}
\Delta \sigma^2&=\sigma_n^2(k_\text{dec})-\sigma_i^2(k_\text{dec}) \\
&= L^{-2/3}V_L^2(k_{\text{dam}, \perp}^{-2/3}-k_{\nu}^{-2/3}),
\end{aligned}
\end{equation}
which stems from both their different energy spectra and different turbulence damping scales. 
As $1/k_\nu$ is negligibly small, $\Delta \sigma^2$ can be written as 
\begin{equation}
\label{eq: supc2appvd}
\Delta \sigma^2 \sim L^{-2/3}V_L^2k_{\text{dam}, \perp}^{-2/3}.
\end{equation}
Given the turbulence driving, it only depends on the perpendicular component of the damping scale of ions.

(3) \emph{$1/k_\text{dec}>l_A>1/k_\text{dam}$}~~~~
This case is similar to Case (2). But the energy spectra of ions and neutrals begin to diverge only from $l_A$. So we just need to replace $k_\text{dec}$ with 
$l_A^{-1}$ in Eq. \eqref{eq: supsvdi} and \eqref{eq: supsvdn}, and get the same $\Delta \sigma^2$ as expressed in Eq. \eqref{eq: supsvddf}.

\begin{figure*}[htbp]
\centering
\subfigure[]{
 \includegraphics[width=6.4cm]{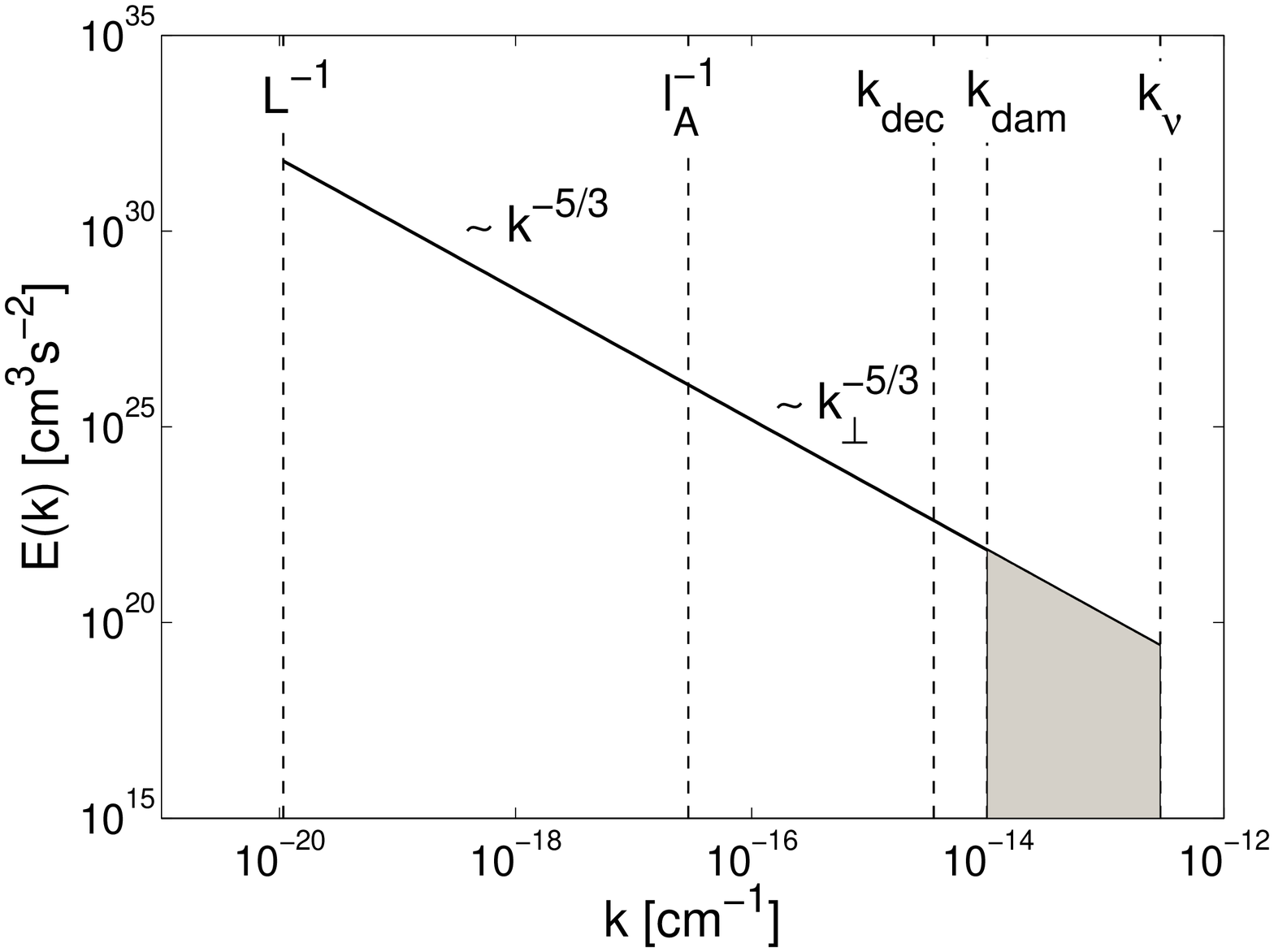} \label{fig:supspec}}
\subfigure[]{
  \includegraphics[width=6.9cm]{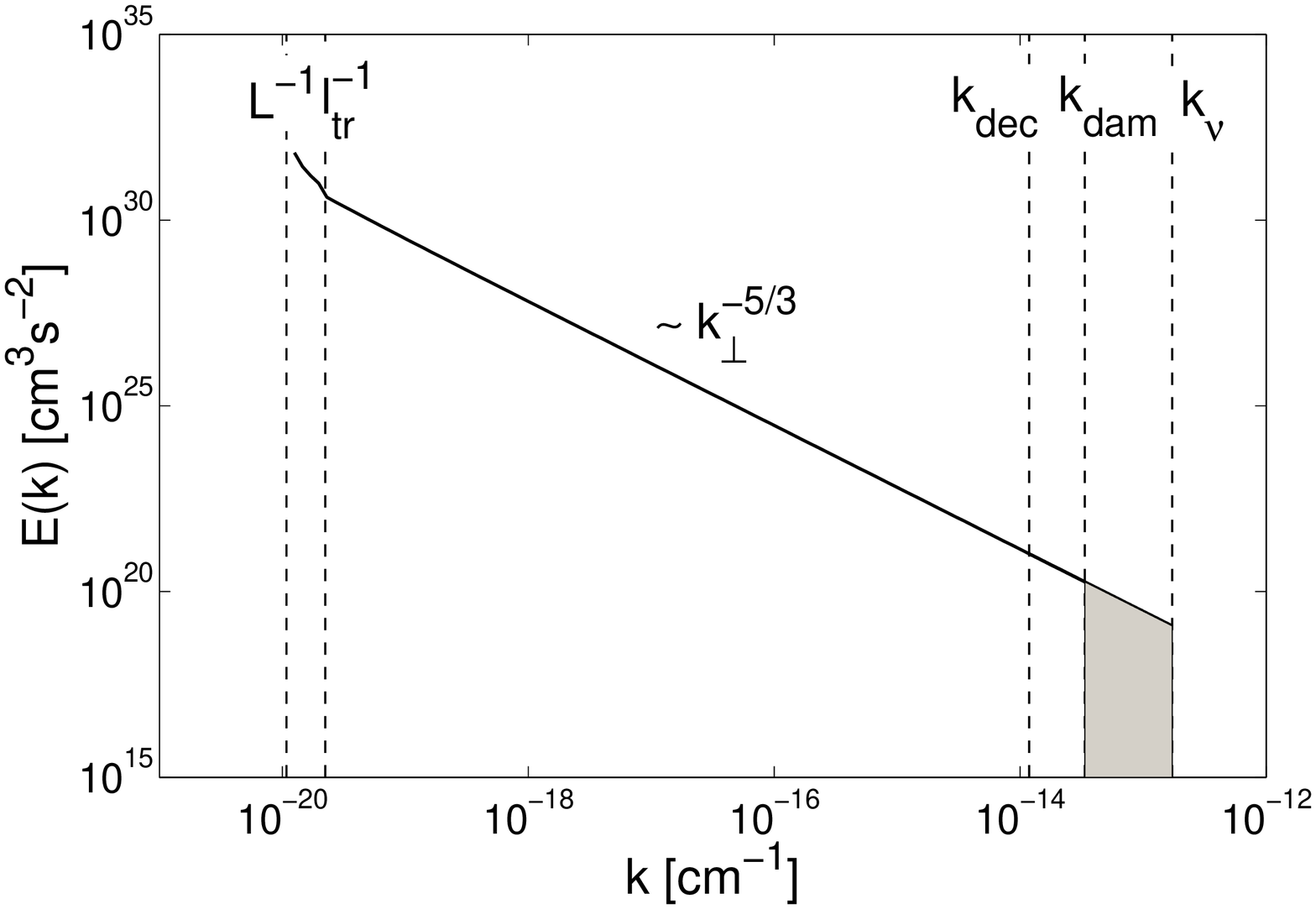} \label{fig:subspec}}
\caption{(a) Energy spectrum of Alfv\'{e}n modes in \emph{Model 1}. The scales $L^{-1}$, $l_A^{-1}$, $k_{dec}$, $k_{dam}$, and $k_\nu$ are 
indicated. The shaded area corresponds to the difference between the squares of the neutral and ion velocity dispersions. (b) Same as (a) but for sub-Alfv\'{e}nic molecular cloud using parameters of \emph{Model 2}. $l_{tr}$ is the transition scale from weak to strong MHD turbulence. } 
\end{figure*}

Fig. \ref{fig:supspec} displays $E(k)$ as a function of $k$ using the parameters of \emph{Model 1}, corresponding to Case (2).
The shaded area illustrates $\Delta \sigma^2$. 
Although $k$ and $k_\perp$ cannot be distinguished from the figure due to strong anisotropy, we will show in Section \ref{subsec: applica} that weather $\Delta \sigma^2$  depends 
on $k_{\text{dam}, \perp}$ or $k_\text{dam}$ plays a crucial role in determining magnetic field. 
Fig. \ref{fig:subspec} shows $E(k)$ of sub-Alfv\'{e}nic turbulence as a comparison, which we will discuss in the next subsection. 

\begin{figure*}[htbp]
\centering
\subfigure[]{
   \includegraphics[width=6.5cm]{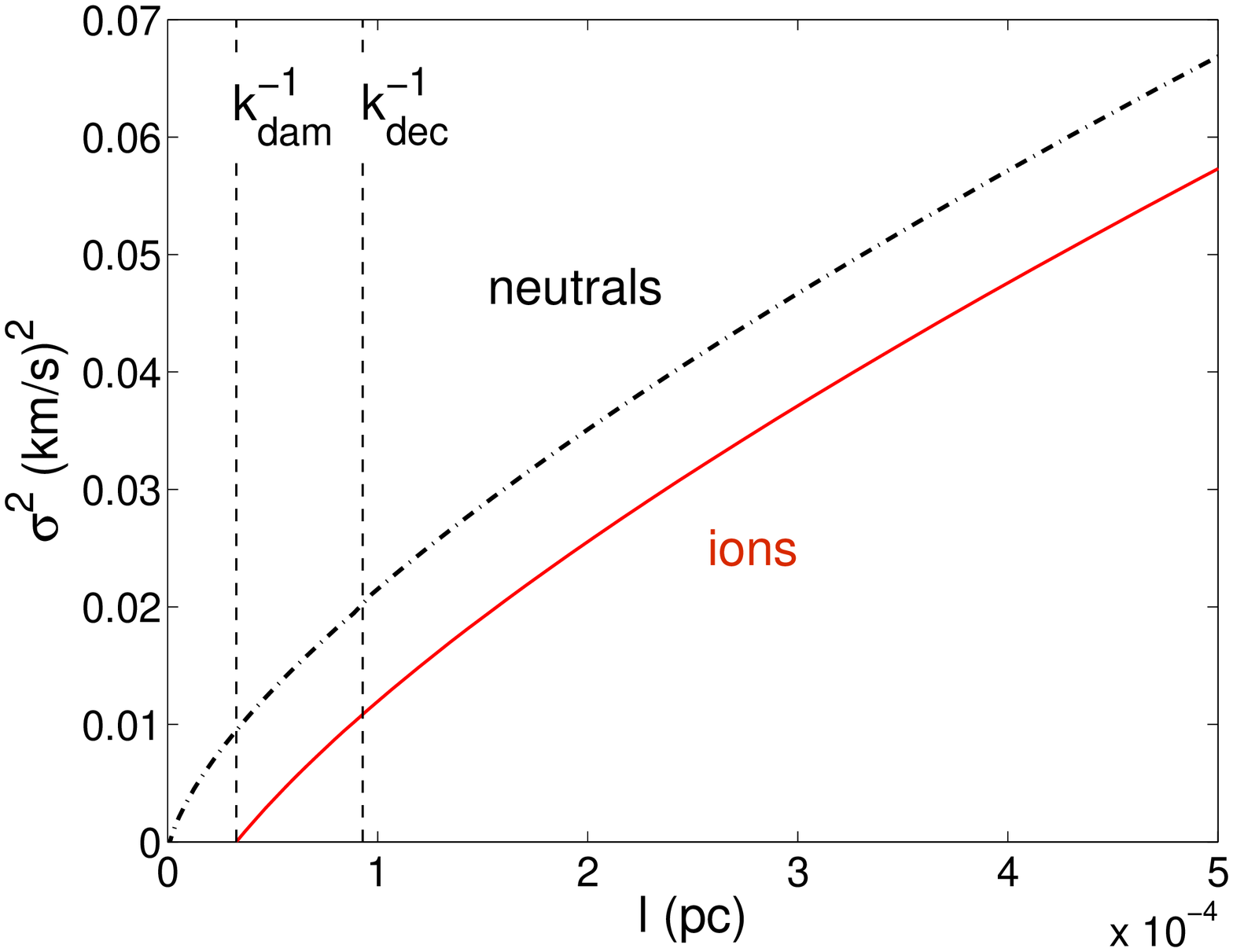}\label{fig:supvdda}}
\subfigure[]{
   \includegraphics[width=6.5cm]{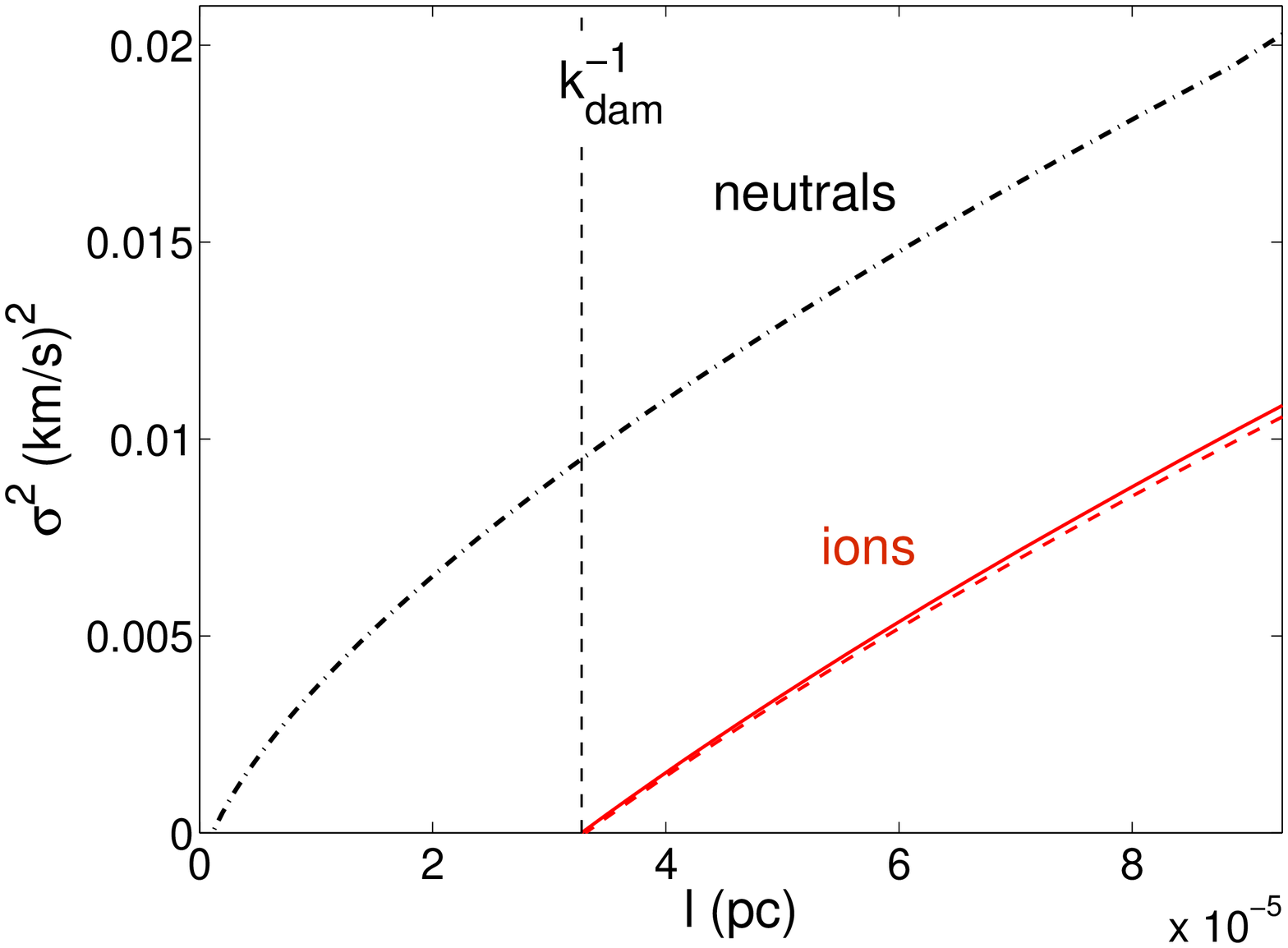}\label{fig:supvddb}}
\caption{ Squared velocity dispersion for neutrals (dash-dotted lines) and ions (solid lines) vs. linear length scale, (a) $[l_A, k_\nu^{-1}]$, (b) $[k_\text{dec}^{-1}, k_\nu^{-1}]$.
Parameters used are taken from \emph{Model 1}. 
The dash-dotted line is calculated by using \eqref{eq: supc2genn}. Solid line is Eq. \eqref{eq: supc2geni}. 
The dashed line in (b) corresponds to Eq. \eqref{eq: supc2appni}. }
\label{fig:supvdd}
\end{figure*}

Fig. \ref{fig:supvdd} simulates the observed $\sigma_n^2(k)$ and $\sigma_i^2(k)$ as a function of length scale (i.e. $k^{-1}$) using the parameters of \emph{Model 1}. 
It shows neutrals 
have larger velocity dispersion compared to that of ions, due to its smaller turbulence damping scale. That results in a wider line width of neutrals than ions 
from an observational point of view. 
Furthermore, since viscous scale $k_\nu^{-1}$ is relatively 
small with the parameters used for a typical molecular cloud, the curve for neutrals can be considered as passing (0, 0) point. It corroborates we can safely neglect the 
term with $k_\nu$ in expressions of $\Delta \sigma^2$. 

Fig. \ref{fig:supvdda} display the results for scales $[l_A, k_\nu^{-1}]$. Fig. \ref{fig:supvddb} zooms in on $[k_\text{dec}^{-1}, k_\nu^{-1}]$. They are derived 
by using Eq. \eqref{eq: supc2geni} (solid line) and \eqref{eq: supc2genn} (dash-dotted line). The dashed line in Fig. \ref{fig:supvddb} is calculated by taking the approximation,
\begin{equation}
\label{eq: supc2appni}
\sigma_i^2(k) \sim L^{-2/3}V_L^2k^{-2/3}-L^{-2/3}V_L^2k_{\text{dam}, \perp}^{-2/3}.
\end{equation}
It replaces $k_\perp$ in Eq. \eqref{eq: supc2geni} by $k$. The dashed line almost overlaps the solid one. It shows at small scales, this change doesn't make a 
significant difference for the velocity dispersion spectrum. Therefore, we can use Eq. \eqref{eq: supc2appni} and \eqref{eq: supc2genn} to compare with the observed velocity 
dispersion spectra and get the same expression of $\Delta \sigma^2$ as Eq. \eqref{eq: supc2appvd}.

\subsection{Sub-Alfv\'{e}nic turbulence}
\label{sec: subvd}
In strong turbulence regime, i.e., $k^{-1}<l_{tr}$, the 3-D energy spectrum is 
\citep{CLV_incomp}
\begin{equation} 
\label{eq: 3denessub}
E(k_\perp, k_\|)= \frac{1}{3\pi} V_L^2 l_{tr}^{-1/3}k_\perp^{-10/3}\exp{(-L^{1/3}\frac{k_\|}{M_A^{4/3}k_\perp^{2/3}})},
\end{equation}
where $l_{tr}$ is the injection scale of strong MHD turbulence. 
The corresponding 1-D energy spectrum can be obtained by integrating over $k_\|$, 
\begin{equation}
E(k_\perp)=\frac{2}{3}L^{-2/3}V_L^2M_A^{2/3}k_\perp^{-5/3}, ~~~~1/k<l_{tr}.
\end{equation}

We consider the case where $l_{tr}>1/k_\text{dec}>1/k_\text{dam}$.
At scales larger than the decoupling scale, Alfv\'{e}nic turbulence cascade proceeds in the strongly coupled two fluids. At $k_\text{dec}$, 
the squared velocity dispersion of ions is given by 
\begin{equation}
\begin{aligned}
\sigma_i^2(k_\text{dec}) & \sim \int_{k_\text{dec}}^{k_{\text{dam}, \perp}} E(k_\perp)dk_\perp= \\
& L^{-2/3}V_L^2M_A^{2/3}k_\text{dec}^{-2/3}-L^{-2/3}V_L^2M_A^{2/3}k_{\text{dam}, \perp}^{-2/3},
\end{aligned}
\end{equation}
which is different from Eq. \eqref{eq: supsvdi} by a $M_A^{2/3}$ factor. 

The energy spectrum of neutrals with their hydrodynamic cascade starting in the strong sub-Alfv\'{e}nic turbulence regime takes the form, 
\begin{equation}
E_n(k)=\frac{2}{3}L^{-2/3}V_L^2M_A^{2/3}k^{-5/3}.
\end{equation}
The squared velocity dispersion for neutrals is then 
\begin{equation}
\begin{aligned}
\sigma_n^2(k_\text{dec}) & \sim \int_{k_\text{dec}}^{k_{\nu}} E_n(k)dk \\
&=L^{-2/3}V_L^2M_A^{2/3}k_\text{dec}^{-2/3}-L^{-2/3}V_L^2M_A^{2/3}k_{\nu}^{-2/3}.
\end{aligned}
\end{equation}

Similar to super-Alfv\'{e}nic turbulence case, in practice, we can use 
\begin{equation}
\sigma_i^2(k) \sim L^{-2/3}V_L^2M_A^{2/3}k^{-2/3}-L^{-2/3}V_L^2M_A^{2/3}k_{\text{dam}, \perp}^{-2/3}
\end{equation}
to approximate 
\begin{equation}
\sigma_i^2(k) \sim L^{-2/3}V_L^2M_A^{2/3}k_\perp^{-2/3}-L^{-2/3}V_L^2M_A^{2/3}k_{\text{dam}, \perp}^{-2/3}
\end{equation}
at small scales due to high anisotropy when comparing with observations.
The difference of the squared velocity dispersions of ions and neutrals is 
\begin{equation}
\label{eq: subca2dsori}
\Delta \sigma^2= L^{-2/3}V_L^2M_A^{2/3}(k_{\text{dam}, \perp}^{-2/3}-k_{\nu}^{-2/3}),
\end{equation}
It can also be written as 
\begin{equation}
\label{eq: subca2ds}
\Delta \sigma^2 \sim  L^{-2/3}V_L^2M_A^{2/3}k_{\text{dam}, \perp}^{-2/3},
\end{equation}
when $k_\nu^{-1}$ is much smaller than $k_\text{dam}^{-1}$.

Fig. \ref{fig:subspec} shows the energy spectrum corresponding to \emph{Model 2}. The shade region in Fig. \ref{fig:subspec} shows 
$\Delta \sigma^2$ expressed by Eq. \eqref{eq: subca2dsori}. 
Different from Super-Alfv\'{e}nic turbulence, anisotropy applies over all scales in both weak and strong turbulence regimes. 

At last we discuss a particular case where $k_\text{dam}^{-1}>k_\text{dec}^{-1}$. We take the situation of \emph{Model 3} as an example. 
For \emph{Model 3}, the turbulence is damped when neutrals and ions are still strongly coupled and behave 
as one fluid. So the Alfv\'{e}nic turbulence is truncated in both neutrals and ions at $k_\text{dam}$.
But the turbulence cascade in ions may resume at small scales 
(see \citealt{LVC04})
as we discussed earlier. 
This will lead to a larger velocity dispersion and wider line width of ions than neutrals. Although this model contradicts the existing observational facts, it still 
deserves special attention since this particular turbulent regime may not be covered by current limited observational data. 

We found among all the cases discussed above for both super- and sub-Alfv\'{e}nic turbulence, only in the Kolmogorov turbulence regime of super-Alfv\'{e}nic turbulence, $\Delta \sigma^2$ has a dependence on total $k_\text{dam}$. For all the other cases, $\Delta \sigma^2$ is only related to the 
perpendicular component of damping scale $k_{\text{dam}, \perp}$.

\subsection{Compressible turbulence}
Compared with Alfv\'{e}n modes, fast modes are more severely damped, having the turbulence truncation in strongly coupled regime or 
critically at the decoupling scale, which depends on the environment parameters (Eq. \eqref{eq: ctnfdf}). 
In any case, the turbulent energy spectra of fast modes dissipate at the same scale for the coupled neutrals and ions,  
so damping of fast modes does not contribute to the difference of squared velocity dispersions between neutrals and ions.

The problem is less straightforward for slow modes. 
The crucial point is the relation between $k_c^+$ and $k_\text{dec}$. 
By equaling $k_{c,\|}^+$ (Eq. \eqref{eq: slcflbsup}, Eq. \eqref{eq: slcflbsub}) and $k_{\text{dec}, \|}=\nu_{ni}/V_A$, we get 
a parameter-dependent critical ionization degree 
\begin{equation} \label{eq: slcrifsup}
   \xi_{i,cr}=\frac{2V_A^2}{\nu_{in} c_s l_A  }
\end{equation}
for super-Alfv\'{e}nic turbulence, and 
\begin{equation}
   \xi_{i,cr}=\frac{2V_A^2}{\nu_{in} c_s LM_A^{-4}  }
\end{equation}
for sub-Alfv\'{e}nic turbulence. 
Given the parameters in \emph{Model 1}, Eq. \eqref{eq: slcrifsup} gives $\xi_{i,cr}\approx 2.15 \times 10^{-4}$.
When $\xi_i>\xi_{i,cr}$, $k_c^+$ is smaller than $k_\text{dec}$ (see Fig. \ref{fig:comexa2}). 
Similar to fast modes, the turbulent energy spectra of two fluids dissipate together and do not cause linewidth differences. 
But in the other situation, when $\xi_i<\xi_{i,cr}$, the damping scale of ions $1/k_c^+$ is smaller than $1/k_\text{dec}$. 
Since the damping scale of Alfv\'{e}n modes is also smaller than $1/k_\text{dec}$, according to Eq. \eqref{eq: slofinds}, 
slow modes have $k_\text{dec}<k_\text{dam}$ in this situation. 
Accordingly, the analysis present in Case (2) in Section \ref{subsec: supvd} and Section \ref{sec: subvd} also applies to 
slow modes.

\section{A summary of important results}
Because of the multitude of turbulent regimes and damping effects, we have provided the expressions of $k_\text{dam}$ and $\Delta \sigma^2$ in a wide variety of situations. 
From an observational point of view, a recapitulation of the results in most typical situations might be useful. We summarize them as follows. 

\emph{Super-Alfv\'{e}nic, Kolmogorov, neutral-ion collisions}------
In the case of Kolmogorov turbulence and damping dominated by neutral-ion collisions, 
the damping scale is (Eq. \eqref{eq: supkdama})
\begin{equation}
k_\text{dam}=2^{\frac{3}{2}}\nu_{ni}^{\frac{3}{4}}\xi_n^{-\frac{3}{4}}L^{-\frac{1}{4}}V_L^{\frac{3}{4}}V_A^{-\frac{3}{2}}.
\end{equation}
The squared velocity dispersions are (Case (1) in Section \ref{subsec: supvd})
 \begin{subequations}\label{eq: imp1ni}
 \begin{align}
    & \sigma_n^2(k) \sim L^{-2/3}V_L^2k^{-2/3}-L^{-2/3}V_L^2k_\nu^{-2/3}, \\
    & \sigma_i^2(k) \sim L^{-2/3}V_L^2k^{-2/3}-L^{-2/3}V_L^2k_\text{dam}^{-2/3}.
 \end{align}
 \end{subequations}
Their difference depends on the total $k_\text{dam}$.
\begin{equation}
   \Delta \sigma^2\sim L^{-2/3}V_L^2k_\text{dam}^{-2/3}. 
\end{equation}

\emph{Super-Alfv\'{e}nic, MHD, neutral-ion collisions}------The damping scale perpendicular to magnetic field is (Eq. \eqref{eq: supkdamperp})
\begin{equation}
\label{eq: kdamperpim2}
  k_{\text{dam}, \perp}=(\frac{2\nu_{ni}}{\xi_n})^{\frac{3}{2}}L^{\frac{1}{2}}V_L^{-\frac{3}{2}}, 
\end{equation}
which is independent of $B$. The squared velocity dispersions are (Case (2) in Section \ref{subsec: supvd})
 \begin{subequations}\label{eq: imp2ni}
 \begin{align}
    & \sigma_n^2(k) \sim L^{-2/3}V_L^2k^{-2/3}-L^{-2/3}V_L^2k_\nu^{-2/3}, \\
    & \sigma_i^2(k) \sim L^{-2/3}V_L^2k^{-2/3}-L^{-2/3}V_L^2k_{\text{dam}, \perp}^{-2/3}.
 \end{align}
 \end{subequations}
Their difference depends on $k_{\text{dam}, \perp}$.
\begin{equation}
\label{eq: supc2appvdim1}
   \Delta \sigma^2\sim L^{-2/3}V_L^2k_{\text{dam},\perp}^{-2/3}. 
\end{equation}

\emph{Sub-Alfv\'{e}nic, strong, neutral-ion collisions}------In strong turbulence regime, 
the perpendicular damping scale can be expressed as a function of $B$ (or $V_A$) (Eq. \eqref{eq: subnidamapp})
\begin{equation}
   k_{\text{dam},\perp}=(\frac{2\nu_{ni}}{\xi_n})^{\frac{3}{2}}L^{\frac{1}{2}}V_L^{-2}V_A^{\frac{1}{2}}.
\end{equation}
The squared velocity dispersions and their difference are (Section \ref{sec: subvd}) 
 \begin{subequations} \label{eq: imp3ni}
 \begin{align}
&  \sigma_n^2(k) \sim L^{-2/3}V_L^2M_A^{2/3}k^{-2/3}-L^{-2/3}V_L^2M_A^{2/3}k_{\nu}^{-2/3}, \\
&  \sigma_i^2(k) \sim L^{-2/3}V_L^2M_A^{2/3}k^{-2/3}-L^{-2/3}V_L^2M_A^{2/3}k_{\text{dam}, \perp}^{-2/3}, \\
&  \Delta \sigma^2 \sim  L^{-2/3}V_L^2M_A^{2/3}k_{\text{dam}, \perp}^{-2/3}.
  \end{align}
 \end{subequations}

In fact, the first situation with Kolmogorov turbulence is not common in molecular clouds. We listed here in order to have a comparison with earlier work assuming isotropic turbulence 
(e.g. LH08) in our next section. 
The dependence of $\Delta \sigma^2$ on $k_{\text{dam}, \perp}$ instead of $k_\text{dam}$ is a direct consequence of turbulence anisotropy, which is critical in determining magnetic field 
from observed molecular clouds. 

\section{Determination of magnetic field }
\label{sec: detemf}
Attention to obtaining magnetic field in molecular clouds has been paid exclusively in earlier studies. 
Determination of magnetic field strength is of fundamental importance in understanding dynamics and 
processes, e.g. star formation, arising in molecular clouds.
LH08 
proposed a new technique of measuring the strength of the plane-of-the-sky component of magnetic field embedded in molecular clouds, by using the 
observed differences between the velocity dispersion spectra of the ions and the neutrals. 
In preceding sections, we provided analytical expressions for damping scales of ions in different turbulent regimes. They are not only essential in explaining 
different line widths of neutrals and ions, but also serve as a possible tool for measuring magnetic field strength through their explicit dependence on magnetic field. 
Although there is a possibility that slow modes can also induce the differences in line widths for certain parameters, 
we restrict ourselves to Alfv\'{e}n modes in this section since they carry most of the MHD turbulence energy  
\citep{CL05}.

\subsection{Dependence of magnetic field on the damping scale}
\label{subsec: magdam}
As discussed earlier, damping due to neutral viscosity may lead to a wider line width of ions than neutrals, which hasn't been found by any current observation. 
So here we only discuss the case where neutral viscosity is negligible. 

(1) \emph{Super-Alfv\'{e}nic turbulence}~~~~
We rewrite Eq. \eqref{eq: supkdam} and get 
\begin{subnumcases}
   {B=\label{eq: magsupl}}
   4(\pi \nu_{ni})^{\frac{1}{2}} \rho^{\frac{1}{2}} \xi_n ^{-\frac{1}{2}} L^{-\frac{1}{6}} V_L^{\frac{1}{2}} k_\text{dam}^{-\frac{2}{3}},\nonumber \\
   ~~~~~~~~~~~~~~~~~~~~~~~~~~~~~~~~~~l_A<1/k_\text{dam}<L, \label{eq: magsupla} \\ 
   \sqrt{\frac{4\pi\rho (\frac{2\nu_{ni}}{\xi_n})^2}{k_\text{dam}^2-(\frac{2\nu_{ni}}{\xi_n})^3LV_L^{-3}} },~1/k_\text{dam}< l_A. \label{eq: magsuplb} 
\end{subnumcases}
We recall that there is an alternative way of getting $k_\text{dam}$ when $k_\text{dam}^{-1}<l_A$. 
Lower limit wave number of the "cutoff" region
$k^{+}$ (Eq. \eqref{eq: supcfkp}) gives 
\begin{equation}
\label{eq: magsuplbcf}
B=\left[\frac{4\pi\rho(f(\chi))^{-1}\nu_{ni}^2 \xi_i}{k_\text{dam}^2-(f(\chi))^{-\frac{3}{2}}\nu_{ni}^3\xi_i^{\frac{3}{2}}LV_L^{-3}}\right]^{\frac{1}{2}},~~~1/k_\text{dam}< l_A,
\end{equation}
The difference of the magnetic field values derived from Eq. \eqref{eq: magsuplb} and \eqref{eq: magsuplbcf} is marginal. When $k_\text{dam}^{-1}>l_A$,  $B$ depends on total $k_\text{dam}$. 
Differently, for $k_\text{dam}^{-1}<l_A$, $B$ is only related to the parallel component of $k_\text{dam}$. As is known, anisotropy starts at $l_A$ and 
increases rapidly with decreasing scales. That leads to a quite weak dependence of $B$ on $k_\text{dam}$. 
If only $k_{\text{dam},\perp}$ can be determined observationally, 
there is no way to evaluate $B$ when $k_\text{dam}^{-1}<l_A$.

(2) \emph{Sub-Alfv\'{e}nic turbulence}~~~~
For $k_\text{dam}^{-1}<l_{tr}$, we can write $B$ in terms of $k_\text{dam}$ from Eq. \eqref{eq: subkdamb}. However, 
given the simple expression of $k_{\text{dam}, \perp}$ by Eq. \eqref{eq: subnidamapp}, we can reach a convenient form of $B$ as 
\begin{equation}
\label{eq: subnis}
 B=\sqrt{4\pi\rho}(\frac{2\nu_{ni}}{\xi_{n}})^{-3}L^{-1}V_L^4k_{\text{dam}, \perp}^2,~~~1/k_\text{dam}< l_{tr}. \\ 
\end{equation}

Here we show the dependence of $B$ on the damping scale. 
In observations, in order to evaluate magnetic field strength, $k_\text{dam}$ should be first 
estimated from the measurements on $\Delta \sigma^2$. By combining $\Delta \sigma^2$ expressed by $k_\text{dam}$ and expressions of $B$ presented here, $B$ can be 
rewritten in terms of observational parameters. 
We will discuss the application on measuring $B$ from the
observations of velocity dispersion spectra in Section \ref{subsec: applica}. 

\subsection{Comparison with LH08}
LH08 suggested an expression to estimate the strength of the plane-of-the-sky component of magnetic field in molecular clouds. 
It is instructive to perform a comparison between 
LH08
method and this work. 
We start by revisiting their theoretical investigation. 
The ambipolar Reynolds number is defined as 
( \citealt{Zwei02}), 
\begin{equation}
  R_{AD}=\frac{lv_l}{\eta_{amb}},
\end{equation}
i.e. "effective magnetic Reynolds number" in 
LH08,
despite the minor change they made. 
Here 
\begin{equation}
\eta_{amb}=\frac{B^2}{4\pi \rho_n\nu_{ni}}, 
\end{equation}
is the effective magnetic diffusivity
\citep{Zwei02, Bisk03}.
LH08
claim that when the condition $R_{AD}\sim1$ holds, magnetic field, along with ions, easily decouples from neutrals. 
In fact, we can rewrite $R_{AD}=1$ as 
\begin{equation}
\label{eq: lh08rad1}
\frac{l^2\nu_{ni}}{V_A^2}=\frac{l}{v_l},
\end{equation}
under the assumption $\rho_i\ll\rho_n$.
The left-hand side is the characteristic diffusion time for magnetic field at a scale of $l$, given by 
\citet{Zwei02}.
It means that $R_{AD}\sim1$ indicates an equilibrium between magnetic field diffusion time and the turbulent eddy turn over time (the right-hand side).
 
The corresponding scale is called the "ambipolar diffusion scale" $L_{AD}$ in 
LH08 (termed as $L^\prime$ in their paper).
The scalings that Eq. \eqref{eq: lh08rad1} deals with are isotropic.
If adopting Kolmogorov scaling for velocity, i.e., $v_l=V_L(l/L)^{1/3}$, one can get $L_{AD}$ from Eq. \eqref{eq: lh08rad1},
\begin{equation}
 L_{AD}=\nu_{ni}^{-3/4}L^{1/4}V_L^{-3/4}V_A^{3/2}.
\end{equation}
Hence, magnetic field strength is given by 
\begin{equation}
B=2(\pi\nu_{ni})^{1/2}\rho^{1/2}L^{-1/6}V_L^{1/2}k_{AD}^{-2/3},\\
\end{equation}
where $k_{AD}=1/L_{AD}$. We found that, under the assumptions (a) $\xi_n \sim 1$, (b) Kolmogorov scaling, and (c) $k_{AD}=k_\text{dam}$, 
the analytical expression of $B$ given by 
LH08 
is basically the same as our result for super-Alfv\'{e}nic turbulence with negligible neutral viscosity and $k_\text{dam}^{-1}>l_A$ (Eq. \eqref{eq: magsupla}).
It is necessary to clarify that the dissipation scale of turbulence in ions is the damping scale, instead of the 
decoupling scale claimed in LH08. 
Notice $k_\text{dam}^{-1}$ is smaller than $k_\text{dec}^{-1}$ in most cases. 
The remaining turbulent energy spectrum of neutrals below the damping scale of ions account for the 
difference in line widths between neutrals and ions.

The main differences from our work is the missing essential information about MHD turbulence in LH08.
The results reported in this paper reveal that damping scales, as well as magnetic field, vary with different turbulent regimes. 
We claim that only when using the actual model of MHD turbulence can one obtain the analytical expressions for the magnetic
field. As a result, the formula of $B$
in LH08 is applicable only when the damping scale is in one very special regime of Alfv\'{e}nic turbulence, 
i.e. $1/k_\text{dam}>l_A$. When $1/k_\text{dam}<l_A$, the linewidth differences do not depend on magnetic field strength. 

Below we will follow the same general approach outlined in 
LH08 
for analyzing the observational velocity dispersion spectra, but apply the theory of turbulence damping 
developed in this work to determine the magnetic field embedded in molecular clouds. The obvious differences are 
that, first of all, our study shows that the linewidth for neutrals and ions are different for different regimes of magnetized turbulence.
This calls for defining the regime of turbulence and checking the self-consistency of the results obtained with the
technique at hand. Second, the anisotropy of Alfv\'{e}nic turbulence is in most cases very prominent at small scales.
Thus this anisotropy must be explicitly accounted for if quantitative measures of magnetic field are sought. In addition,
the explicit expressions of the turbulent velocity dispersion spectra that we obtained allow additional ways of studying magnetic fields.

\subsection{Applications with velocity dispersion spectra in molecular clouds}
\label{subsec: applica}

In Section \ref{subsec: magdam} we expressed magnetic field strength $B$ in terms of $k_\text{dam}$. 
Also we showed in Section \ref{sec: vdin} that
$k_\text{dam}$ (or $k_{\text{dam},\perp}$ in most cases) can be observationally obtained through measurement of $\Delta \sigma^2$.
This provides us with the way of evaluating $B$ using the respective formula in different turbulent regimes. 

The essential base of magnetic field determination is to identify the turbulence properties. Our study indicates that 
to relate the linewidth difference with the magnetic field one must know the media magnetization. This does not eliminate the utility of the
technique of obtaining magnetic field strength, as the initial value may be only approximate based on measurements using other techniques, e.g. Chandrasekhar-Fermi technique\footnote{The Chandreasekhar-Fermi technique assumes perfect alignment of dust grains, which may not be the case for
cloud cores in question 
(see \citealt{Lazarian07rev} for a review).} 
(see  \citealt{ChanFer53, Falce08, Houd13}), 
Zeeman observations 
\citep{Crut10},
and measurements of turbulence anisotropy 
\citep{LazP01, EL05, EL11, Toff11}.
Then the magnetic field strength obtained through the differences of the squared velocity dispersions can be used to correct the values of magnetization and also test the consistency of the results. 

In what follows we
pursue a limited aim of exemplifying the use of our analytical description on the difference of neutral-ion linewidths for studying
magnetic field, showing the similarities and differences in the procedures that we propose with those in LH08.  
In this work, we will generally show how to identify the turbulent regime and extract information on magnetic field from velocity dispersion spectra 
in both super- and sub-Alfv\'{e}nic turbulent molecular clouds. 
The application with observational data and results on magnetic field evaluation will be provided in a forthcoming paper.
The relative importance of neutral viscosity can be determined by $r$ (Section \ref{sec: rat}) with environment parameters provided. Here we only discuss the situation with negligible neutral viscosity for 
simplicity.

(1) \emph{Super-Alfv\'{e}nic turbulent cloud}~~~~
We follow the approach in LH08 for analyzing the parameters of velocity dispersion spectra. 
The expressions of the squared velocity dispersions we obtained in Section \ref{subsec: supvd} can be compared with the fits to observed data
(LH08). 
That is, 
 \begin{subequations}\label{eq: obsfitpa}
 \begin{align}
& \sigma_n^2(k)=bp^n, \\
&\sigma_i^2(k)=a+bp^n. 
\end{align}
\end{subequations}
Here $a$, $b$ are fit parameters to the lower envelope of the observed $\sigma^2$ of neutrals and ions. 
And $p$ is beam size in unit of angular measurement, which can be rewritten in terms of the distance $d$ and subcloud length scale $k^{-1}$, by $p=206265  / (kd)$.
This fit is obtained by changing the resolution of the telescope. 
Moreover, 
We also neglect the term containing $k_\nu^{-2/3}$ in $\sigma_n^2(k)$ due to its relatively small value. Hence, from Eq. \eqref{eq: obsfitpa}, \eqref{eq: imp1ni}, and \eqref{eq: imp2ni}, we can arrive at 
\begin{subequations} \label{eq: fitabsup}
 \begin{align}
& n=2/3, \\
& b\left(\frac{206265}{d}\right)^{2/3}=L^{-2/3}V_L^2, \label{eq: superbla} \\
& \left(-\frac{b}{a}\right)^{3/2}\left(\frac{206265}{d}\right)= \label{eq: superbkdam}
\begin{cases} 
k_{\text{dam}}, ~~~~~~1/k_{\text{dam}} > l_A, \\
k_{\text{dam},\perp} ,~~~1/k_{\text{dam}} < l_A.\\
\end{cases}
 \end{align}
 \end{subequations}
We assume that if we deal with Alfv\'{e}nic turbulence, $n$ value should be $2/3$ 
(GS95).
The discrepancy between this value and the observed data in LH08 
and the subsequent publications
\citep{Hez10, Hez14}
may indicate that other processes apart from Alfv\'{e}nic turbulence interfere with the linewidths. 
For instance, compressible motions also impact the observed spectral index. 
The energy spectrum of fast modes can lead to a shallower turbulent velocity dispersion spectrum with $n=1/2$
\citep{CL02_PRL}. 
It can also come from an Alfv\'{e}nic turbulent spectrum different from GS95 model. 
The scaling relations and scale-dependent anisotropy of Alfv\'{e}nic turbulence with an arbitrary spectrum index are presented in the 
appendix in 
\citet{LV99}, 
while considering magnetic reconnection in GS59 type turbulence. 
Assuming the observationally measured "n" corresponds to the actual turbulence power spectrum, the above analysis can be easily validated 
by adjusting the scaling relations shown in Section \ref{sec: proscal} accordingly.
But it is more likely 
due to the limitations related to recovering of the 3-D dispersion from the observed 2-D values 
\citep{FalLaz10}. 
The "n" obtained from the lower envelope of the line of sight (LOS) velocity dispersions may not exactly follow the actual velocity dispersion spectrum (see more discussions 
in Section \ref{subsec: magmeadis}).
Notice that the spectral index $n$ does not depend on the LOS direction, since it arises from 
MHD turbulence in the global frame of reference, i.e. the only reference frame available for observations integrated along the LOS 
(see \citealt{CL03, EL05}).
While this issue deserves a separate investigation, for our limited purposes of exemplifying possibilities of our approach, we
assume that $n=2/3$, which corresponds well to the theoretical expectation of strong Alfv\'{e}nic cascade (GS95) and numerical studies 
(e.g. \citealt{CL03, KowL10}).

Combining Eq. \eqref{eq: superbla}, \eqref{eq: superbkdam} and $V_A$ value obtained from the estimated magnetic field, we obtain 
\begin{equation} \label{eq: selconc}
    k_\text{dam} l_A(\text{or } k_{\text{dam}, \perp}l_A )=|a|^{-3/2}V_A^3.
\end{equation}

When $ k_\text{dam} (\text{or } k_{\text{dam}, \perp}) l_A <1$, we are in isotropic Kolmogorov turbulence regime. Accordingly, Eq. \eqref{eq: magsupla} applies. 
Taking advantage of Eq. \eqref{eq: superbla} and \eqref{eq: superbkdam}, Eq. \eqref{eq: magsupla} 
can be expressed in terms of the parameters $a$ and $b$, so $B$ can be observationally obtained 
\begin{equation}
B=4(\pi \nu_{ni})^{\frac{1}{2}} \rho^{\frac{1}{2}} \xi_n ^{-\frac{1}{2}} |a| b^{-\frac{3}{4}}\left(\frac{206265}{d}\right)^{-\frac{1}{2}}.
\end{equation}
With Kolmogorov scaling used, the formula employed by 
LH08
is very close to Eq. \eqref{eq: magsupla}. 
They both can be used to evaluate magnetic field strength. 

When $ k_\text{dam} (\text{or } k_{\text{dam}, \perp}) l_A >1$, we are in anisotropic MHD turbulence regime. As is discussed earlier, unlike isotropic turbulence, 
$B$ in this regime depends on the parallel component of $k_\text{dam}$. 
Eq. \eqref{eq: superbkdam} takes the form 
\begin{equation}
  \left(-\frac{b}{a}\right)^{3/2}\left(\frac{206265}{d}\right)=k_{\text{dam},\perp},
\end{equation}
and $k_{\text{dam},\perp}$ is independent to $B$ (Eq. \eqref{eq: kdamperpim2}).
Therefore, the approach of evaluating $B$ in this regime is not practically feasible.
In this case, either Eq. \eqref{eq: magsupla} or LH08 formula can not reach a correct evaluation of $B$. 

\begin{figure*}[htbp]
\centering
\subfigure[]{
   \includegraphics[width=6.5cm]{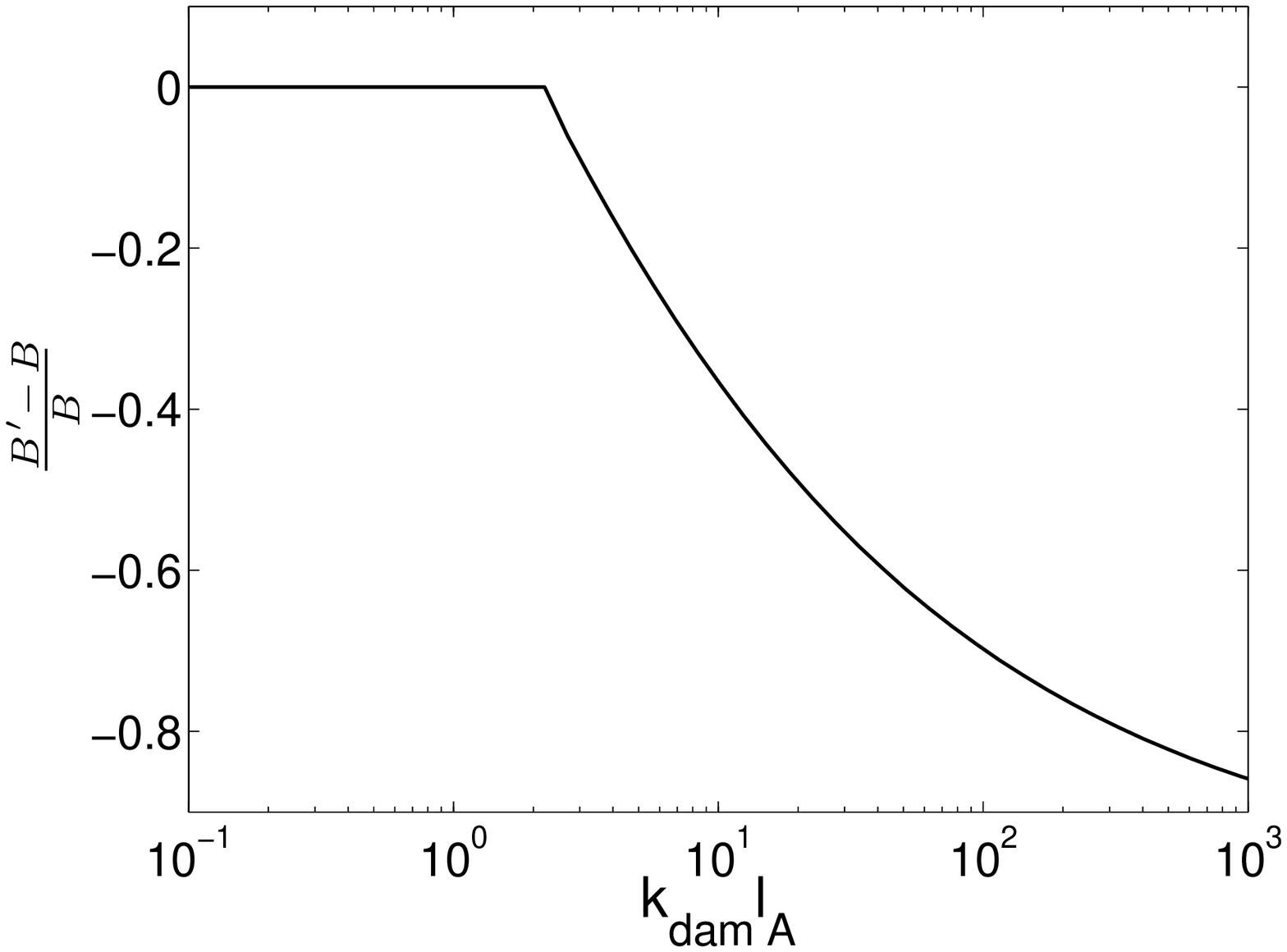}\label{fig:twomag}}
\subfigure[]{
   \includegraphics[width=6.5cm]{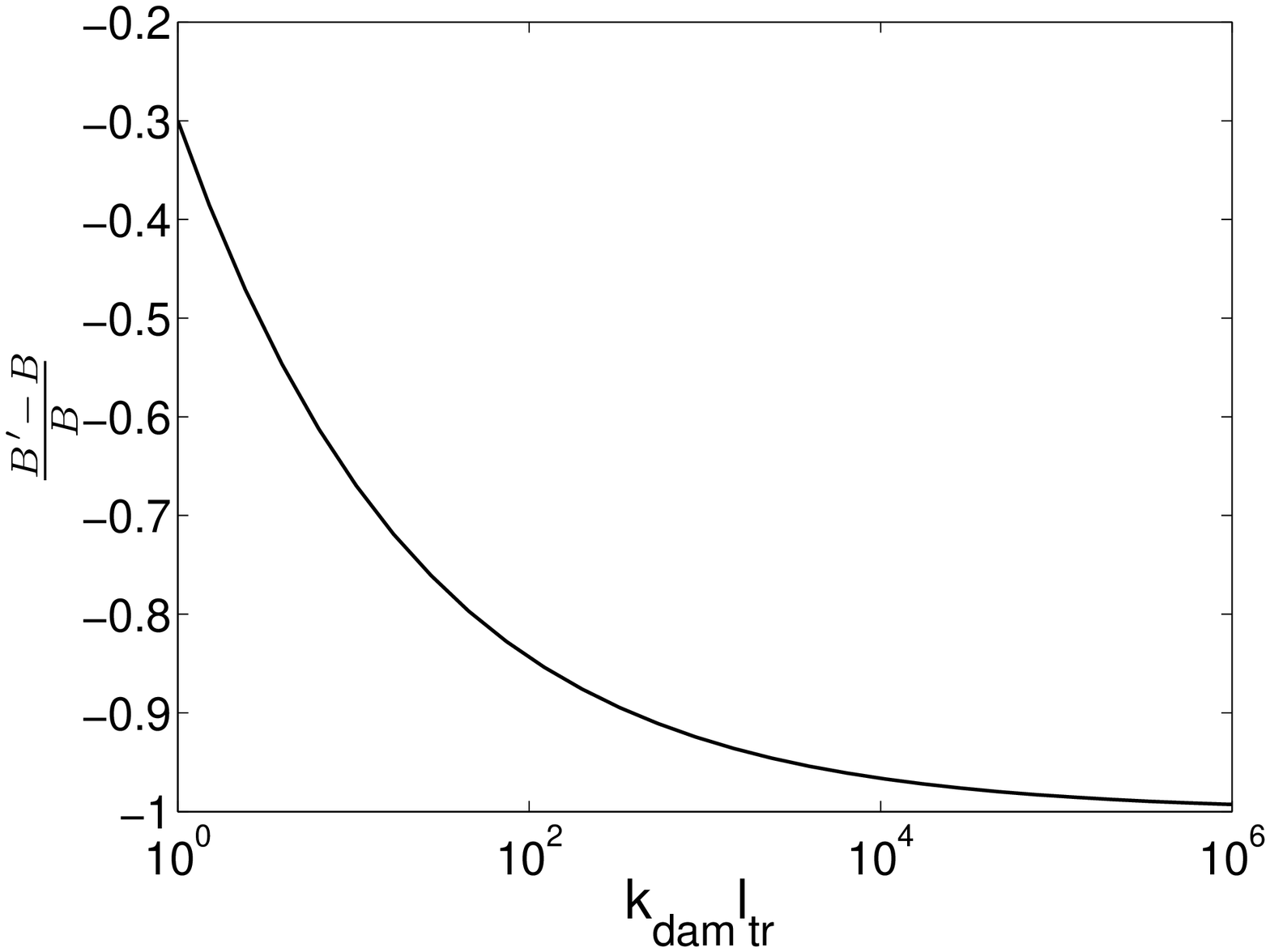}\label{fig:twomagsub}}
\caption{Error of the measured magnetic field using Eq. \eqref{eq: magsupla} 
(a) as a function of $k_\text{dam}l_A$ in a super-Alfv\'{e}nic cloud, (b) as a function of $k_\text{dam}l_{tr}$ in a sub-Alfv\'{e}nic cloud.  }
\label{fig:twomagto}
\end{figure*}

To examine the validity of Eq. \eqref{eq: magsupla} in both Kolmogorov and MHD turbulent regimes, 
we carry out a numerical test using the parameters of \emph{Model 1}, but keep $M_A$ as a free parameter by adjusting $V_L$. 
Fig. \ref{fig:twomag} exhibits the error of the measured magnetic field, $B^\prime$, using Eq. \eqref{eq: magsupla} as a function of $k_\text{dam}l_A$. 
We found at scales smaller than $l_A$, $B^\prime$ underestimates the real $B$. 
This shows in a sense how much Eq. \eqref{eq: magsupla} or LH08 formula deviates from the real magnetic field strength when it is applied in 
anisotropic MHD turbulent regime. 
It indicates that in an observed super-Alfv\'{e}nic cloud, when $k_\text{dam}^{-1}$ is slightly smaller than $l_A$, the effect of turbulence anisotropy 
can be included in error bars. 
But when $k_\text{dam}^{-1}$ is sufficiently smaller than $l_A$, turbulence 
local anisotropy develops to such an extent that the magnetic field strength given by LH08 formula will be far away from the actual value. 
Take the cloud given by \emph{Model 1} for example, $k_\text{dam}l_A \sim 344$, the corresponding 
error is as large as $\sim 80\%$.

(2) \emph{Sub-Alfv\'{e}nic turbulent cloud}~~~~
With the help of Zeeman measurement or turbulence anisotropy technique (see above), a cloud can be 
preliminarily determined as a sub-Alfv\'{e}nic turbulent source.

The observed larger velocity dispersion of neutrals compared to that of ions rules out the situation where $k_\text{dam}^{-1}>k_\text{dec}^{-1}$ (see the discussion at the end of Section \ref{sec: subvd}).
Hence, by comparing Eq. \eqref{eq: obsfitpa} with \eqref{eq: imp3ni}, we can get 
 \begin{subequations} \label{eq: subobs} 
 \begin{align}
& n=2/3, \\
& b\left(\frac{206265}{d}\right)^{2/3}=L^{-2/3}V_L^2M_A^{2/3},\label{eq: subobsb} \\
& \left(-\frac{b}{a}\right)^{3/2}\left(\frac{206265}{d}\right)=k_{\text{dam}, \perp}. \label{eq: subobsc}
\end{align}
\end{subequations}
Note, that the difference from super-Alfv\'{e}nic case is the $M_A^{2/3}$ factor in Eq. \eqref{eq: subobsb}. 
Consequently, Eq. \eqref{eq: subobsb} and \eqref{eq: subobsc} are not sufficient to derive $B$ through Eq. \eqref{eq: subnis}.
If additional information on the value of $L$ is given, together with $k_{\text{dam}, \perp}$ obtained from Eq. \eqref{eq: subobsc}, Eq. \eqref{eq: subnis} 
can then give an estimate of $B$.

On the other hand, we find a straightforward way to derive $B$ from the fitted velocity dispersion spectrum. Eq. \eqref{eq: subobsb} alone can give, 
\begin{equation}
\label{eq: subgenb}
B=\sqrt{4\pi \rho}~ b^{-3/2}\left(\frac{d}{206265}\right) L^{-1}V_L^4. \\
\end{equation}
If we assume the injection scale $L$ is comparable to the cloud's diameter, then the above expression becomes
\begin{equation}
 B=\sqrt{4\pi \rho}~ b^{-3/2}P^{-1}V_L^4,
\end{equation}
where $P=\frac{206265L}{d}$ is the angular diameter of the cloud. 
In particular, $V_L$ is especially important for determination of $B$ due to 
its $4$th power dependence. 

It is necessary to point out that 
LH08
method, with the assumption of isotropic super-Alfv\'{e}nic turbulence, can not apply to sub-Alfv\'{e}nic molecular clouds. 
We again perform a numerical test by using the parameters from \emph{Model 2}, but keep $L$ as a free parameter. 
Fig. \ref{fig:twomagsub} presents the error of the measured $B^\prime$ by employing Eq. \eqref{eq: magsupla} in comparison with 
the real $B$, as a function of $k_\text{dam} l_{tr}$. 
It shows in strong turbulence regime in a sub-Alfv\'{e}nic cloud, LH08 approach can underestimate the magnetic field strength 
considerably even when $1/k_\text{dam}$ is comparable to $l_{tr}$. 
At a smaller scale, for instance, $k_\text{dam}l_{tr} \sim 1.5 \times 10^6$ in the \emph{Model 2} cloud, the error comes to the 
maximum $\sim 100\%$.

Fig. \ref{fig:twomagto} informs us LH08 technique for measuring magnetic field in real clouds, indeed, can give reliable estimates 
when the damping takes place in Kolmogorov turbulence regime, i.e. $1/k_\text{dam}>l_A$. 
But, more commonly, $1/k_\text{dam}<l_A$ in a super-Alfv\'{e}nic cloud and $1/k_\text{dam}<l_{tr}$ in a sub-Alfv\'{e}nic cloud, 
LH08 technique can lead to an severe underestimate of the real $B$. For the former case, if $1/k_\text{dam}$ is relatively large, 
e.g. $k_\text{dam} l_A \sim 10$, LH08 estimate can be considered as a lower limit of the real $B$ and a zeroth order approximation which 
needs to be tested for the consistency through Eq. \eqref{eq: selconc} and corrected.
But when $1/k_\text{dam}$ is much smaller than $l_A$ ($M_A>1$) or $l_{tr}$ ($M_A<1$), LH08 method is not applicable.

An alternative method for determining $B$ is given in Appendix \ref{sec: altm}. 
In consideration of the possible discrepancy between the lower envelope of the 2-D velocity dispersions and the actual 3-D one, 
instead of using the turbulent velocity dispersion spectra, we directly use the constant difference between the squared velocity dispersions of neutrals and ions. 
We will apply these methods to observational data in our following work. 
It is worthwhile to notice that all these measurements suffer uncertainties due to LOS effect, which will be discussed in the following section.

\subsection{Effects of angle between magnetic field and line of sight}
\label{subsec: los}
A limitation of our analysis is that we assumed the LOS direction is perpendicular to the mean magnetic field for the measurements in the sub-Alfv\'{e}nic regime. 
When we consider Alfv\'{e}nic turbulence, the turbulent motions that account for the observed velocity dispersions are in the direction perpendicular to the magnetic field. 
But in practice, 
an observer can only measure the projected velocity dispersion along LOS, i.e. $\sigma_\text{LOS}$. We next briefly discuss the effect of this observational limitation on evaluation of $B$. 

Consider first sub-Alfv\'{e}nic turbulence.
All eddies of different sizes basically align with the global mean magnetic field $B_0$. Then the LOS component of $\sigma$ is 
\begin{equation}
 \sigma_\text{LOS}=\sigma \sin \alpha, 
\end{equation}
where $\alpha$ is the angle between $B_0$ and LOS. 
Accordingly, 
\begin{equation}
\Delta \sigma^2_\text{LOS}=\Delta \sigma^2 \sin^2\alpha.
\end{equation}
Therefore, the measured $B$ can be biased.
We take the case when $k_\text{dam}^{-1}<l_{tr}$ as an example. When employing Eq. \eqref{eq: alttecsubmg},  we will get an underestimated $B$, 
\begin{equation}
B^\prime=B  \sin^2\alpha.
\end{equation}
Here we estimate $V_L$ using the global $\sigma$ measured at the cloud size.

The measurement of turbulent velocity (i.e., velocity dispersion) in super-Alfv\'{e}nic turbulence does not depend on LOS orientation, since the fluid motions are isotropic above the scale $l_{A}$ (Eq. \eqref{eq: scarsuphy}).  
If a telescope's beam size is larger than $l_A$, most contribution in the observed $\sigma$ is from hydro-like motions, 
so LOS direction is irrelevant. 
But if the beam size is smaller than $l_A$, within each $l_A$-size eddy, all smaller eddies can be considered aligning along the local mean magnetic field of the $l_A$-size eddy, $B(l_A)$. 
Similar to above analysis, the turbulent velocity at scale $l$ ($l<l_A$) has a projection on LOS direction as 
$v_{l, \text{LOS}}=v_l\sin \alpha$. Here $\alpha$ is the angle between $B(l_A)$ and LOS. 
Since each LOS crosses regions with random $B(l_A)$ orientations, the observed $\sigma_\text{LOS}^2$ is an average over all orientations, 
\begin{equation}
  \left \langle \sigma_\text{LOS}^2 \right \rangle = \sigma^2 \left \langle \sin^2 \alpha \right \rangle =\sigma^2 \frac{\int^{\alpha_2}_{\alpha_1} \sin^2\alpha d\alpha}{\alpha_2-\alpha_1},
\end{equation}
where $(\alpha_1, \alpha_2)$ is the range of angles between the $l_A$-size eddies and LOS. Accordingly, $\Delta \sigma^2_\text{LOS}=\Delta \sigma^2  \left \langle \sin^2 \alpha \right \rangle$.
We see for MHD turbulence in super-Alfv\'{e}nic case, although $\Delta \sigma^2$ is independent of magnetic field strength, the observed LOS component of $\Delta \sigma^2$ depends on the 
orientations of local mean magnetic field at $l_A$ scale. 

Since the fit parameter $a(=-\Delta \sigma^2_\text{LOS})$ and $b$ in Eq. \eqref{eq: fitabsup} and \eqref{eq: subobs} are also subject to the LOS effect, 
the orientation of magnetic field to LOS introduces additional uncertainties for
determining magnetic field from the difference between the velocity dispersion spectra of neutrals and ions.

\section{Discussions}
\subsection{Turbulence in partially ionized gas: comparison with selected earlier works}
Waves in partially ionized gas are the subject well studied 
(e.g., \citealt{Braginskii:1965, Kum03, Zaqa11, Soler13, Sol13}). 
However, MHD turbulence is
different from linear waves. Difference in the properties of compressible and incompressible motions,
and anisotropy of MHD turbulence must be taken into account. Numerical studies in 
\citet{CL02_PRL, CL03}
quantified the coupling between the fast, slow and Alfv\'{e}n modes and provided the basis for
our present study. In particular, these studies show that it is legitimate to consider the Alfv\'{e}nic cascade
independently from the cascades of other modes 
(see also GS95, \citealt{LG01, CLV_incomp}).
To quantify the differences between the linewidths, we calculated the differences
of the squared velocity dispersions of neutrals and ions arising from the differences of the damping acting on
these species. 
In this paper we focused our attention on the Alfv\'{e}nic cascade, since the other fundamental components, fast modes and slow 
modes are more efficiently damped and do not have differential damping in neutrals and ions, 
unless under a particular condition, slow modes are damped in weakly coupled regime.  
This potentially opens a possibility of studying the ratio of the intensities of the latter components to that of Alfv\'{e}nic component. Indeed,
the expressions that we obtained for the linewidth differences depend on the properties of the Alfv\'{e}nic component of the turbulence. 
At the same time, the velocity dispersion at a sufficiently large scale depends on the amplitudes of all three components of turbulent cascade.
Note, that the separation into modes and studying the effects of individual modes has been proved to be a fruitful way for cosmic ray studies 
(see \citealt{YL02, YL04, YL08, YL11, Brunetti_Laz, BruLaz11, XY13, LY14}).

From the observational point of view, only the large-scale anisotropy is attainable
(\citealt{Hey08, Hey12}, also shown by the synthetic observations in \citealt{EL05}). 
That makes our formulae expressed in terms of the parameters at the driving scale of turbulence, i.e. $L$ and $V_L$, 
more valuable in practice. 
We would like to stress the fact that the super-Alfv\'{e}nic (or trans-Alfv\'{e}nic) turbulence observed at a cloud scale does not mean the anisotropies of turbulence are negligible also at the damping scale. 
As the amplitude of velocity perturbations decreases with scale, turbulence transfers from Kolmogorov-like at large scales to GS95 type 
at scales smaller than $l_A$. 
Our study shows the scale of interest $k_\text{dam}^{-1}$ is much smaller than $l_A$ in a typical molecular cloud (\emph{Model 1}, see 
Fig. \ref{fig:supdam}), where the approach described in this paper instead of LH08 is required. 

On the other hand, 
the equilibrium state in the linear analysis is assumed as a homogeneous partially ionized plasma. But in a real molecular cloud, 
when the inhomogeneity in density is significant, local $M_A$ may differ considerably from the global value at large scales. 
In particular, substantial density fluctuations expected in high-$M_A$ turbulence can appreciably change the local value of $V_A$, 
affecting the locally measured $M_A$. In this situation turbulence which is globally sub-Alfv\'{e}nic can have super-Alfv\'{e}nic 
dense clumps 
\citep{Burk09}.
Considering this complexity, for such high-density subvolumes our analysis should be modified with the local value of $M_A$ to be used 
rather than that for the entire cloud.

We consider MHD turbulence in partially ionized gas. This type of turbulence has been studied both numerically and analytically 
(see \citealt{LG01, TilBal10, MacL02, LVC04}).
In particular, \citet{LG01} (henceforth LG01) studied super-Alfv\'{e}nic compressible MHD turbulence in an ion-dominated medium with $\beta\gtrsim1$. 
In that work neutral-neutral collisions were neglected, and only neutral-ion collisional damping were considered. 
Through a different approach by including the force that neutrals exert on ions in the momentum equation in the derivation of dispersion relation, 
LG01 obtained the damping rates in asymptotic limits (equation (39) in LG01), 
\begin{subnumcases} 
 {\omega_I=-\frac{1}{2}\frac{m_n n_n}{m_i n_i} \nu_{ni}}
    \frac{1}{2}(kL_N)^2,~~~ kL_N \ll 1, \label{eq: LG01a} \\ 
    1 ,~~~~~~~~~~~~~~~~~ kL_N \gg 1. \label{eq: LG01b}
\end{subnumcases}
Here $m_n$ and $m_i$ are neutral and ion mass. $L_N=\frac{c_s}{\nu_{ni}}(\frac{2m_i}{m_n})^{1/2}$ is the neutral 
mean free path (equation (37) in LG01), but has different definition from our $l_n$. 
In comparison, the damping rates presented in this paper are general. Specifically, 
in decoupling regime, Eq. \eqref{eq: damrwd} in this paper agrees with the above expression at $kL_N \gg 1$. 
In the other limit when viscosity of neutrals dominates damping ($kL_N \ll 1$), Eq. \eqref{eq: nvsappb} in this paper recovers Eq. \eqref{eq: LG01a} in ion dominated environment, 
where neutral mean free path is determined by their interactions with ions.
An important point to be stressed is that, in LG01, the damping rate in strong coupling regime is expressed in terms of $k$. Namely, damping is isotropic in their consideration. 
They also adopt the approximation $k_\perp \approx k$ throughout their paper. 
In comparison, our analysis shows damping rate depends on the angle between $k$ and $B$ (see e.g., Eq. \eqref{eq: anasolsc}). Our analysis of the anisotropic damping and cascade 
allows us to decompose $k_\text{dam}$ into $k_{\text{dam}, \perp}$ and $k_{\text{dam}, \|}$, which turns out to be crucial in explaining the observed difference between the velocity dispersion spectra of the neutrals and ions.

A different treatment of MHD turbulence in partially ionized gas was given in 
\citet{LVC04}. 
The authors predicted interesting new effects, e.g. the resurrection of MHD cascade in ions. However, their paper
considers only the effects of neutral drag, which corresponds to $r>1$ in the present study. In the present paper we consider both neutral viscosity and neutral-ion damping. However, we did not go into considering all the cases of media covered in 
\citet{LVC04}. 

The possible influence of compressible modes as well as more cases of partially ionized media we are going to study in subsequent publications. 
Interpretation of linewidth differences
requires the detailed knowledge of the energy distribution in different modes. This potentially can be available through 
the analysis of observations 
(see \citealt{BurL13}).

\citet{BurLB14} investigated the behavior of Alfv\'{e}n modes in neutrals and ions by performing two-fluid MHD simulations. They confirmed that neutrals and ions form hydro and MHD cascades
separately as they decouple at the decoupling scale. 
They found Alfv\'{e}n modes in super-Alfv\'{e}nic turbulence damps below the decoupling scale. That is consistent with our findings on the relation between decoupling and damping scales 
(see Eq. \eqref{eq: damdecrel}).
A detailed quantitative comparison of the dependence of damping on turbulence parameters, e.g., $L$, $V_L$, $M_A$, presented in this work with their simulations 
will be a good test of our theoretical work.

\subsection{Differences between the squared velocity dispersions in neutrals and ions}
The difference between the turbulent velocity dispersion spectra of coexistent neutrals and ions is a known
observational fact  
(\citealt{Houde00a, Houde00b, Lai03} and references therein).
The work by 
\citet{Houde04}
related this difference to
ambipolar diffusion and the later work
LH08 related this to the differential damping of turbulence
in the fluids of neutrals and ions. The latter work is different from our study in an important way, namely,
we employ our knowledge of the properties of MHD turbulent cascade to calculate the damping.
On the basis of the up-to-date understanding of MHD turbulence, 
we found the variations of damping are determined by turbulent regimes and anisotropy. 
Our investigation confronts the essence of Alfv\'{e}nic turbulence 
and provides solutions which can be extensively applied in diverse ISM conditions. 

Our analysis did not take into account the possible differences in masses of different ionized and neutral species and its effect on 
differential damping. 
This is based on the assumption that the frequencies of magnetic perturbations that we study are much smaller than the resonance 
frequencies of ions with different masses. This approximation is well motivated for the problem of linewidth differences that we address.

\subsection{Magnetic field measurements}
\label{subsec: magmeadis}
Our study results in revealing several regimes which entail different dependence of the differences of squared
velocity dispersions on the underlying magnetic field. 
It shows the ability of determining magnetic field strength from the observed linewidth differences is limited.
This is in contrast to LH08 where a single universal expression relating the magnetic field with the measured 
difference between the squared velocity dispersion spectra of neutrals and ions was suggested. 
In practice, we identified the limited parameter space for which LH08 technique can be applied. 
It is restricted by the stringent condition that the turbulence should be super-Alfv\'{e}nic for the entire inertial range, from injection to damping, 
which is not common in typical molecular clouds. In all other cases different expressions should be employed. 
We found that when the damping scale is smaller than $l_A$ in super-Alfv\'{e}nic turbulence, the difference between the squared velocity 
dispersions of neutrals and ions is uncorrelated with magnetic field strength. 
Therefore evaluating magnetic field from the linewidth differences in this regime is not feasible. 
In cases when the linewidth differences can tell us the strength of the underlying magnetic field, we obtained expressions applicable  
in different observational situations.
The accuracy of this technique will be tested by further research.

To distinguish the regimes of turbulence we need additional data. Fortunately, new 
techniques have been developed to distinguish reliably super- and sub-Alfv\'{e}nic 
turbulence. Those include studies of anisotropy from centroids 
\citep{LP00, EL05, EL10}
as well as more complex techniques like Tsallis analysis 
(e.g., \citealt{EL11}).
With this information one then can use the corresponding expressions to obtain the actual value of the 
strength of magnetic field. A comparison of the approximate $M_A$ obtained by the above techniques
and one obtained with the linewidth differences can make the evaluation of magnetic field strength more reliable.

In other words, instead of the direct interpretation of the linewidth differences in terms of magnetic field,
as this is suggested in LH08, we propose a phased approach, where one first identifies the regime of turbulence and then uses the difference in neutral-ion linewidths to obtain the magnetization of the media with higher precision. 

Our study shows that the interpretation of the difference of linewidths of neutrals and ions in terms of magnetic field is not so straightforward as it follows from LH08 paper. 
The main complication augmented in this paper is related to the necessity of knowing the regime of turbulence in order 
to apply the appropriate expression for evaluating magnetic field, as discussed above.
The second set of complications is related to
determining 3-D velocity dispersions from the observed 2-D velocity dispersion distribution as was discussed by 
\citet{FalLaz10}.
Especially, we are not convinced that one should so much rely on the value of the turbulence spectral index that follows from relating 2D and 3D velocity dispersions. 
The theoretically motivated relation between the observed spectral indexes and that of the underlying spectrum of Alfv\'{e}nic turbulence 
(see \citealt{LP04,LP06,LP08}), 
cannot account for the aforementioned relation between the observed 2D and true 3D statistics. 
Instead, the empirical relation between them is only approximate, as explained in 
\citet{FalLaz10}.
The measurement of the turbulence spectral index near the damping scale is an additional complication, which is expected to distort the 
index compared to its value at the inertial range.\footnote{Additional distortions can occur in some range due to transfer of Alfv\'{e}nic turbulence from weak to strong regime, but they are expected to be smaller.}
Moreover, only Alfv\'{e}nic turbulence is considered in interpreting the different velocity dispersion spectra, 
but in fact both compressible and incompressible motions influence 
the spectral index measured in a way suggested in LH08. 
The potential effect of slow modes in contributing to linewidth differences can also decrease the accuracy of the magnetic field measurement with only Alfv\'{e}n modes taken into account.

Nevertheless, one should not dismiss the utility of the new way of obtaining
magnetic fields. It is well known that magnetic fields are notoriously difficult to measure in astrophysics.
For instance, one of the popular ways of measuring magnetic field is based on the Chandrasekhar Fermi technique
which is known to have limited accuracy related both to the technique 
(see \citealt{ChanFer53, Falce08, Houd13} and references therein) 
and to the assumption that grains are equally aligned at different depths in the cloud.
The variations of grain alignment degree that follow from the modern theory of grain alignment 
(see \citealt{Lazarian07rev} for a review)
introduce an additional complication for the quantitative study of magnetic fields with the
Chandrasekhar-Fermi technique.

In many branches of astrophysical research, including the application of Chandrasekhar-Fermi technique, 
it is frequently assumed that MHD turbulence is isotropic. 
For some applications, e.g. cosmic ray propagation and acceleration, it has been shown that this improper assumption 
results in substantial errors 
(see \citealt{YL02, YL04, YL08}). 
Chandrasekhar-Fermi Method was first proposed for estimating magnetic field in low-density sub-Alfv\'{e}nic regions
\citep{ChanFer53}, 
where turbulence anisotropy can be prominent 
(see e.g., \citealt{Houde13}).
But the issue on turbulence anisotropy has not been studied yet. 
Our present study probed one of the astrophysical problems associated with MHD turbulence, and  
demonstrates that the key to an accurate quantitative explanation on the observed linewidth differences  
is a proper comprehensive description of MHD turbulence.

\section{Summary}
Motivated by the observed linewidth differences between molecular neutral and ion species, 
we performed a thorough analysis on Alfv\'{e}nic turbulence cascade in partially ionized medium.
Meanwhile, fast modes are found to be irrelevant in interpreting linewidth differences, since they are damped out when neutrals and ions are still coupled. 
While the damping of slow modes strongly depends on parameters. Within a particular set of parameters, the differential damping can exist and result in 
linewidth differences of neutrals and ions.

Following the modern prescription of MHD turbulence, we acquired different expressions of damping scales in various turbulent regimes.
With the aim to fully capture the variety of astrophysical situations, we have considered both 
super- and sub-Alfv\'{e}nic turbulence. In terms of damping we considered
both neutral-ion collisions and neutral viscosity, and defined the conditions when either of the effects is dominant. 

We confirmed the observed linewidth differences can be explained by the different turbulent damping scales of neutral and ion fluids. However, we found that the differences depend on the regime of magnetic turbulence. 
We provided analytical expressions for the difference between the squared velocity dispersions of neutrals and ions for different regimes of turbulence. It is important that for some regimes the explicit expressions of 
damping scales allow us to deduce magnetic field strength in astrophysical applications.

By comparing with LH08 model, 
we found that their approach provides expression that is correct only when 
turbulence anisotropy on the damping scales can be discarded. Our work provides the expressions for
different regimes of Alfv\'{e}nic turbulence with different turbulence properties and varying degrees of anisotropy. 

On the basis of our study we propose new ways of studying magnetic field for both super-and sub-Alfv\'{e}nic
turbulence. These new techniques are intended to synergistically augment the existing ways of studying magnetic
fields in turbulent molecular clouds and interstellar media. 
\\
\\
S. X. acknowledges the support from China Scholarship Council during her stay in University of Wisconsin-Madison. 
A. L. acknowledges the support of the NSF grant AST-1212096.
S. X. and A. L. are grateful for the warm hospitality of the International Institute of Physics (Natal). 
S. X. and H. Y. are supported by the NSFC grant AST-11073004.
We thank Blakesley Burkhart for many stimulating discussions on turbulence. 
We are grateful to an anonymous referee for the very helpful comments which improve the quality of the paper.

\appendix

\section{Turbulent energy dissipation}
\label{app:a}
The anisotropy of turbulence damping has been studied earlier in 
\citet{YL04}.
However, their study was focused on the fast modes for which the effect of anisotropic damping was found extremely important. The cascading of 
fast modes is radically different from that of Alfv\'{e}n modes which we mainly deal with in this paper. Therefore the approach in 
\citet{YL04}
should be modified to take into account the efficient redistribution of energies over different directions in $k$ space as the Alfv\'{e}n modes cascade. 

In what follows we use the explicit form of the Alfv\'{e}n modes' energy tensor and calculate the damping for different directions of wave vector at a particular $k_\perp$. 
Then since in the process of Alfv\'{e}nic cascading, the directions of wavevectors are randomized, it is reasonable to consider the total energy dissipation integrated over all angles between 
$k$ and $B$. In a sense this is the limit case corresponding to the most efficient energy dissipation. 

Here we only discuss the case where the energy is dissipated through neutral-ion collisions. 

(1) \emph{Super-Alfv\'{e}nic turbulence}~~~~
In Kolmogorov turbulence regime, i.e. $l_A<k^{-1}<L$, the turbulent energy spectrum is isotropic. The energy transfer rate along the cascade is 
\begin{equation} 
  \eta_c(k)=E(k) \tau_{cas}^{-1}=\frac{2}{3} V_L^3 L^{-1} k^{-1}, 
\end{equation}
where $\tau_{cas}^{-1}$ is the cascading rate from Eq. \eqref{eq: supcaraa}, and $E(k)$ is given by Eq. \eqref{eq: supmhdspecb}. 
The energy loss due to neutral-ion collisions during one eddy turn over time, namely, the energy dissipation rate is 
\begin{equation} 
   \eta_d(k)=E(k) |\omega_I|=\frac{\xi_n}{3 \nu_{ni}}V_L^2V_A^2 L^{-2/3} k^{1/3} \langle \cos^2\theta \rangle =\frac{\xi_n}{6 \nu_{ni}}V_L^2V_A^2 L^{-2/3} k^{1/3} 
\end{equation}
where $\omega_I$ is taken from Eq. \eqref{eq: anasolsc}, and $\langle \cos^2\theta \rangle$ is a statistical average over all angles.
The scale on which neutral-ion collisions are efficient enough to cut off the Alfv\'{e}nic turbulence is the damping scale. If we set 
\begin{equation} 
      \eta_d(k) = \eta_c(k), 
\end{equation}
we arrive at 
\begin{equation} 
k_\text{dam}=2^{\frac{3}{2}}\nu_{ni}^{\frac{3}{4}}\xi_n^{-\frac{3}{4}}L^{-\frac{1}{4}}V_L^{\frac{3}{4}}V_A^{-\frac{3}{2}}, 
\end{equation}
the same as Eq. \eqref{eq: supkdama}.

In MHD turbulence regime ($k^{-1}<l_A$), turbulent energy has anisotropic distribution in $k$ space. Following the above method, the energy cascading rate is 
\begin{equation} 
  \eta_c(k_\perp, k_\|)=E(k_\perp, k_\|) \tau_{cas}^{-1}=\frac{1}{3\pi} V_A^3 l_A^{-2/3} k_\perp^{-8/3} \exp{(-l_A^{1/3}\frac{k_\|}{k_\perp^{2/3}})},
\end{equation}
where $E(k_\perp, k_\|)$ is the 3-D energy density given by Eq. \eqref{eq: 3denessup}. 
We can see $ \eta_c(k_\perp, k_\|)$ has the largest value when $k_\| \sim 0$. It means 
the energy cascade is highly anisotropic and mainly acts in the direction perpendicular to the magnetic field.
At a given $k_\perp$, instead of using the scaling indicated by Eq. \eqref{eq: supscal}, we integrate the above equation over all directions, and get 
\begin{equation} 
\begin{split}
     &  \eta_c(k_\perp)=\frac{1}{3\pi} V_A^3 l_A^{-2/3} k_\perp^{-8/3} \int_0^{k_\text{max}} \exp{(-l_A^{1/3}\frac{k_\|}{k_\perp^{2/3}})} dk_\| \\
     & ~~~~~~~~~~~\approx \frac{1}{3\pi} V_L^3 L^{-1}k_\perp^{-2}.
\end{split}
\end{equation}
Here $k_\|^{-1}$ extends to a sufficiently small scale, so that we can neglect the second term in the integral. 
The 3-D energy dissipation rate is 
\begin{equation} 
   \eta_d(k_\perp, k_\|)=E(k_\perp, k_\|) |\omega_I|=\frac{\xi_n}{6\pi \nu_{ni}}V_A^4 l_A^{-1/3} k_\perp^{-10/3} k_\|^2 \exp{(-l_A^{1/3}\frac{k_\|}{k_\perp^{2/3}})}.
\end{equation}
It decreases at both limit directions $k_\| \sim 0$ and $k_\| \sim k_\text{max}$, and reaches the largest value when 
$k_\| \sim  l_A^{-1/3} k_\perp^{2/3}$. 
Similarly, by integrating over all directions, we have 
\begin{equation} 
\label{eq: suporietad}
\begin{split}
 &  \eta_d(k_\perp)=\frac{\xi_n}{6\pi \nu_{ni}}V_A^4 l_A^{-1/3} k_\perp^{-10/3} \int_0^{k_\text{max}} k_\|^2 \exp{(-l_A^{1/3}\frac{k_\|}{k_\perp^{2/3}})} dk_\|. \\
 & ~~~~~~~~~~~~\approx \frac{\xi_n}{3\pi \nu_{ni}}V_L^4 L^{-4/3} k_\perp^{-4/3}.
\end{split}
\end{equation}
By equaling $\eta_d(k_\perp)$ and $\eta_c(k_\perp)$,  we get the perpendicular damping scale
\begin{equation} 
\label{eq: supenedams}
   k_{\text{dam},\perp} = (\frac{\nu_{ni}}{\xi_n})^{3/2}L^{1/2}V_L^{-3/2},
\end{equation}
only different from Eq. \eqref{eq: supkdamperp} by a constant. 

\begin{figure*}[h]
\centering
\subfigure[]{
   \includegraphics[width=6.5cm]{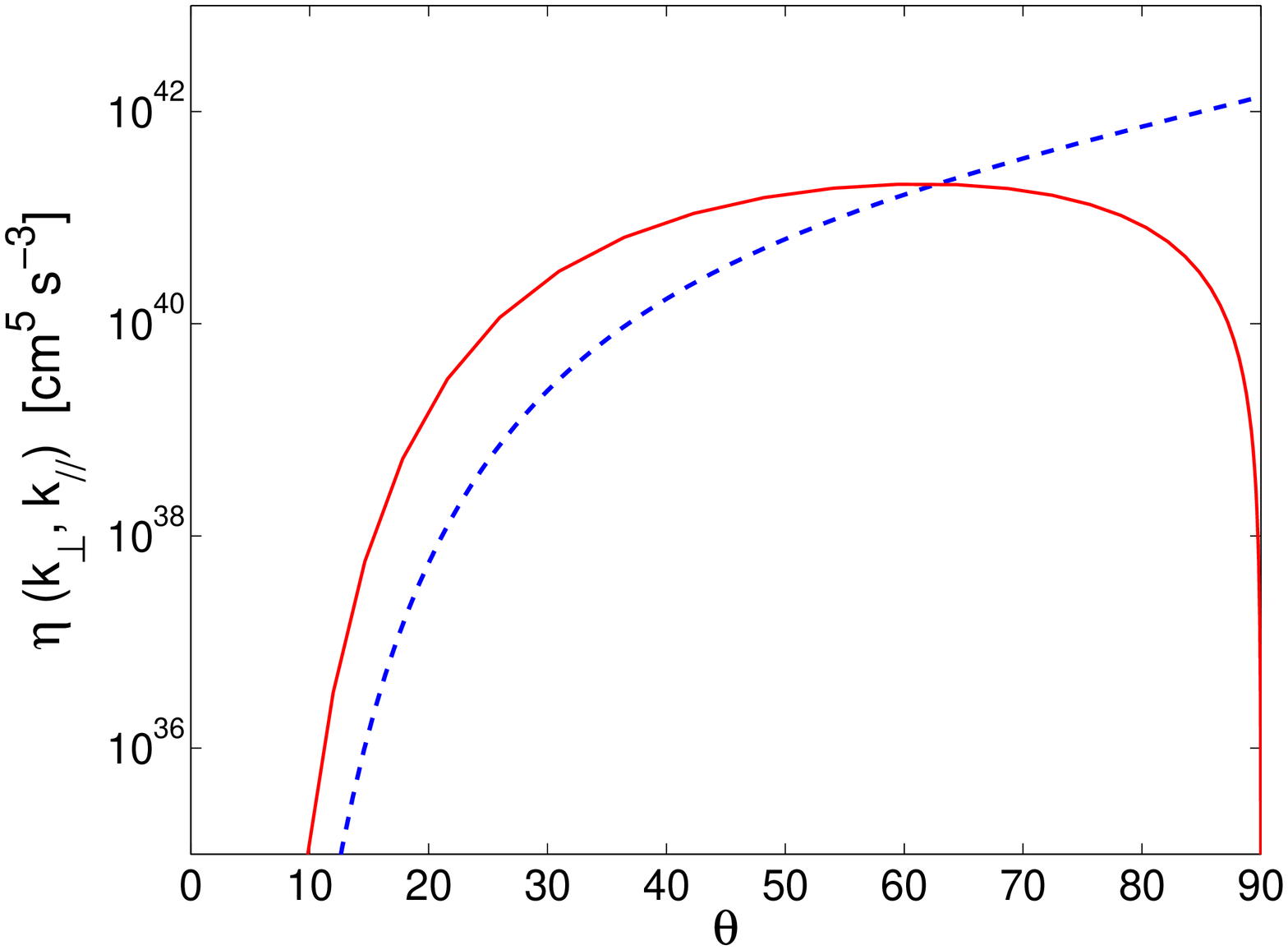}\label{fig:supener1}}
\subfigure[]{
   \includegraphics[width=6.5cm]{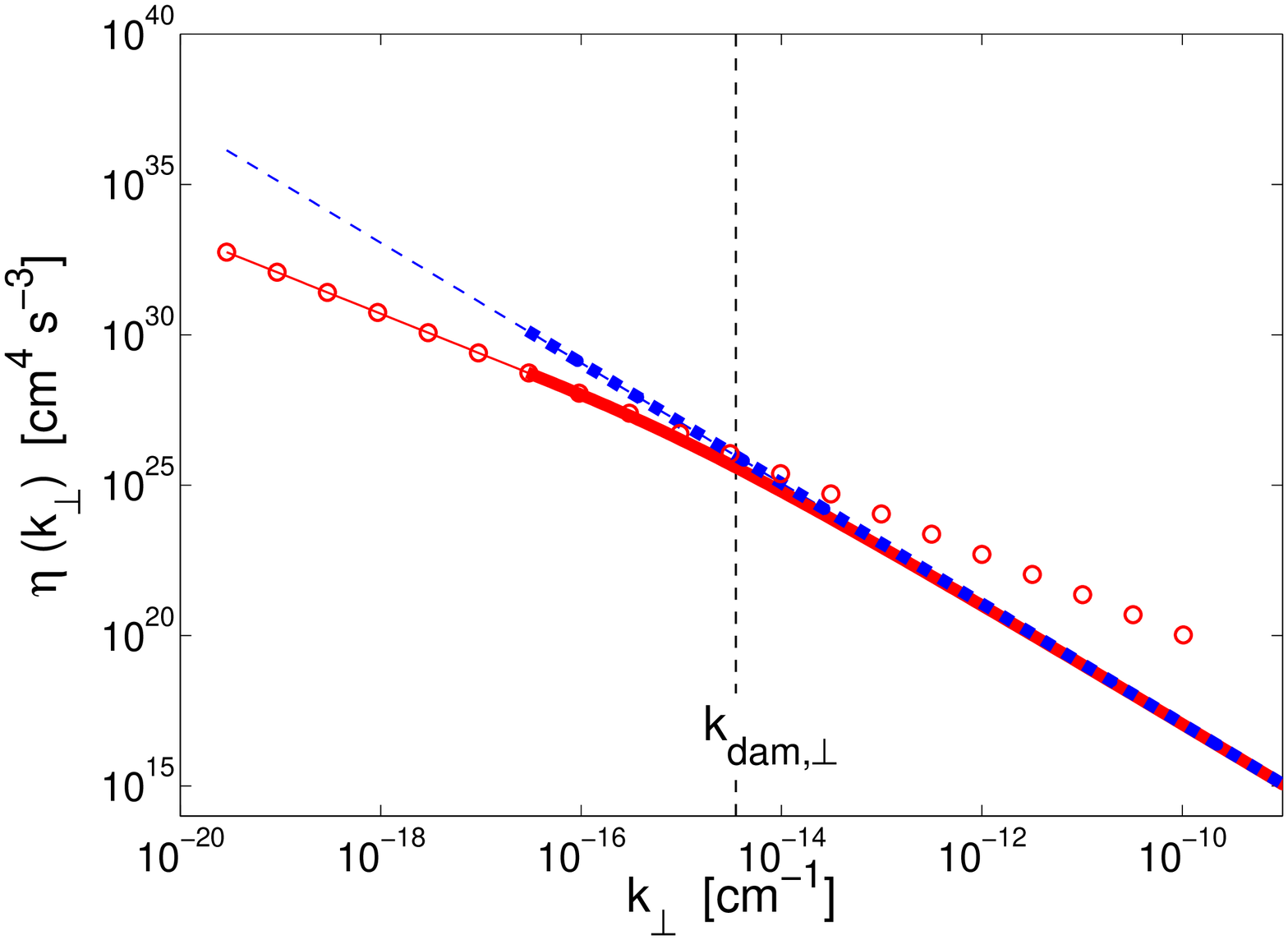}\label{fig:supener2}}
\caption{(a) $\eta_c(k_\perp, k_\|)$ (dashed line) and $\eta_d(k_\perp, k_\|)$ (solid line) vs. $\theta$ at $k_\perp l_A=50$. Parameters are taken from \emph{Model 1}. 
(b) $\eta_c(k_\perp)$ (dashed line) and $\eta_d(k_\perp)$ (solid line) vs. $k_\perp$ at scales where $k_\perp l_A>1$. Open circles are the original $\eta_d(k_\perp)$ (Eq. \eqref{eq: suporietad}). 
Its interaction with the dashed line gives $k_{\text{dam}, \perp}$ (vertical dashed line, Eq. \eqref{eq: supenedams}). Thinner lines correspond to the results with a larger $B$ ($=86.6$ $\mu$ G). }
\label{fig:supener}
\end{figure*}

We notice in fact, the energy dissipation rate is limited by the energy cascading rate. 
By using the parameters of \emph{Model 1}, 
Fig. \ref{fig:supener1} shows $\eta_c(k_\perp, k_\|)$ (dashed line) and $\eta_d(k_\perp, k_\|)$ (solid line) as a function of $\theta$ at $k_\perp l_A=50$ as an example. 
There are a range of directions in which $\eta_d(k_\perp, k_\|)$ is larger than $\eta_c(k_\perp, k_\|)$,  which cannot happen in reality. 
By taking this effect into account, we set $\eta_c(k_\perp, k_\|)$ as the actual $\eta_d(k_\perp, k_\|)$ when $\eta_d(k_\perp, k_\|)$ exceeds $\eta_c(k_\perp, k_\|)$ in a 
certain propagation direction. Then we numerically integrate the modified $\eta_d(k_\perp, k_\|)$ over all directions. 
Fig. \ref{fig:supener2} displays the resulting $\eta_c(k_\perp)$ (dashed line) and $\eta_d(k_\perp)$ (solid line) at different $k_\perp$ in strong MHD turbulence regime . 
Unlike the original $\eta_d(k_\perp)$ given by Eq. \eqref{eq: suporietad} (open circles), the modified $\eta_d(k_\perp)$ is always lower or equal to $\eta_c(k_\perp)$. 
The scale where $\eta_d(k_\perp)$ becomes comparable to $\eta_c(k_\perp)$ is the damping scale, which can be represented by $k_{\text{dam}, \perp}$ given by Eq. 
\eqref{eq: supenedams}, indicated by the vertical dashed line. 
It also corresponds to the intersection between the original $\eta_d(k_\perp)$ and $\eta_c(k_\perp)$. 

If we increase $B$ by $10$ times, the results (thinner lines) overlaps the ones with smaller $B$. It confirms our earlier conclusion that $k_{\text{dam}, \perp}$ is independent 
of $B$ in strong MHD turbulence for super-Alfv\'{e}nic case. 

(2) \emph{Sub-Alfv\'{e}nic turbulence}~~~~
In strong turbulence at $k^{-1}<l_{tr}$, the 3-D energy density is expressed in Eq. \eqref{eq: 3denessub}.
Similar to super-Alfv\'{e}nic case, we have the energy cascading rate 
\begin{equation} 
  \eta_c(k_\perp, k_\|)=E(k_\perp, k_\|) \tau_{cas}^{-1}=\frac{1}{3\pi} V_L^3 L^{-2/3}M_A^{-1/3} k_\perp^{-8/3} \exp{(-L^{1/3}\frac{k_\|}{M_A^{4/3}k_\perp^{2/3}})}. 
\end{equation}
It becomes 
\begin{equation} 
\begin{split}
     &  \eta_c(k_\perp)=\frac{1}{3\pi} V_L^3 L^{-2/3}M_A^{-1/3} k_\perp^{-8/3}  \int_0^{k_\text{max}} \exp{(-L^{1/3}\frac{k_\|}{M_A^{4/3}k_\perp^{2/3}})} dk_\| \\
     & ~~~~~~~~~~~\approx \frac{1}{3\pi} V_L^3 L^{-1}M_A k_\perp^{-2}.
\end{split}
\end{equation}
at a certain $k_\perp$. 
The energy dissipation rate is 
\begin{equation} 
   \eta_d(k_\perp, k_\|)=E(k_\perp, k_\|) |\omega_I|=\frac{\xi_n}{6\pi \nu_{ni}}V_L^2 L^{-1/3} M_A^{-2/3} V_A^2 k_\perp^{-10/3} k_\|^2 \exp{(-L^{1/3}\frac{k_\|}{M_A^{4/3}k_\perp^{2/3}})}. 
\end{equation}
By integrating over all directions, we get 
\begin{equation} 
\begin{split}
 &  \eta_d(k_\perp)=\frac{\xi_n}{6\pi \nu_{ni}}V_A^4 l_A^{-1/3} k_\perp^{-10/3} \int_0^{k_\text{max}} k_\|^2 \exp{(-l_A^{1/3}\frac{k_\|}{k_\perp^{2/3}})} dk_\| \\
 & ~~~~~~~~~~~~\approx \frac{\xi_n}{3\pi \nu_{ni}}V_L^{16/3} L^{-4/3}V_A^{-4/3} k_\perp^{-4/3}.
\end{split}
\end{equation}
The equality between $\eta_c(k_\perp)$ and $\eta_d(k_\perp)$ leads to
\begin{equation} 
\label{eq: subenedams}
   k_{\text{dam},\perp} = (\frac{\nu_{ni}}{\xi_n})^{3/2}L^{1/2}V_L^{-2}V_A^{1/2}, 
\end{equation}
only different from Eq. \eqref{eq: subnidamapp} by a constant. 

\begin{figure}[h]
\centering
 \includegraphics[width=6.5cm]{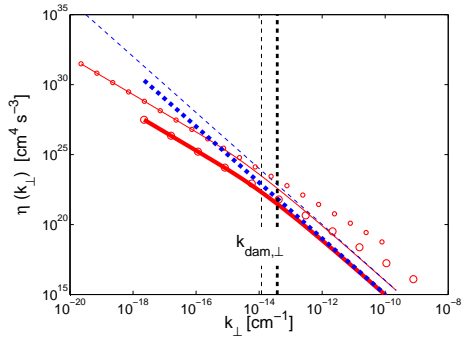}
\caption{Same as Fig. \ref{fig:supener2}, but for sub-Alfv\'{e}nic turbulence at $k_\perp l_{tr}>1$. Thicker lines and larger circles are results corresponding to larger $B$ ($=0.87$ mG).}
\label{fig:subener}
\end{figure}

We again modify $\eta_d(k_\perp, k_\|)$ to be not larger than $\eta_c(k_\perp, k_\|)$ in any direction. 
Fig. \ref{fig:subener} shows the integrals $\eta_c(k_\perp)$ and $\eta_d(k_\perp)$ by using parameters from \emph{Model 2}.  The same symbols as Fig. \ref{fig:supener2} are used here. 
Different from the super-Alfv\'{e}nic case, when we increase $B$ by 10 times to $0.87$ mG, both $\eta_c(k_\perp)$ and $\eta_d(k_\perp)$ (thicker lines) shift, together with $k_{\text{dam}, \perp}$ 
given by Eq. \eqref{eq: subenedams}. It shows in strong sub-Alfv\'{e}nic turbulence, $k_{\text{dam}, \perp}$ indeed depends on $B$. And the dependence is well described by 
Eq. \eqref{eq: subenedams} (or Eq. \eqref{eq: subnidamapp}). 

Assuming that the randomization of wave vectors in the cascading process is so fast that we get an upper estimate of the dissipation which provides
a order of unity correspondence with the results in the main text of the paper. 
One can argue that the dissipation of energy at angles close to $90$ degree (see Figure \ref{fig:supener1}) may happen faster than the replenishment of energy during the cascade. 
However, if this happens, we expect that the Alfv\'{e}nic cascade with $k_\perp \gg k_\|$ stops. The consistency of the estimates in the Appendix and main text supports this idea. 
Naturally, numerical studies of the cascade in ion-electron and neutral fluids can test the accuracy of our assumption.

\section{The cutoff region of MHD waves in different turbulent regimes}
\label{app:b}
Cutoff region is introduced in MHD turbulence regime in super-Alfv\'{e}nic turbulence and strong turbulence regime in sub-Alfv\'{e}nic 
turbulence. 

(1) \emph{Alfv\'{e}n modes}~~~~
By setting the discriminant of Eq. \eqref{eq:dp} as zero, the lower and upper limit wave numbers of the non-propagation region are given by 
\citep{Sol13}, 
\begin{equation}
\label{eq: oricff}
k^{\pm}=\frac{\nu_{ni}}{V_{Ai}\cos\theta}\left[\frac{\chi^2+20\chi-8}{8(1+\chi)^3}\pm\frac{\chi^{1/2}(\chi-8)^{3/2}}{8(1+\chi)^3}\right]^{-1/2}.
\end{equation}
Notice that $\cos\theta$ is also a function of $k$. Under the consideration of scale-dependent anisotropy, we arrive at new expressions for the 
limit wave numbers of the cutoff region. 
By applying Eq. \eqref{eq: supscal}, 
the above expression becomes 
\begin{subequations}
 \label{eq: supcfkptot} 
 \begin{align}
&k^{+}=(f(\chi))^{-\frac{3}{4}}\nu_{ni}^{\frac{3}{2}}\xi_i^{\frac{3}{4}}L^{\frac{1}{2}}V_L^{-\frac{3}{2}} \sqrt{1+(f(\chi))^{\frac{1}{2}}\nu_{ni}^{-1}\xi_i^{-\frac{1}{2}}L^{-1}V_L^3V_A^{-2}}, \label{eq: supcfkp} \\
&k^{-}=(g(\chi))^{-\frac{3}{4}}\nu_{ni}^{\frac{3}{2}}\xi_i^{\frac{3}{4}}L^{\frac{1}{2}}V_L^{-\frac{3}{2}} \sqrt{1+(g(\chi))^{\frac{1}{2}}\nu_{ni}^{-1}\xi_i^{-\frac{1}{2}}L^{-1}V_L^3V_A^{-2}},
\end{align}
 \end{subequations}
for super-Alfv\'{e}nic turbulence. And in combination with Eq. \eqref{eq: subscal}, Eq. \eqref{eq: oricff} becomes 
\begin{subequations}
 \label{eq: subcfkptot} 
 \begin{align}
&k^{+}=(f(\chi))^{-\frac{3}{4}}\nu_{ni}^{\frac{3}{2}}\xi_i^{\frac{3}{4}}L^{\frac{1}{2}}V_L^{-2}V_A^{\frac{1}{2}} \sqrt{1+(f(\chi))^{\frac{1}{2}}\nu_{ni}^{-1}\xi_i^{-\frac{1}{2}}L^{-1}V_L^4V_A^{-3}}, \label{eq: subcfkp} \\
&k^{-}=(g(\chi))^{-\frac{3}{4}}\nu_{ni}^{\frac{3}{2}}\xi_i^{\frac{3}{4}}L^{\frac{1}{2}}V_L^{-2}V_A^{\frac{1}{2}} \sqrt{1+(g(\chi))^{\frac{1}{2}}\nu_{ni}^{-1}\xi_i^{-\frac{1}{2}}L^{-1}V_L^4V_A^{-3}}, \label{eq: subcfkpb} 
\end{align}
 \end{subequations}
for sub-Alfv\'{e}nic turbulence,
where 
\begin{equation}
\begin{split}
&f(\chi)=\frac{\chi^2+20\chi-8}{8(1+\chi)^3}+\frac{\chi^{\frac{1}{2}}(\chi-8)^{\frac{3}{2}}}{8(1+\chi)^3}, \\
&g(\chi)=\frac{\chi^2+20\chi-8}{8(1+\chi)^3}-\frac{\chi^{\frac{1}{2}}(\chi-8)^{\frac{3}{2}}}{8(1+\chi)^3},
\end{split}
\end{equation}
and $\xi_i=\rho_i/\rho$. 
We can clearly see the expressions of $k^{+}$ given by Eq. \eqref{eq: supcfkp} and \eqref{eq: subcfkp} are very similar to the damping scales in strong MHD turbulence given by Eq. \eqref{eq: supkdamb} and \eqref{eq: subkdamb}. 

In the case of Alfv\'{e}n modes, as long as cutoff region $(k_c^+, k_c^-)$ resides in the strong MHD turbulence regime, 
we always have $k_c^+=k_\text{dam}$, its simplified forms in different situations can be found in Table \ref{Tab: Dams}.

For $k_c^-$ in super-Alfv\'{e}nic turbulence, Eq. \eqref{eq: cbaltw} and \eqref{eq: supscal} together give 
\begin{equation}
\label{eq: alfkcmsup}
  k_c^-=\frac{\nu_{in}}{2V_{Ai}} \sqrt{1+l_A \frac{\nu_{in}}{2V_{Ai}}}.
\end{equation}
$k_c^-$ in sub-Alfv\'{e}nic turbulence is then (Eq. \eqref{eq: cbaltw} and \eqref{eq: subscal})
\begin{equation}\label{eq: alfkcm}
  k_c^-=\frac{\nu_{in}}{2V_{Ai}} \sqrt{1+L M_A^{-4} \frac{\nu_{in}}{2V_{Ai}}}.
\end{equation}

(2) \emph{Slow modes}~~~~
The limit wavenumbers of the cutoff are 
\begin{equation} \label{eq: slcfbdsup}
\begin{split}
 &  k_c^+=\Bigg[\frac{2\nu_{ni}}{c_s \xi_n l_A} \bigg(1+\sqrt{\frac{2\nu_{ni}l_A}{c_s \xi_n }}\bigg)\Bigg]^{1/2}, \\
 &  k_c^-=\frac{\nu_{in}}{2c_{si}} \sqrt{1+l_A \frac{\nu_{in}}{2c_{si}}},
\end{split}
\end{equation}
for super-Alfv\'{e}nic turbulence, and 
\begin{equation}
\begin{split}
   & k_c^+=\Bigg[\frac{2\nu_{ni}}{c_s \xi_n L M_A^{-4}} \bigg(1+\sqrt{\frac{2\nu_{ni} LM_A^{-4}}{c_s \xi_n }}\bigg)\Bigg]^{1/2}, \\
   & k_c^-=\frac{\nu_{in}}{2c_{si}} \sqrt{1+LM_A^{-4} \frac{\nu_{in}}{2c_{si}}},
\end{split}
\end{equation}
for sub-Alfv\'{e}nic turbulence.

\section{Alternative observation technique on magnetic field determination }
\label{sec: altm}
The main limitations of present observations (Section \ref{subsec: applica}) is that notable uncertainties can be introduced 
when employing the lower envelope of the 2-D velocity dispersions as the real 3-D one. 
We next introduce an alternative technique with the aim of achieving more accurate observational data and reliable determination of magnetic field. 

\subsection{Super-Alfv\'{e}nic case}
For a molecular cloud within super-Alfv\'{e}nic turbulence regime, our analysis showed the difference of the squared velocity dispersions of neutrals and ions is
(Section \ref{subsec: supvd})
\begin{equation}
\Delta \sigma^2 \sim 
\begin{cases}
 L^{-2/3}V_L^2 k_\text{dam}^{-2/3},~~~~~~~~~~~~~~~1/k_{\text{dam}} > l_A, \\
 L^{-2/3}V_L^2 k_{\text{dam}, \perp}^{-2/3},~~~~~~~~~~~~~~1/k_{\text{dam}} < l_A. \\
\end{cases}
\end{equation}
If we take the case with neglected neutral viscosity as an example, in combination with Eq. \eqref{eq: magsupl}, $B$ can be expressed 
in terms of $\Delta \sigma^2$, only for $k_{\text{dam}}^{-1} > l_A$, namely 
\begin{equation}
 B=4(\pi \nu_{ni})^{1/2} \rho^{1/2} \xi_n ^{-1/2} L^{1/2} V_L^{-3/2}\Delta \sigma^2.
\end{equation} 
We see that, to determine $k_\text{dam}$ and $B$, parameters $\Delta \sigma^2$, $V_L$, and $L$ are needed.
 $V_L$ can be taken as the global turbulent velocity measured at the cloud size $\sim L$. 
 $\Delta \sigma^2$ is the difference between the neutral and ion squared velocity dispersions, independent of length scales. 
 Thus we can avoid the uncertainties from multiple-scale measurements. 

\subsection{Sub-Alfv\'{e}nic case}
We consider $k_\text{dam}$ in strong turbulence regime. Similar to super-Alfv\'{e}nic case, with $V_L$, $\Delta \sigma^2$, and $L$ measured using aforementioned method, on the ground of 
\begin{equation}
\Delta \sigma^2=L^{-2/3}V_L^2M_A^{2/3}k_{\text{dam}, \perp}^{-2/3},
\end{equation}
$k_{\text{dam}, \perp}$ becomes a function of a single variable $B$. Together with the expressions of $B$, both $k_{\text{dam}, \perp}$($\sim k_\text{dam}$) and $B$ can be 
evaluated. In combination with Eq. \eqref{eq: subnis}, we obtain 
\begin{equation}
\label{eq: alttecsubmg}
B=\sqrt{4\pi\rho}(\frac{2\nu_{ni}}{\xi_{n}})^{-1}L^{-1}V_L^4 (\Delta \sigma^2)^{-1}
\end{equation}
for $k_\text{dam}^{-1}< l_{tr}$.
Compared with Eq. \eqref{eq: subgenb}, the new method (Eq. \eqref{eq: alttecsubmg}) has the advantage of single-scale measurement.

The prominent advantage of the technique is by performing measurements at one scale, we can avoid the uncertainties from the fitted velocity dispersion spectra. 
With more observational sources attained, it will be of great value to test this new technique in our future work.

\section{A summary of the notations used in this paper}
\label{app:c}
\begin{longtable}{|l|c|}
\hline
wave number & $k$ \\
$k$ component parallel to local magnetic field & $k_\|$ \\
$k$ component perpendicular to local magnetic field & $k_\perp$ \\
injection scale of turbulence & $L$ \\
injection scale of strong turbulence & $l_A$ \\
transition scale from weak to strong turbulence & $l_{tr}$ \\
turbulent velocity at $L$ & $V_L$ \\
turbulent velocity at a scale $l$  & $v_l$ \\
turbulent velocity at $l_{tr}$  & $v_{tr}$ \\
magnetic field & $\mathbf{B}$ \\
Alfv\'{e}n speed & $V_A$ \\
Alfv\'{e}n speed of ion-electron gas & $V_{Ai}$ \\
Alfv\'{e}nic Mach number & $M_A$ \\
cascading rate & $\tau_{cas}^{-1}$ \\
neutral-ion collision frequency & $\nu_{ni}$ \\
ion-neutral collision frequency & $\nu_{in}$ \\
drag coefficient & $\gamma_d$ \\
ion density & $\rho_i$ \\
neutral density & $\rho_n$ \\
total density & $\rho$ \\
decoupling scale & $k_\text{dec}^{-1}$ \\
damping scale & $k_\text{dam}^{-1}$ \\
viscous scale & $k_\nu^{-1}$ \\
wave frequency & $\omega$ \\
real part of wave frequency & $\omega_R$ \\
imaginary part of wave frequency & $\omega_I$ \\
wave propagation angle with regard of magnetic field & $\theta$ \\
ion fraction & $\xi_i$ \\
neutral fraction & $\xi_n$ \\
ratio of neutral and ion densities & $\chi$ \\
collision frequency of neutrals & $\tau_\upsilon^{-1}$ \\
mass of hydrogen atom & $m_H$ \\
ion number density & $n_i$ \\
neutral number density & $n_n$ \\
total number density & $n$ \\
mean free path for a neutral particle & $l_n$ \\
cross section for a neutral-neutral collision & $\sigma_{nn}$ \\
pressure of neutral gas & $P_n$ \\
adiabatic constant & $\gamma$ \\
sound speed & $c_s$ \\
sound speed in neutrals & $c_{sn}$ \\
sound speed in ions & $c_{si}$ \\
limit wave numbers of non-propagation region & $k^+$, $k^-$ \\
limit wave numbers of cutoff region & $k_c^+$, $k_c^-$ \\
temperature & $T$ \\
kinematic viscosity & $\nu_n$ \\
ratio of gas pressure to magnetic pressure & $\beta$ \\
ratio of gas pressure to magnetic pressure in ions & $\beta_i$ \\
turbulent energy spectra density & $E(k)$ \\
energy dissipation rate & $\eta_d$ \\
energy cascading rate & $\eta_c$ \\
velocity dispersion & $\sigma$ \\
squared velocity dispersion of ions & $\sigma_i^2$ \\
squared velocity dispersion of neutrals & $\sigma_n^2$ \\
difference of the squared velocity dispersions of ions and neutrals & $\Delta \sigma^2$ \\
ambipolar Reynolds number & $R_{AD}$ \\
effective magnetic diffusivity & $\eta_{amb}$ \\
ambipolar diffusion scale  & $L_{AD}$ \\
fit parameters to observational  velocity dispersion spectrum & $a$, $b$ \\
beam size & $p$ \\
distance & $d$ \\

\hline

\end{longtable}

\bibliographystyle{apj.bst}
\bibliography{yan}

\begin{thebibliography}{81}
\expandafter\ifx\csname natexlab\endcsname\relax\def\natexlab#1{#1}\fi

\bibitem[{{Armstrong} {et~al.}(1995){Armstrong}, {Rickett}, \&
  {Spangler}}]{Armstrong95}
{Armstrong}, J.~W., {Rickett}, B.~J., \& {Spangler}, S.~R. 1995, \apj, 443, 209

\bibitem[{{Beresnyak} \& {Lazarian}(2008)}]{BL08}
{Beresnyak}, A., \& {Lazarian}, A. 2008, \apj, 682, 1070

\bibitem[{{Beresnyak} \& {Lazarian}(2009)}]{BL09}
---. 2009, \apj, 702, 1190

\bibitem[{{Biskamp}(2003)}]{Bisk03}
{Biskamp}, D. 2003, {Magnetohydrodynamic Turbulence}

\bibitem[{{Braginskii}(1965)}]{Braginskii:1965}
{Braginskii}, S.~I. 1965, Reviews of Plasma Physics, 1, 205

\bibitem[{{Brandenburg} \& {Lazarian}(2013)}]{Brad13}
{Brandenburg}, A., \& {Lazarian}, A. 2013, \ssr

\bibitem[{{Brandenburg} \& {Lazarian}(2014)}]{BraL14}
---. 2014, {Astrophysical Hydromagnetic Turbulence}, ed. A.~{Balogh},
  A.~{Bykov}, P.~{Cargill}, R.~{Dendy}, T.~{Dudok de Wit}, \& J.~{Raymond}, 87

\bibitem[{{Brunetti} \& {Lazarian}(2007)}]{Brunetti_Laz}
{Brunetti}, G., \& {Lazarian}, A. 2007, \mnras, 378, 245

\bibitem[{{Brunetti} \& {Lazarian}(2011)}]{BruLaz11}
---. 2011, \mnras, 412, 817

\bibitem[{{Burkhart} {et~al.}(2009){Burkhart}, {Falceta-Gon{\c c}alves},
  {Kowal}, \& {Lazarian}}]{Burk09}
{Burkhart}, B., {Falceta-Gon{\c c}alves}, D., {Kowal}, G., \& {Lazarian}, A.
  2009, \apj, 693, 250

\bibitem[{{Burkhart} \& {Lazarian}(2013)}]{BurL13}
{Burkhart}, B., \& {Lazarian}, A. 2013, in IAU Symposium, Vol. 294, IAU
  Symposium, ed. A.~G. {Kosovichev}, E.~{de Gouveia Dal Pino}, \& Y.~{Yan},
  325--336

\bibitem[{{Burkhart} {et~al.}(2014){Burkhart}, {Lazarian}, {Balsara}, {Meyer},
  \& {Cho}}]{BurLB14}
{Burkhart}, B., {Lazarian}, A., {Balsara}, D., {Meyer}, C., \& {Cho}, J. 2014,
  arXiv: 1412.3452

\bibitem[{{Chandran}(2008)}]{ChaB08}
{Chandran}, B.~D.~G. 2008, Physical Review Letters, 101, 235004

\bibitem[{{Chandrasekhar} \& {Fermi}(1953)}]{ChanFer53}
{Chandrasekhar}, S., \& {Fermi}, E. 1953, \apj, 118, 113

\bibitem[{{Chepurnov} \& {Lazarian}(2010)}]{CheL10}
{Chepurnov}, A., \& {Lazarian}, A. 2010, \apj, 710, 853

\bibitem[{{Cho} \& {Lazarian}(2002)}]{CL02_PRL}
{Cho}, J., \& {Lazarian}, A. 2002, Physical Review Letters, 88, 245001

\bibitem[{{Cho} \& {Lazarian}(2003)}]{CL03}
---. 2003, \mnras, 345, 325

\bibitem[{{Cho} \& {Lazarian}(2005)}]{CL05}
---. 2005, Theoretical and Computational Fluid Dynamics, 19, 127

\bibitem[{{Cho} {et~al.}(2002{\natexlab{a}}){Cho}, {Lazarian}, \&
  {Vishniac}}]{CLV_newregime}
{Cho}, J., {Lazarian}, A., \& {Vishniac}, E.~T. 2002{\natexlab{a}}, \apjl, 566,
  L49

\bibitem[{{Cho} {et~al.}(2002{\natexlab{b}}){Cho}, {Lazarian}, \&
  {Vishniac}}]{CLV_incomp}
---. 2002{\natexlab{b}}, \apj, 564, 291

\bibitem[{{Crutcher} {et~al.}(2010){Crutcher}, {Wandelt}, {Heiles},
  {Falgarone}, \& {Troland}}]{Crut10}
{Crutcher}, R.~M., {Wandelt}, B., {Heiles}, C., {Falgarone}, E., \& {Troland},
  T.~H. 2010, \apj, 725, 466

\bibitem[{{Dobrowolny} {et~al.}(1980){Dobrowolny}, {Mangeney}, \&
  {Veltri}}]{Dob80}
{Dobrowolny}, M., {Mangeney}, A., \& {Veltri}, P. 1980, \aap, 83, 26

\bibitem[{{Draine} {et~al.}(1983){Draine}, {Roberge}, \& {Dalgarno}}]{Drai83}
{Draine}, B.~T., {Roberge}, W.~G., \& {Dalgarno}, A. 1983, \apj, 264, 485

\bibitem[{{Esquivel} \& {Lazarian}(2005)}]{EL05}
{Esquivel}, A., \& {Lazarian}, A. 2005, \apj, 631, 320

\bibitem[{{Esquivel} \& {Lazarian}(2010)}]{EL10}
---. 2010, \apj, 710, 125

\bibitem[{{Esquivel} \& {Lazarian}(2011)}]{EL11}
---. 2011, \apj, 740, 117

\bibitem[{{Eyink} {et~al.}(2013){Eyink}, {Vishniac}, {Lalescu}, {Aluie},
  {Kanov}, {B{\"u}rger}, {Burns}, {Meneveau}, \& {Szalay}}]{Eyin13}
{Eyink}, G., {et~al.} 2013, \nat, 497, 466

\bibitem[{{Falceta-Gon{\c c}alves} {et~al.}(2010){Falceta-Gon{\c c}alves},
  {Lazarian}, \& {Houde}}]{FalLaz10}
{Falceta-Gon{\c c}alves}, D., {Lazarian}, A., \& {Houde}, M. 2010, \apj, 713,
  1376

\bibitem[{{Falceta-Gon{\c c}alves} {et~al.}(2008){Falceta-Gon{\c c}alves},
  {Lazarian}, \& {Kowal}}]{Falce08}
{Falceta-Gon{\c c}alves}, D., {Lazarian}, A., \& {Kowal}, G. 2008, \apj, 679,
  537

\bibitem[{{Ferriere} {et~al.}(1988){Ferriere}, {Zweibel}, \& {Shull}}]{Ferr88}
{Ferriere}, K.~M., {Zweibel}, E.~G., \& {Shull}, J.~M. 1988, \apj, 332, 984

\bibitem[{{Goldreich} \& {Sridhar}(1995)}]{GS95}
{Goldreich}, P., \& {Sridhar}, S. 1995, \apj, 438, 763

\bibitem[{{Heyer} {et~al.}(2008){Heyer}, {Gong}, {Ostriker}, \&
  {Brunt}}]{Hey08}
{Heyer}, M., {Gong}, H., {Ostriker}, E., \& {Brunt}, C. 2008, \apj, 680, 420

\bibitem[{{Heyer} \& {Brunt}(2012)}]{Hey12}
{Heyer}, M.~H., \& {Brunt}, C.~M. 2012, \mnras, 420, 1562

\bibitem[{{Hezareh} {et~al.}(2014){Hezareh}, {Csengeri}, {Houde}, {Herpin}, \&
  {Bontemps}}]{Hez14}
{Hezareh}, T., {Csengeri}, T., {Houde}, M., {Herpin}, F., \& {Bontemps}, S.
  2014, \mnras, 438, 663

\bibitem[{{Hezareh} {et~al.}(2010){Hezareh}, {Houde}, {McCoey}, \&
  {Li}}]{Hez10}
{Hezareh}, T., {Houde}, M., {McCoey}, C., \& {Li}, H.-b. 2010, \apj, 720, 603

\bibitem[{{Houde} {et~al.}(2000{\natexlab{a}}){Houde}, {Bastien}, {Peng},
  {Phillips}, \& {Yoshida}}]{Houde00a}
{Houde}, M., {Bastien}, P., {Peng}, R., {Phillips}, T.~G., \& {Yoshida}, H.
  2000{\natexlab{a}}, \apj, 536, 857

\bibitem[{{Houde} {et~al.}(2004){Houde}, {Dowell}, {Hildebrand}, {Dotson},
  {Vaillancourt}, {Phillips}, {Peng}, \& {Bastien}}]{Houde04}
{Houde}, M., {Dowell}, C.~D., {Hildebrand}, R.~H., {Dotson}, J.~L.,
  {Vaillancourt}, J.~E., {Phillips}, T.~G., {Peng}, R., \& {Bastien}, P. 2004,
  \apj, 604, 717

\bibitem[{{Houde} {et~al.}(2013{\natexlab{a}}){Houde}, {Fletcher}, {Beck},
  {Hildebrand}, {Vaillancourt}, \& {Stil}}]{Houde13}
{Houde}, M., {Fletcher}, A., {Beck}, R., {Hildebrand}, R.~H., {Vaillancourt},
  J.~E., \& {Stil}, J.~M. 2013{\natexlab{a}}, \apj, 766, 49

\bibitem[{{Houde} {et~al.}(2013{\natexlab{b}}){Houde}, {Hezareh}, {Jones}, \&
  {Rajabi}}]{Houd13}
{Houde}, M., {Hezareh}, T., {Jones}, S., \& {Rajabi}, F. 2013{\natexlab{b}},
  \apj, 764, 24

\bibitem[{{Houde} {et~al.}(2000{\natexlab{b}}){Houde}, {Peng}, {Phillips},
  {Bastien}, \& {Yoshida}}]{Houde00b}
{Houde}, M., {Peng}, R., {Phillips}, T.~G., {Bastien}, P., \& {Yoshida}, H.
  2000{\natexlab{b}}, \apj, 537, 245

\bibitem[{{Houde} {et~al.}(2002){Houde}, {Bastien}, {Dotson}, {Dowell},
  {Hildebrand}, {Peng}, {Phillips}, {Vaillancourt}, \& {Yoshida}}]{Houde02}
{Houde}, M., {et~al.} 2002, \apj, 569, 803

\bibitem[{{Kamaya} \& {Nishi}(1998)}]{KalN98}
{Kamaya}, H., \& {Nishi}, R. 1998, \apj, 500, 257

\bibitem[{{Kowal} \& {Lazarian}(2010)}]{KowL10}
{Kowal}, G., \& {Lazarian}, A. 2010, \apj, 720, 742

\bibitem[{{Kowal} {et~al.}(2009){Kowal}, {Lazarian}, {Vishniac}, \&
  {Otmianowska-Mazur}}]{KowL09}
{Kowal}, G., {Lazarian}, A., {Vishniac}, E.~T., \& {Otmianowska-Mazur}, K.
  2009, in Revista Mexicana de Astronomia y Astrofisica Conference Series,
  Vol.~36, Revista Mexicana de Astronomia y Astrofisica Conference Series,
  89--96

\bibitem[{{Kulsrud} \& {Pearce}(1969)}]{Kulsrud_Pearce}
{Kulsrud}, R., \& {Pearce}, W.~P. 1969, \apj, 156, 445

\bibitem[{{Kumar} \& {Roberts}(2003)}]{Kum03}
{Kumar}, N., \& {Roberts}, B. 2003, \solphys, 214, 241

\bibitem[{{Lai} {et~al.}(2003){Lai}, {Velusamy}, \& {Langer}}]{Lai03}
{Lai}, S.-P., {Velusamy}, T., \& {Langer}, W.~D. 2003, \apjl, 596, L239

\bibitem[{{Lazarian}(2006)}]{Lazarian06}
{Lazarian}, A. 2006, \apjl, 645, L25

\bibitem[{{Lazarian}(2007)}]{Lazarian07rev}
---. 2007, Journal of Quantitative Spectroscopy and Radiative Transfer, 106,
  225

\bibitem[{{Lazarian} \& {Pogosyan}(2000)}]{LP00}
{Lazarian}, A., \& {Pogosyan}, D. 2000, \apj, 537, 720

\bibitem[{{Lazarian} \& {Pogosyan}(2004)}]{LP04}
---. 2004, \apj, 616, 943

\bibitem[{{Lazarian} \& {Pogosyan}(2006)}]{LP06}
---. 2006, \apj, 652, 1348

\bibitem[{{Lazarian} \& {Pogosyan}(2008)}]{LP08}
---. 2008, \apj, 686, 350

\bibitem[{{Lazarian} {et~al.}(2001){Lazarian}, {Pogosyan},
  {V{\'a}zquez-Semadeni}, \& {Pichardo}}]{LazP01}
{Lazarian}, A., {Pogosyan}, D., {V{\'a}zquez-Semadeni}, E., \& {Pichardo}, B.
  2001, \apj, 555, 130

\bibitem[{{Lazarian} \& {Vishniac}(1999)}]{LV99}
{Lazarian}, A., \& {Vishniac}, E.~T. 1999, \apj, 517, 700

\bibitem[{{Lazarian} {et~al.}(2004){Lazarian}, {Vishniac}, \& {Cho}}]{LVC04}
{Lazarian}, A., {Vishniac}, E.~T., \& {Cho}, J. 2004, \apj, 603, 180

\bibitem[{{Lazarian} {et~al.}(2012){Lazarian}, {Vlahos}, {Kowal}, {Yan},
  {Beresnyak}, \& {de Gouveia Dal Pino}}]{Lazssrv12}
{Lazarian}, A., {Vlahos}, L., {Kowal}, G., {Yan}, H., {Beresnyak}, A., \& {de
  Gouveia Dal Pino}, E.~M. 2012, \ssr, 173, 557

\bibitem[{{Lazarian} \& {Yan}(2014)}]{LY14}
{Lazarian}, A., \& {Yan}, H. 2014, \apj, 784, 38

\bibitem[{{Li} \& {Houde}(2008)}]{LH08}
{Li}, H.-b., \& {Houde}, M. 2008, \apj, 677, 1151

\bibitem[{{Lithwick} \& {Goldreich}(2001)}]{LG01}
{Lithwick}, Y., \& {Goldreich}, P. 2001, \apj, 562, 279

\bibitem[{{Lithwick} \& {Goldreich}(2003)}]{LG03}
---. 2003, \apj, 582, 1220

\bibitem[{{Lithwick} {et~al.}(2007){Lithwick}, {Goldreich}, \&
  {Sridhar}}]{LGS07}
{Lithwick}, Y., {Goldreich}, P., \& {Sridhar}, S. 2007, \apj, 655, 269

\bibitem[{{Mac Low}(2002)}]{MacL02}
{Mac Low}, M.-M. 2002, in APS Meeting Abstracts, 1005

\bibitem[{{McKee} \& {Ostriker}(2007)}]{Mckee_Ostriker2007}
{McKee}, C.~F., \& {Ostriker}, E.~C. 2007, \araa, 45, 565

\bibitem[{{Mouschovias}(1987)}]{Mous87}
{Mouschovias}, T.~C. 1987, in NATO ASIC Proc. 210: Physical Processes in
  Interstellar Clouds, ed. G.~E. {Morfill} \& M.~{Scholer}, 453--489

\bibitem[{{Perez} \& {Boldyrev}(2008)}]{PB08}
{Perez}, J.~C., \& {Boldyrev}, S. 2008, \apjl, 672, L61

\bibitem[{{Piddington}(1956)}]{Pidd56}
{Piddington}, J.~H. 1956, \mnras, 116, 314

\bibitem[{{Shu}(1992)}]{Shu92}
{Shu}, F.~H. 1992, {The physics of astrophysics. Volume II: Gas dynamics.}

\bibitem[{{Soler} {et~al.}(2013{\natexlab{a}}){Soler}, {Carbonell}, \&
  {Ballester}}]{Soler13}
{Soler}, R., {Carbonell}, M., \& {Ballester}, J.~L. 2013{\natexlab{a}}, \apjs,
  209, 16

\bibitem[{{Soler} {et~al.}(2013{\natexlab{b}}){Soler}, {Carbonell},
  {Ballester}, \& {Terradas}}]{Sol13}
{Soler}, R., {Carbonell}, M., {Ballester}, J.~L., \& {Terradas}, J.
  2013{\natexlab{b}}, \apj, 767, 171

\bibitem[{{Stone} {et~al.}(1998){Stone}, {Ostriker}, \& {Gammie}}]{Stone98}
{Stone}, J.~M., {Ostriker}, E.~C., \& {Gammie}, C.~F. 1998, \apjl, 508, L99

\bibitem[{{Tilley} \& {Balsara}(2010)}]{TilBal10}
{Tilley}, D.~A., \& {Balsara}, D.~S. 2010, \mnras, 406, 1201

\bibitem[{{Tofflemire} {et~al.}(2011){Tofflemire}, {Burkhart}, \&
  {Lazarian}}]{Toff11}
{Tofflemire}, B.~M., {Burkhart}, B., \& {Lazarian}, A. 2011, \apj, 736, 60

\bibitem[{{Xu} \& {Yan}(2013)}]{XY13}
{Xu}, S., \& {Yan}, H. 2013, \apj, 779, 140

\bibitem[{{Yan}(2015)}]{Yan15}
{Yan}, H. 2015, in Astrophysics and Space Science Library, Vol. 407,
  Astrophysics and Space Science Library, ed. A.~{Lazarian}, E.~M. {de Gouveia
  Dal Pino}, \& C.~{Melioli}, 253

\bibitem[{{Yan} \& {Lazarian}(2002)}]{YL02}
{Yan}, H., \& {Lazarian}, A. 2002, Physical Review Letters, 89, B1102+

\bibitem[{{Yan} \& {Lazarian}(2004)}]{YL04}
---. 2004, \apj, 614, 757

\bibitem[{{Yan} \& {Lazarian}(2008)}]{YL08}
---. 2008, \apj, 673, 942

\bibitem[{{Yan} \& {Lazarian}(2011)}]{YL11}
---. 2011, \apj, 731, 35

\bibitem[{{Zaqarashvili} {et~al.}(2011){Zaqarashvili}, {Khodachenko}, \&
  {Rucker}}]{Zaqa11}
{Zaqarashvili}, T.~V., {Khodachenko}, M.~L., \& {Rucker}, H.~O. 2011, \aap,
  529, A82

\bibitem[{{Zweibel}(2002)}]{Zwei02}
{Zweibel}, E.~G. 2002, \apj, 567, 962

\end{thebibliography}

\end{document}